\documentclass[iop,apj]{emulateapj}
\usepackage{times}
\usepackage{epsfig}
\usepackage{amsmath, amsthm, amssymb}
\usepackage[plainpages=false, colorlinks=true, anchorcolor=blue, linkcolor=blue, citecolor=blue, bookmarks=false]
{hyperref}
\usepackage{color}
\usepackage{microtype}
\usepackage{float}
\usepackage{verbatim}
\usepackage{wrapfig}

\newcommand{\beq}{\begin{equation}}
\newcommand{\eeq}{\end{equation}}
\newcommand{\bea}{\begin{eqnarray}}
\newcommand{\eea}{\end{eqnarray}}
\newcommand{\Msun}{$M_\sun$}
\providecommand{\abs}[1]{\lvert#1\rvert}

\newcommand{\fheat}{$f_\textrm{heat}$}

\newcommand{\subdate}{2013 October 21}
\newcommand{\shortauth}{Couch \& O'Connor}
\newcommand{\slugcom}{Submitted to ApJ on \subdate}
\slugcomment{\slugcom}

\lefthead{\sc \footnotesize \slugcom \hfill \shortauth}
\righthead{\sc \footnotesize \slugcom \hfill \shortauth}

\begin{document}

\title{High-Resolution Three-Dimensional Simulations of Core-Collapse Supernovae in Multiple Progenitors} 
\author{Sean M. Couch\altaffilmark{1,3}}
\author{Evan P. O'Connor\altaffilmark{2}}

\altaffiltext{1}{Flash Center for Computational Science, Department of
  Astronomy \& Astrophysics, University of Chicago, Chicago, IL,
  60637; 
\href{mailto:smc@flash.uchicago.edu}{smc@flash.uchicago.edu}} 

\altaffiltext{2}{Canadian Institute for Theoretical Astrophysics,
  University of Toronto, Toronto, Ontario M5S 3H8, Canada}

\altaffiltext{3}{Hubble Fellow}

\begin{abstract}

Three-dimensional simulations of core-collapse supernovae are granting new insight into the as-yet uncertain mechanism that drives successful explosions.
While there is still debate about whether explosions are obtained more easily in 3D than in 2D, it is undeniable that there exist qualitative and quantitative differences between the results of 3D and 2D simulations.
We present an extensive set of high-resolution one-, two-, and three-dimensional core-collapse supernova simulations with multispecies neutrino leakage carried out in two different progenitors.
Our simulations confirm the results of \citet{Couch:2013fh} indicating that 2D explodes more readily than 3D. 
We argue that this is due to the inadequacies of 2D to accurately capture important aspects of the three-dimensional dynamics.  
We find that without artificially enhancing the neutrino heating rate we do not obtain explosions in 3D.
We examine the development of neutrino-driven convection and the standing accretion shock instability and find that, in separate regimes, either instability can dominate.
We find evidence for growth of the standing accretion shock instability for both  15-\Msun\ and 27-\Msun\ progenitors, however, it is weaker in 3D exploding models.
The growth rate of both instabilities is artificially enhanced along the symmetry axis in 2D as compared with our axis-free 3D Cartesian simulations.
Our work highlights the growing consensus that core-collapse supernovae must be studied in 3D if we hope to solve the mystery of how the explosions are powered.


\end{abstract}

\section{Introduction}
\label{sec:Intro}

Nearly 80 years since \citet{Baade:1934ed} introduced the idea that supernovae represent the formation of a ``body of considerably smaller mass'' from an ordinary star, we still lack a complete understanding of the mechanism that drives these luminous explosions. 
The leading candidate for the core-collapse supernova (CCSN) explosion mechanism has been, for several decades, the neutrino-heating mechanism \citep{Colgate:1966cl, Bethe:1985da}.  
For detailed reviews of the neutrino mechanism see:  \citet{Bethe:1990he}; \citet{Janka:2007cz};  \citet{Janka:2012cb}; and \citet{Burrows:2013hp}.
The collapse of a massive stellar core into a proto-neutron star (PNS) of supra-nuclear density liberates an enormous sum of gravitational binding energy, about 2-3$\times10^{53}$ erg.
Over the course of 10s of seconds following core collapse, this energy will be radiated away as neutrinos as the PNS cools and contracts \citep{Burrows:1986kx}.
The neutrino mechanism hinges on a small fraction of this energy being deposited in the hot, dense accretion flow bounded from below by the PNS and from above by the shock launched upon core-bounce.  
This turns out to be a tall order; detailed 1D simulations of the neutrino mechanism \citep[e.g.,][]{Liebendorfer:2001fl} fail for all but the lowest-mass progenitors \citep[e.g.,][]{Kitaura:2006gm}.  

The situation is better in 2D wherein a number of simulations have found successful explosions \citep[e.g.,][]{ {Marek:2009kc}, {Muller:2012gd}, {Muller:2012kq}, {Bruenn:2013es}}.
These two-dimensional simulations have shown the critically important role that non-spherically-symmetric phenomena such as convection and turbulence can have on the development of a successful CCSN.
While these intrinsically multi-dimensional fluid motions are present and boost the effectiveness of the neutrino mechanism in two-dimensional simulations, there is still a forced symmetry present.
Just as spherically-symmetric simulations are incorrect and unphysical because of the forced symmetry, so too are 2D simulations.
Well-known artifacts that arise from the imposition of 2D symmetry include the amplification of hydrodynamic instability growth along the axis and the ``inverse'' turbulent energy cascade \citep[][and Section \ref{sec:turb}]{Kraichnan:1967jk}, which erroneously pumps turbulent energy to large scales in 2D.

The advent of fully three-dimensional simulations of core-collapse supernovae has brought with it important insights into this complex problem. 
By being able to simulate CCSN with no forced symmetries, we can begin to evaluate the deficiencies associated with such approximations.  
Three-dimensional hydrodynamic simulations are computationally very expensive and three-dimensional neutrino-radiation hydrodynamic simulations are even more challenging.  
To reduce computational cost, a number of studies comparing two- and three-dimensional CCSN simulations used parameterized local heating and cooling to model the effect of neutrinos \citep{Nordhaus:2010ct, Hanke:2012dx, Burrows:2012gc, Murphy:2013eg, Dolence:2013iw, Couch:2013fh}.
These studies, which all used the 15-\Msun\ progenitor model of \citet{Woosley:1995jn}, generally found that the standing accretion shock instability (SASI) was sub-dominant to neutrino-driven convection, particularly for energetic explosions.
Similar evidence for reduced SASI activity in successful 3D explosions is found in the general-relativistic (GR) simulations \citet{Ott:2013gz}.
Although, \citet{Muller:2012kq} find strong SASI activity in the successful explosion of a 27-\Msun\ star in 2D GR neutrino transport simulations \citep[see also,][]{Fernandez:2013um}.

Important discrepancies arose in the results of parameterized studies as to whether 3D explosions were obtained more easily than 2D.
The Princeton group \citep{Nordhaus:2010ct, Burrows:2012gc, Murphy:2013eg, Dolence:2013iw} found that going to 3D reduced the critical luminosity for explosion by as much as 25\% \citep{Nordhaus:2010ct}, though their later simulations showed a somewhat more mild favorability for explosion in 3D \citep{Dolence:2013iw}.
\footnote{The cause of the disparate results of \citet{Nordhaus:2010ct} and \citet{Dolence:2013iw} has been attributed to significant ``inaccuracies'' in the original implementation of the 3D monopole gravity solver in CASTRO \citep{Burrows:2012gc}.  \citeauthor{Dolence:2013iw} report that these inaccuracies have been eliminated.}
The Garching group was unable to reproduce this result \citep{Hanke:2012dx}.
Instead, they found that there was little difference between the critical luminosities between 2D and 3D, though their results showed strong resolution dependence.
\citet{Couch:2013fh} came to yet a third conclusion, finding that explosions are {\it harder} to obtain in 3D than in 2D, i.e. the critical luminosity is higher in 3D.
When considering only the highest-resolution results of \citet{Hanke:2012dx}, however, there is arguably good agreement between their results and those of \citet{Couch:2013fh}.
More recently, \citet{Hanke:2013kf} performed two- and three-dimensional simulations with sophisticated neutrino transport and were the first to isolate the standing accretion shock instability (SASI) in 3D neutrino transport simulations.
Similar to \citet{Couch:2013fh}, they find that with all other physics and computational techniques held fixed, 2D simulations explode more readily than their 3D counterparts.
\citet{Takiwaki:2013ui} also find that 2D explodes with more vigor than 3D for an 11.2-\Msun progenitor, though their earlier, lower-resolution work shows just the opposite \citep{Takiwaki:2012ck}.

In this article, we present a detailed analysis and comparison of 1D, 2D, and 3D CCSN simulations covering the first $\sim400$ ms after bounce in 15-\Msun and 27-\Msun progenitors.
Our simulations employ high resolution and an efficient multispecies neutrino leakage scheme that accurately captures the {\it global} effects of neutrino transport.
We find that: 1) in all cases 2D simulations explode more vigorously and more readily than 3D simulations; 
2) both the SASI and neutrino-driven convection are present in both progenitors, and both may be dominant under certain conditions in our simulations.  
In 3D simulations that result in explosions, the strength of the SASI is reduced as compared with failed explosions and is generally sub-dominant to neutrino-driven convection, in agreement with \citet{Ott:2013gz}.
At late times in 3D failed explosions, following a period of marked shock retraction, the spiral-mode of the SASI becomes strong and dominant.
3) Without additional artificially-enhanced neutrino heating, our 3D simulations fail to explode while comparable 2D simulations explode robustly.
This is apparently in agreement with the full-transport simulations of \citet{Hanke:2013kf}, thus supporting the efficacy of our neutrino treatment to mimic the results of full transport.

This paper is organized as follows.  
In Section \ref{sec:methods} we describe our numerical approach, giving details of our hydrodynamic solver, grid structure, and self-gravity treatment.  
We also describe the salient aspects of our neutrino leakage scheme and the progenitors used.
In Section \ref{sec:results} we present and analyze our results.
We apply rigorous diagnostics to our simulations to quantify phenomena such as neutrino-driven convection and the SASI, as well as turbulence.
In Section \ref{sec:conclusion} we discuss the implications of our results within the context of other 3D simulations and conclude.

\section{Computational Methods}
\label{sec:methods}

\subsection{Hydrodynamics, Grid, and Gravity}

Our basic hydrodynamic scheme is the same as that described in detail by \citet{Couch:2013fh}.
Our core-collapse application is built in the FLASH simulation framework \citep{{Fryxell:2000em}, Dubey:2009wz, Lee:2013flash}.\footnote{Available at \url{http://flash.uchicago.edu}.} 
We solve the Eulerian equations of hydrodynamics on an adaptive refinement mesh in a directionally-unsplit fashion \citep{Lee:2009kq,Lee:2013cd}.
We use third-order piecewise parabolic method \citep[PPM,][]{Colella:1984cg} spatial reconstruction and second-order temporal integration via a predictor-corrector approach.
We employ a hybrid Riemann solver for interface flux calculations that uses HLLC in smooth flow and the more diffusive HLLE within shocks. 
We use a {\tt minmod} slope limiter for self-steepening characteristic wave families, such as pressure, and a monotonized central difference ({\tt mc}) limiter for all others.

The system of hydrodynamic equations is closed using an equation of state (EOS) that accounts for thermodynamic contributions from baryons, photons, and pairs of arbitrary relativity and degeneracy.
\footnote{Our tabular EOS data, and routines for reading and interpolating the data, are available at \url{http://stellarcollapse.org}.
This EOS is also now available as part of the public FLASH4 release distribution.}
At high densities, we use the EOS model of \citet{Lattimer:1991fz} with incompressibility parameter $K = 220$ MeV (hereafter LS220), as is currently favored by experimental and observational constraints \citep[e.g.,][]{Hempel:2012bh, {Steiner:2013hi}}.  
In numerical simulations that isolate the hydrodynamic impact of different high-density EOS models  \citep{Couch:2013df}, LS220 is found to be less favorable to neutrino-driven explosions than the softer $K=180$ MeV variant of the Lattimer \& Swesty EOS (LS180) and more conducive to neutrino-driven explosions than the commonly used model of \citet{Shen:1998kx}.  
Similar EOS dependence is reported by \citet{Suwa:2012ug}. 
\citet{OConnor:2011hk} also find differences between LS180 and LS220 in terms of the critical neutrino heating efficiency required to obtain explosion, as well as in black hole formation time in failed explosions. 
The EOS-dependence in these works, however, may be somewhat exaggerated as compared with more realistic, multidimensional simulations \citep[see, e.g.,][]{Muller:2012kq}.

Adaptive mesh refinement (AMR) is achieved by use of the oct-tree-based PARAMESH library \citep{MacNeice:2000fc}.
We use 3D Cartesian, 2D cylindrical, and 1D spherical coordinate geometries.  
The coarsest level of refinement in our simulations has a grid spacing in each linear direction of $\Delta x_{\rm max}^i = 250$ km, and the finest level of refinement yields $\Delta x_{\rm min}^i = 0.49$ km.
We apply a radial limiter to the maximum allowed refinement that results in a pseudo-logarithmic grid spacing.  
The limiter enforces a typical grid spacing $\Delta x^i (r) \sim \alpha r$, where $r$ is the spherical radius of the zone and $\alpha$ is a number setting the resolution scale.
This approach results in nested, quasi-spherical levels of refinement.
For our fiducial resolution, we set $\alpha = 0.75\%$.
For this $\alpha$, the first forced jump in refinement, that increases $\Delta x^i$ to 0.98 km, occurs at $r \sim 100$ km.
The second refinement decrement occurs at $r \sim 200$ km, the third at $r \sim 400$ km, and so on.

We account for self-gravity of the system by solving Poisson's equation using the new spherical multipole method of \citet{Couch:2013ws}.  
This approach is optimized for arbitrary non-spherical coordinate systems and has been shown to be highly-accurate for the CCSN problem.
The multipole solver of \citeauthor{Couch:2013ws} gives extremely good momentum conservation, even in cases of asymmetric explosions where the PNS receives a substantial kick.
In such cases the center-of-mass position is preserved to better precision than one zone spacing, while the center of the PNS moves by several tens of zone spacings.
In the present simulations, we use a maximum Legendre order, $\ell_{\rm max}$, in the multipole expansion of 16 in 2D, and 8 in 3D, including all non-zero $m$ terms.

\subsection{Neutrino Physics}

Our multidimensional neutrino treatment is the same scheme employed by \citet{Ott:2013gz}, though re-implemented in the FLASH framework.
During the collapse phase prior to core bounce, we employ the parameterized deleptonization scheme of \citet{Liebendorfer:2005ft}.  
This is scheme is based on polynomial fits of the electron fraction as a function of density from 1D GR CCSN simulations including full neutrino transport.
This parametric deleptonization approach is only appropriate before bounce.
Following core bounce, we switch from the \citeauthor{Liebendorfer:2005ft} approach to a more self-consistent method for treating the neutrino effects.
Including neutrino transport in multidimensional simulations and maintaining sufficiently high spatial resolution is still prohibitively computationally expensive.
For this study, we choose to simulate high spatial resolution and approximate the neutrino treatment by incorporating a simple and efficient neutrino leakage scheme that reproduces the global neutrino effects on the matter: postbounce deleptonization and cooling of the postshock material, PNS contraction, and neutrino heating to name a few.

Our leakage scheme determines the rate of energy ($Q_\nu = d\epsilon_\nu/dt$) and lepton ($R_\nu = d{(Y_e)}_\nu/dt$) emission from the matter via an interpolation between two limiting regimes.
In the optically thick regime, neutrinos will escape, or \emph{leak}, to infinity on a diffusion time scale. 
We use this timescale to set the diffusion emission rates ($Q_{\nu, \mathrm{diff}}$ and $R_{\nu,\mathrm{diff}}$).
Neutrinos in optically thin regions will stream away to infinity at a rate equal to the actual emission rate ($Q_{\nu, \mathrm{free}}$ and $R_{\nu, \mathrm{free}}$).
In the optically thick region, the emission rate is much larger than the diffusion rate, as it is almost completely balanced by absorption.
Likewise, in the optically thin region, the diffusion rate is much larger than the free emission rate since it grows as the optical depth goes to zero.
This leads us to the following interpolation for the effective emission rates, 
\begin{equation}
  S_{\nu,\mathrm{eff}} = \frac{S_{\nu,\mathrm{free}} \times
    S_{\nu,\mathrm{diff}}}{S_{\nu,\mathrm{free}} + S_{\nu,\mathrm{diff}}}\,,
\end{equation}
where $S_\nu = \{Q_\nu,R_\nu\}$ is either the energy emission $Q_\nu$ or the lepton number emission $R_\nu$ rate.
To capture the effect of neutrino heating in the postshock region, we also calculate additional factors, $Q_{\nu,\mathrm{heat}}$ and $R_{\nu,\mathrm{heat}}$, and subtract them from $Q_{\nu,\mathrm{eff}}$ and $R_{\nu,\mathrm{eff}}$, respectively.
Below we briefly describe the process to determine each of these $Q_\nu$ and $R_\nu$ terms.
Full details and expressions can be found in \citet{OConnor:2010bi,Rosswog:2003fu, Ruffert:1996te}, and the simulation code.
\footnote{The kernel routines of our leakage calculation are available at \url{http://stellarcollapse.org/ottetal2013} and the FLASH-specific implementation will be part of the next major FLASH release version (v4.1).}

\emph{$Q_{\nu,\mathrm{free}}$ and $R_{\nu,\mathrm{free}}$}: The free emission rates for electron neutrinos and antineutrinos are determined from electron and positron capture on protons and neutrons, respectively.
Additionally, thermal processes that produce pairs of neutrinos of all types are included.
Specifically, we include electron-positron annihilation, nucleon-nucleon bremsstrahlung, and plasmon decay.
We note that the most dominant thermal neutrino-pair production process is electron-positron annihilation.
These free emission rates are all calculated based on the local thermodynamic properties of the matter.

\emph{$Q_{\nu,\mathrm{diff}}$ and $R_{\nu,\mathrm{diff}}$}: The diffusive emission rates require knowledge of how optically deep (denoted by $\chi_{\nu_i}$ in our leakage scheme\footnote{As a clarifying note, $\chi_{\nu_i}$ is an energy-independent optical depth, see \citet{Rosswog:2003fu} for full details.}) particular hydrodynamic grid zones are with respect to infinity.
Using this, and an estimate of the local opacity ($\zeta_{\nu_i}$ in our scheme), it is possible to determine the rate of energy and lepton number leakage to infinity.
Obtaining $\chi_{\nu_i}$ in a spherically symmetric setup is trivial, one simply integrates the opacity $\zeta_{\nu_i}$ from infinity to the desired radius.
The situation is more complicated in multidimensional simulations as the true optical depth would require extensive calculation for each grid point.
Instead, we estimate $\chi_{\nu_i}$ by performing the leakage calculation on a set of radial rays and interpolating back to the hydrodynamic grid.
These rays are located on a spherical grid and spaced $\sim5^\circ$ apart (37 polar rays and 75 azimuthal rays).
Each ray is composed of 1000 radial zones extending out to 3000\,km.
The radial ray sampling is uniform at 0.5 km up to a radius of 150 km, then logarithmic.
The coordinate origin of the rays is adjusted every time step to follow the center of the PNS, defined as the squared-density-weighted mean location \citep[see][]{Couch:2013ws}.
For calculating the opacities, we include elastic scattering of neutrinos of all species on neutrons and protons, as well as absorption opacities for electron neutrinos on neutrons and electron antineutrino absorption on protons.

\emph{$Q_{\nu,\mathrm{heat}}$ and $R_{\nu,\mathrm{heat}}$}: Neutrino charged-current heating and the development of the gain region (where neutrino heating dominates neutrino cooling) is paramount to the neutrino driven explosion mechanism for core-collapse supernovae.  
The gain region naturally arises when neutrinos are transported through the matter, however, the leakage scheme described above does not produce neutrino heating and therefore we must explicitly include it.  
The strength of the neutrino heating in the postshock region depends on several physical quantities.  
It scales with the energy flux of electron neutrinos and antineutrinos passing through the matter; the forward-peakedness of the neutrino and antineutrino distribution functions; the charged-current cross section of the neutrinos and antineutrinos with the surrounding neutrons and protons; and the number density of these target nucleons. 
For the neutrino heating contribution to our leakage scheme we use the following expression for both $\nu_e$ and $\bar{\nu}_e$ \citep{Janka:2001fp}, 
\begin{equation}
 Q_{\nu_i} = f_\mathrm{heat}\frac{L_{\nu_i}(r)}{4\pi r^2} \left \langle
 \frac{1}{F_{\nu_i}} \right \rangle \sigma_{\nu_i} \frac{\rho X_{(n/p)}}{m_\mathrm{amu}} e^{-2\tau_{\nu_i}}\,.
 \label{eq:heating_rate}
\end{equation}
In this expression, $f_\mathrm{heat}$ is a tunable factor that allows us to increase or decrease the amount of neutrino heating.
The nominal value of $f_\mathrm{heat}$ is 1.0, such that values greater (smaller) than unity reflect artificial enhanced (diminished) heating.
$L_{\nu_i}(r) / 4\pi r^2$ is the energy flux of neutrinos impingent on the material from below. 
As this is a non-local quantity, we solve for it on our grid of rays and interpolate back to the hydrodynamic grid. 
On a spherical ray, the luminosity at any given radius is simply the integrated $Q_{\nu_i}$ from the center to that radius.
$\rho X_{(n/p)} / m_\mathrm{amu}$ is the number density of absorbers.  
$X_n$ and $X_p$ are the mass fractions of neutrons and proton, respectively.  
$X_n$ is used for calculating the neutrino heating from electron neutrino, while $X_p$ is for the electron antineutrino reaction. 
The $\exp{(-2\tau_{\nu_i})}$ term is used to suppress heating in the diffusion-dominated regime where charged-current absorption is already taken into account.
$\tau_{\nu_i}$ is an optical depth computed on the rays and interpolated to the hydrodynamic grid.  
$\tau_{\nu_i}$ plays a similar role as $\chi_{\nu_i}$, but for empirical reasons, and following the original implementation of this leakage scheme, $\tau_{\nu_i}$ is computed following \citet{Ruffert:1996te}.  
We also use $\tau_{\nu_i}$ to determine the forward-peakedness of the neutrino radiation field, $\langle F_{\nu_i}^{-1} \rangle$.  
As in \citet{OConnor:2010bi}, we approximate this expression as $4.275\tau_{\nu_i}+1.15$ which is a fit from two-dimensional multi-angle neutrino transport simulations of \citet{Ott:2008gn}. 
In Equation (\ref{eq:heating_rate}), $\sigma_{\nu_i}$ is the charged-current cross section for electron neutrino/antineutrino absorption on neutrons/protons.  
We use the following definitions,
\begin{eqnarray}
  \sigma_{\nu_e} = \frac{1+3g_A^2}{4}\sigma_0 \frac{\langle
  \epsilon^2\rangle^\mathrm{ns}_{\nu_e}}{m_e^2 c^4} B_{\nu_e}\,,\label{eq:sig_nue}\\ 
\sigma_{\bar{\nu}_e} = \frac{1+3g_A^2}{4}\sigma_0 \frac{\langle
  \epsilon^2\rangle^\mathrm{ns}_{\bar{\nu}_e}}{m_e^2 c^4} B_{\bar{\nu}_e}\,,
\label{eq:sig_anue}
\end{eqnarray}
where $g_A = -1.25$, $\sigma_0 = 1.76\times10^{-44}$\,cm$^2$, $\langle \epsilon^2 \rangle_{\nu_i}^\mathrm{ns}$ is the mean squared energy of the neutrinos at the neutrinosphere (where $\tau_i = 2/3$), and $B_{\nu_i}$ is the final state lepton blocking function.
We interpolate the mean squared energy at the neutrinosphere from the grid of rays to our computational domain. 
For complete expressions and more details of the implementation of the heating rates we refer the reader to \citet{OConnor:2010bi}.

\subsection{Progenitors and Initial Conditions}

We use two progenitor stars in our study: the 15-$M_\sun$ model  (s15) of \citet[][hereafter WH07]{Woosley:2007bd} and the 27-$M_\sun$ model (s27) of \citet*[][hereafter WHW02]{Woosley:2002ck}.  
\citet{OConnor:2011hk} suggest the compactness parameter, $\xi_{M} = M/M_\odot [r(M)/1000\ {\rm km}]^{-1} \rvert_{t=t_{\rm bounce}}$, where $r(M)$ is the radius enclosing a mass of $M$, as a useful parameter for estimating the likelihood of explosion by the neutrino mechanism of a given progenitor. 
For s15 we measure a bounce-time compactness of $\xi_{2.5} = 0.179$, and for s27 we find $\xi_{2.5} = 0.233$, in good agreement with the calculations of \citet{OConnor:2011hk} and \citet{Ugliano:2012cr}, respectively.
Both of these systematic studies predict explosions for these progenitors.
All of our multidimensional simulations are initialized from 1D simulations at 2 ms postbounce.
For both progenitors we use the G15 collapse-phase deleptonization parameters of \citet{Liebendorfer:2005ft}.
We measure collapse times of 249 ms for s15 and 297 ms for s27.
\citet{Ott:2013gz}, in their 3D GR simulations using LS220 and the pre-collapse \citet{{Liebendorfer:2005ft}} deleptonization, measured a collapse time for s27 of 299 ms, in remarkable agreement with our Newtonian calculations. 

All of our 3D simulations were run on the BlueGene/Q {\it Mira} at Argonne Leadership Computing Facility.
The cost of our 3D simulations was $\sim$3000 core-hours per millisecond of postbounce evolution.

\section{Results}
\label{sec:results}

\subsection{Overview and Explosion Criteria}

\begin{deluxetable}{cccccc}
\tablecolumns{6}
\tabletypesize{\scriptsize}
\tablecaption{
Basic simulation parameters and results. 
\label{table:results}
}
\tablewidth{0pt}
\tablehead{
\colhead{Progenitor} &
\colhead{$\alpha$ [\%]\tablenotemark{a}} &
\colhead{$f_{\textrm{heat}}$\tablenotemark{b}} &
\colhead{$t_{400}\ [\textrm{ms}]$\tablenotemark{c}} &
\colhead{$\bar{\eta}_{\textrm{heat}}$\tablenotemark{d}} &
\colhead{$E_{\textrm{diag}}\ [B]$\tablenotemark{e}} 
}
\startdata 
\cutinhead{1D\tablenotemark{f}}
s15 & 0.75 & 1.40 & 382.8   & 0.228 & 0.255   \\
s27 & 0.75 & 1.45 & 338.4   & 0.171 & 0.167   \\
\cutinhead{2D}
s15 & 0.75 & 0.90 & \nodata & 0.043 & \nodata \\
s15 & 0.75 & 0.95 & 271.7   & 0.111 & 0.099   \\
s15 & 0.75 & 1.00 & 250.4   & 0.125 & 0.122   \\
s15 & 0.75 & 1.05 & 213.7   & 0.141 & 0.099   \\
s15 & 1.40 & 0.90 & \nodata & 0.056 & \nodata \\
s15 & 1.40 & 0.95 & 340.6   & 0.102 & 0.076   \\
s15 & 1.40 & 1.00 & 275.8   & 0.130 & 0.132   \\
s15 & 1.40 & 1.05 & 231.9   & 0.143 & 0.137   \\
s27 & 0.75 & 0.90 & \nodata & 0.038 & \nodata \\
s27 & 0.75 & 0.95 & 432.8   & 0.070 & 0.020   \\
s27 & 0.75 & 1.00 & 226.1   & 0.094 & 0.027   \\
s27 & 0.75 & 1.05 & 190.9   & 0.115 & 0.026   \\
s27 & 1.40 & 0.90 & \nodata & 0.041 & \nodata \\
s27 & 1.40 & 0.95 & 267.4   & 0.088 & 0.029   \\
s27 & 1.40 & 1.00 & 209.5   & 0.100 & 0.023   \\
s27 & 1.40 & 1.05 & 185.0   & 0.117 & 0.041   \\
\cutinhead{3D}
s15 & 0.75 & 1.00 & \nodata & 0.062 & \nodata \\
s15 & 0.75 & 1.05 & 272.8   & 0.109 & 0.048   \\
s15 & 1.40 & 1.00 & \nodata & 0.062 & \nodata \\
s15 & 1.40 & 1.05 & 250.2   & 0.109 & 0.057   \\  
s27 & 0.75 & 1.00 & \nodata & 0.048 & \nodata \\
s27 & 0.75 & 1.05 & 319.9   & 0.076 & 0.009   \\
s27 & 1.40 & 1.00 & \nodata & 0.047 & \nodata \\
s27 & 1.40 & 1.05 & 269.0   & 0.081 & 0.019  
\enddata
\tablenotetext{a}{Effective angular and radial resolution as a fraction of radial coordinate.  All simulations have a finest grid spacing of 0.49 km.}
\tablenotetext{b}{Neutrino heating factor [see Equation (\ref{eq:heating_rate})].}
\tablenotetext{c}{Postbounce time at which the average shock radius exceeds 400 km.}
\tablenotetext{d}{Time-averaged heating efficiency.}
\tablenotetext{e}{Diagnostic explosion energy defined as the peak value of the sum of kinetic, internal, and gravitational binding energies where that sum is positive, in units of $B=10^{51}$ erg.}
\tablenotetext{f}{For the sake of brevity, we list only the 1D models that explode.}
\end{deluxetable}

\begin{figure*}[htb]
\centering
\begin{tabular}{cc}
  \includegraphics[width=3.4in,trim= 0in 0.15in 0.3in 0.35in,clip]{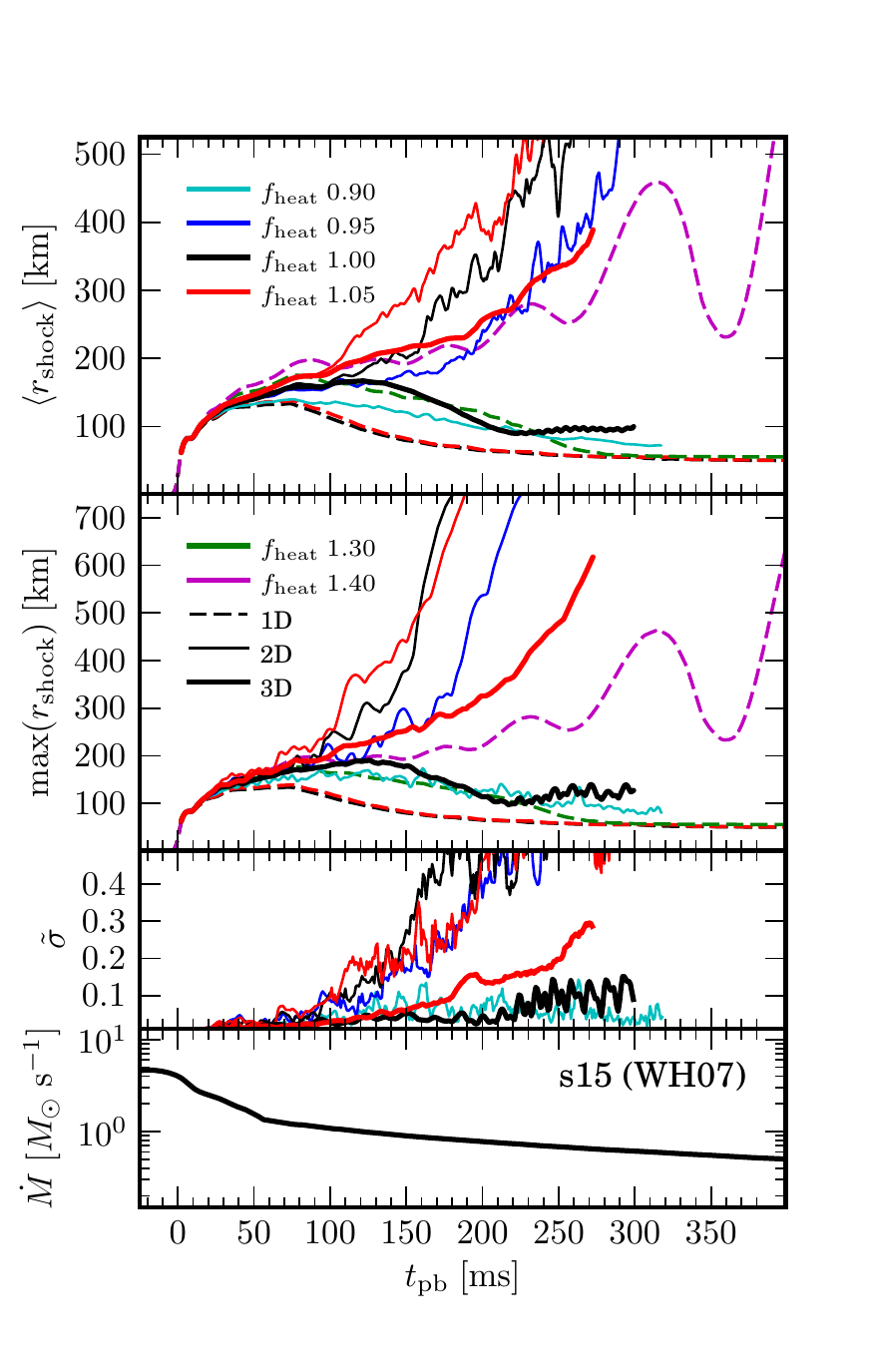} &
  \includegraphics[width=3.4in,trim= 0in 0.15in 0.3in 0.35in,clip]{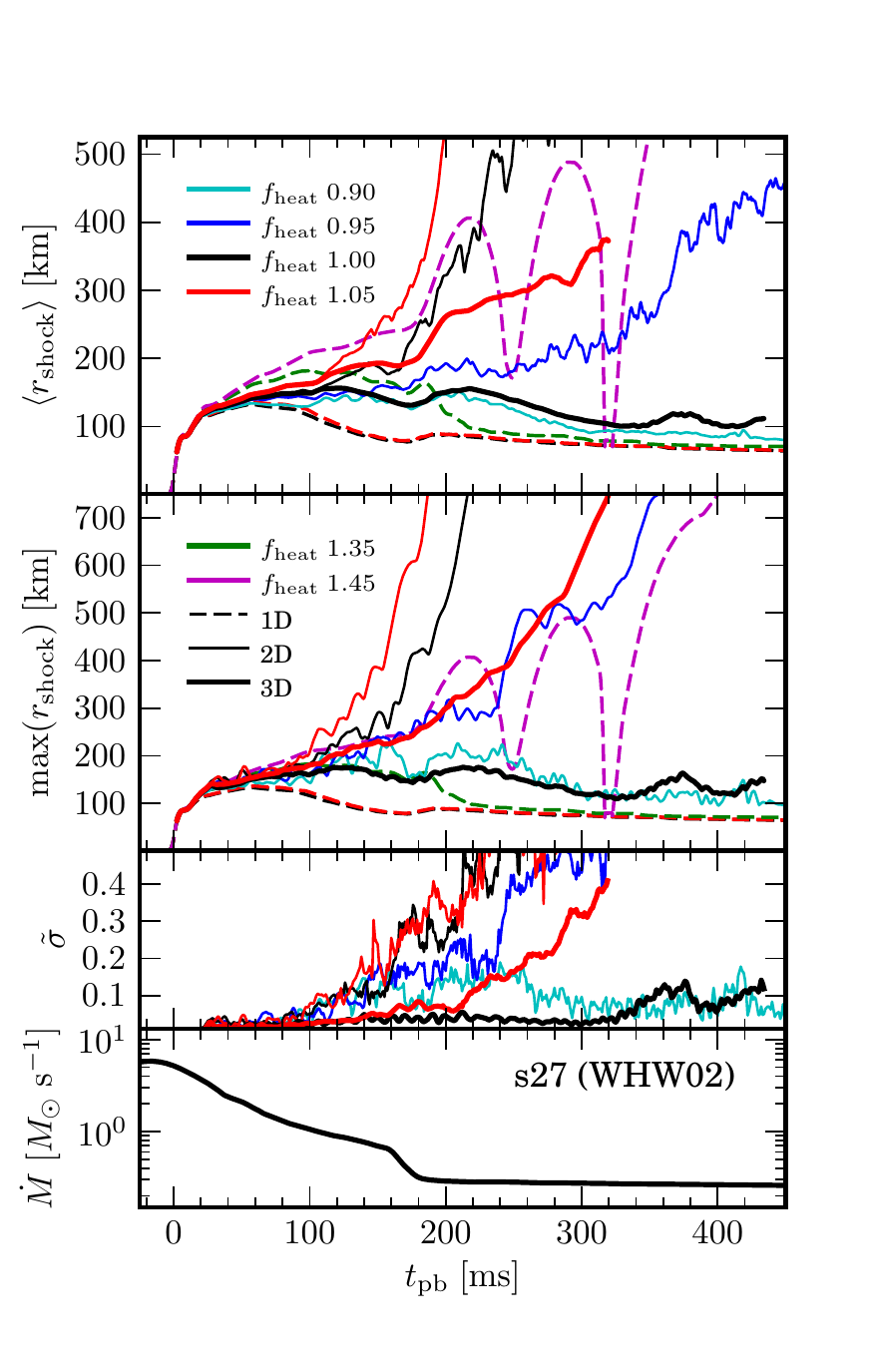}
\end{tabular}
\caption{
  Shock diagnostics for s15 (left) and s27 (right).  
  The top panels display the average shock radii for simulations in 1D, 2D, and 3D for various heat factors, $f_{\rm heat}$.
  The second panels display the maximum shock radii for the same.
  The third panels show the standard deviation of the shock radius, weighted by the average shock radius, $\tilde{\sigma} = \langle r_{\rm shock} \rangle^{-1} [(4 \pi)^{-1} \int d\Omega (r_{\rm shock} - \langle r_{\rm shock} \rangle)^2]^{1/2}$.  
  The bottom panels show the mass accretion rates for each progenitor, evaluated at a fixed radius of 500 km.
  Explosion occurs for all 2D simulations with $f_{\rm heat} \ge 0.95$ while in 3D a heat factor of 1.05 is required to achieve explosion in both s15 and s27.
  Much higher heat factors are required for explosion in 1D, as shown, and s27 requires a slightly higher critical heat factor (1.45) than s15 (1.40).
  The shock structure in 2D is universally more asymmetric, as evidenced by $\tilde{\sigma}$.
  For 3D, the normalized standard deviations of the shock remain relatively small until either an explosion sets in or the SASI develops at late times.
  The periodic oscillations that typify the SASI are evident in the 3D failed explosions, particularly so in s15.
  The relatively higher frequency and amplitude of the SASI oscillations in s15 are due to the greater mass accretion rate at late times as compared with s27 \citep[c.f.][]{Foglizzo:2007cq}.
}
\label{fig:rshock}
\end{figure*}

\begin{figure*}[htb]
\centering
\begin{tabular}{ccc}
  \includegraphics[width=2.3in,trim= 0.3in 0.2in 2.2in .5in,clip] {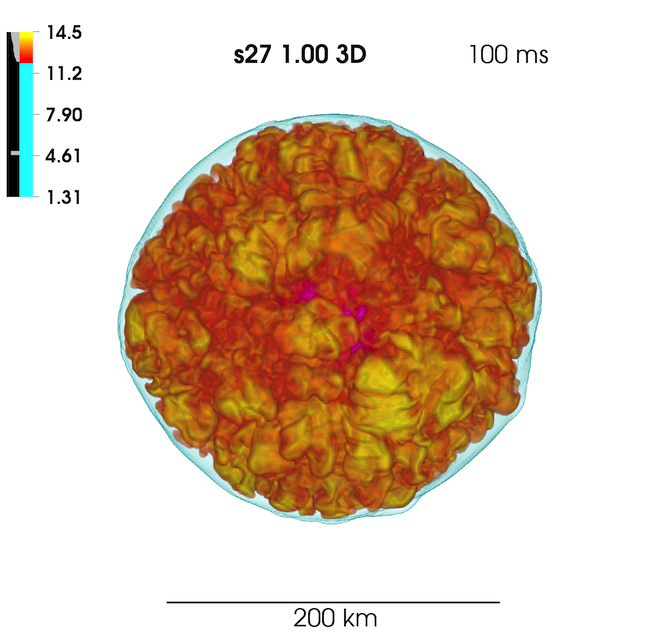} &
  \includegraphics[width=2.3in,trim= 0.3in 0.2in 2.2in .5in,clip] {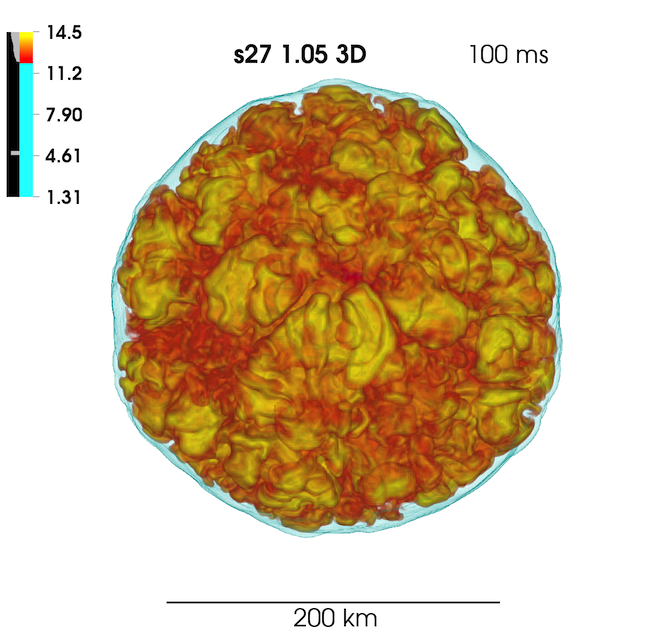} &
  \includegraphics[width=2.3in,trim= 0.3in 0.2in 2.2in .5in,clip] {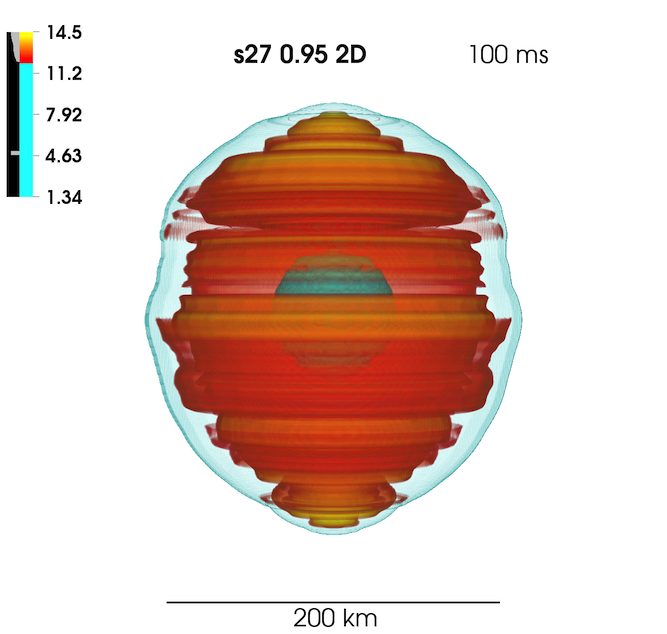} \\
  \includegraphics[width=2.3in,trim= 0.3in 0.2in 2.2in .5in,clip] {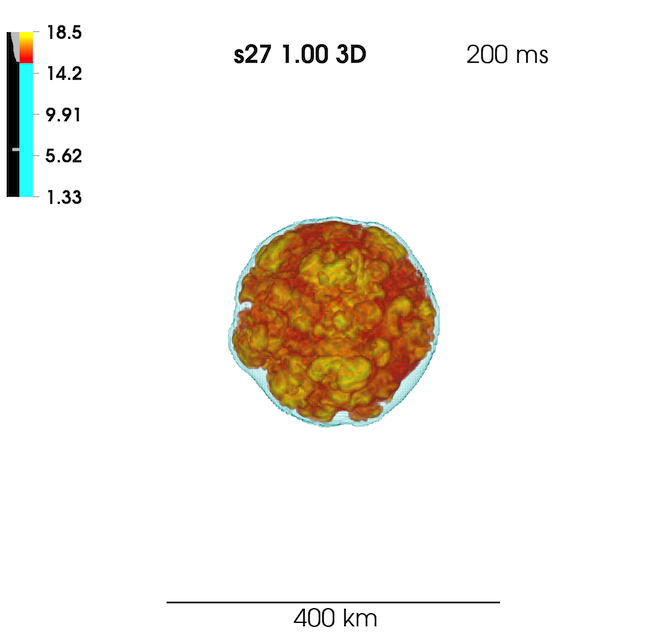} &
  \includegraphics[width=2.3in,trim= 0.3in 0.2in 2.2in .5in,clip] {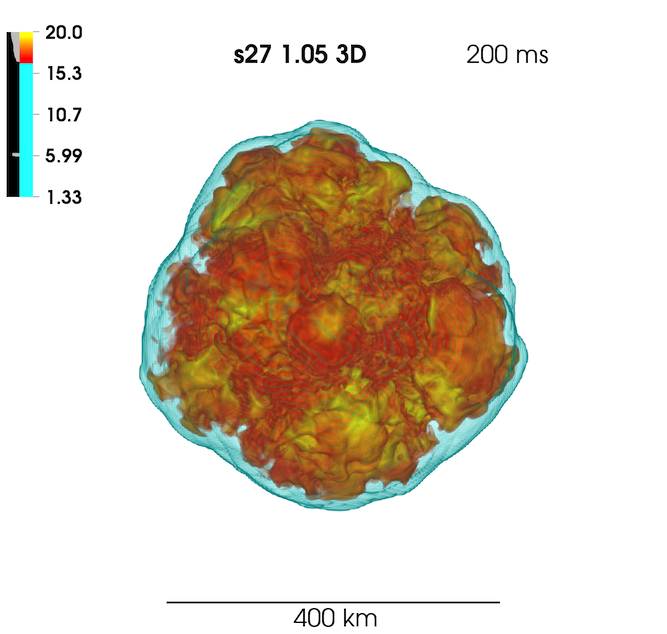} &
  \includegraphics[width=2.3in,trim= 0.3in 0.2in 2.2in .5in,clip] {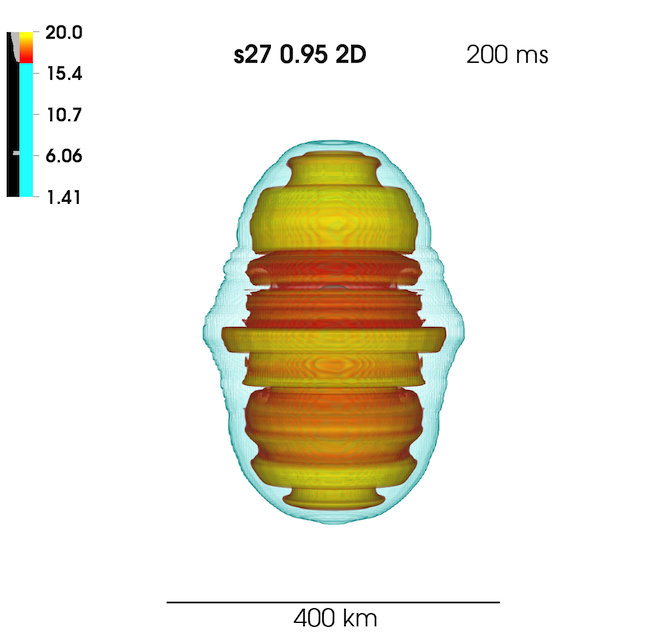} \\
  \includegraphics[width=2.3in,trim= 0.3in 0.2in 2.2in .5in,clip] {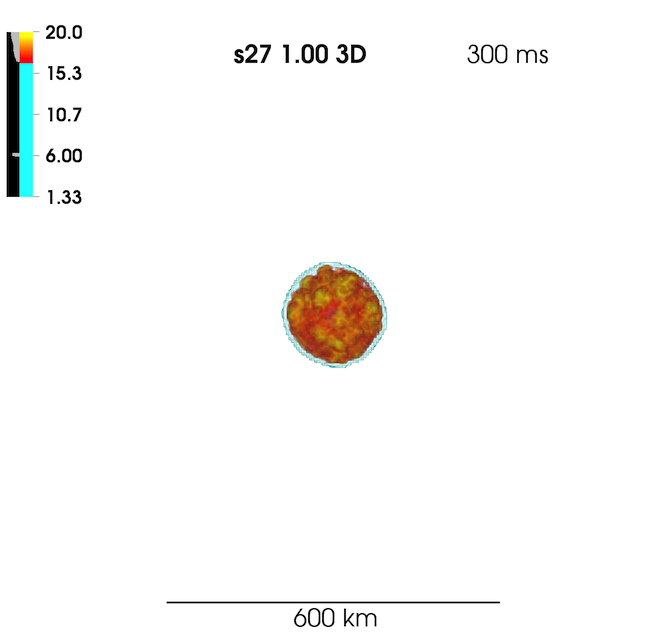} &
  \includegraphics[width=2.3in,trim= 0.3in 0.2in 2.2in .5in,clip] {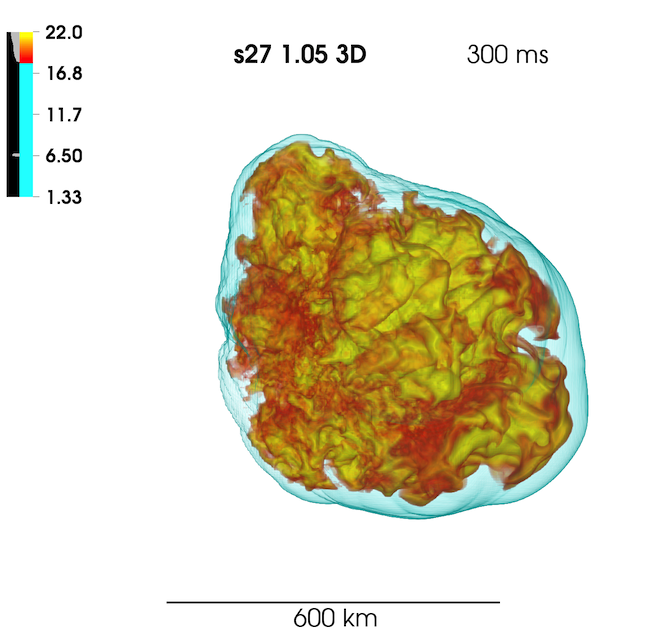} &
  \includegraphics[width=2.3in,trim= 0.3in 0.2in 2.2in .5in,clip] {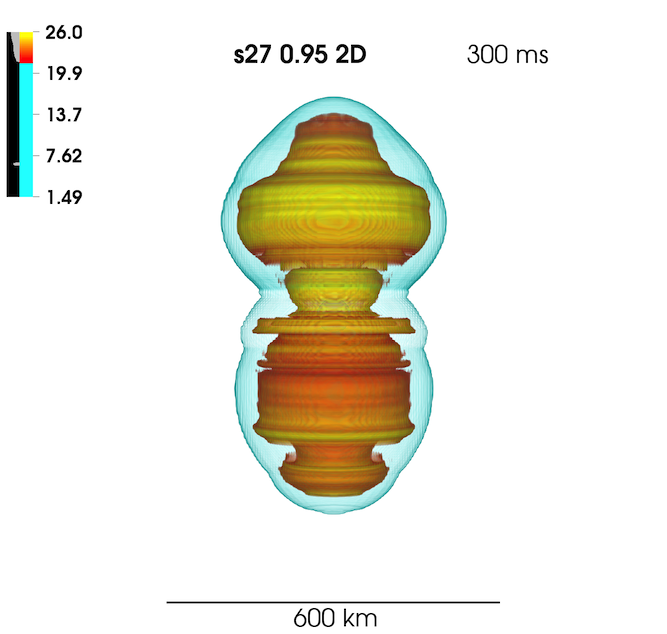}
\end{tabular}
\caption{
  Volume renderings of entropy, in $k_B$ baryon$^{-1}$, for three different s27 simulations at three different postbounce times.
  The left column shows s27 $f_{\rm heat}$ 1.00 3D, the middle column shows s27 $f_{\rm heat}$ 1.05 3D, and the right column shows s27 $f_{\rm heat}$ 0.95 2D.
  The 2D simulation data are wrapped around the axis, reflecting the assumption of axisymmetry made in 2D simulations.
  Time increases from top to bottom.
  The colormap and transfer function are indicated in the top left of each panel, and the postbounce time is displayed in the top right.
  The spatial scale also increases from top to bottom.
  The shock surface is visible in blue.
  High-entropy buoyant convective plumes are evident.
  Model s27 $f_{\rm heat}$ 1.00 3D does not explode and the shock recedes on these time scales.
  Model s27 $f_{\rm heat}$ 1.05 3D is in the process of exploding.
  The 2D simulation clearly presents greater explosion asymmetry, in a characteristic bipolar fashion.
  At these stages of evolution the failed explosion case, s27 $f_{\rm heat}$ 1.00 3D, does not show evidence for substantial SASI development.
  The SASI in this model grows to large amplitudes only after 300 ms postbounce.
} 
\label{fig:volRends}
\end{figure*}

\begin{figure*}[htb]
\centering
\begin{tabular}{cc}
  \includegraphics[width=3.4in,trim= 0in 0.1in 0.3in 0.35in,clip]{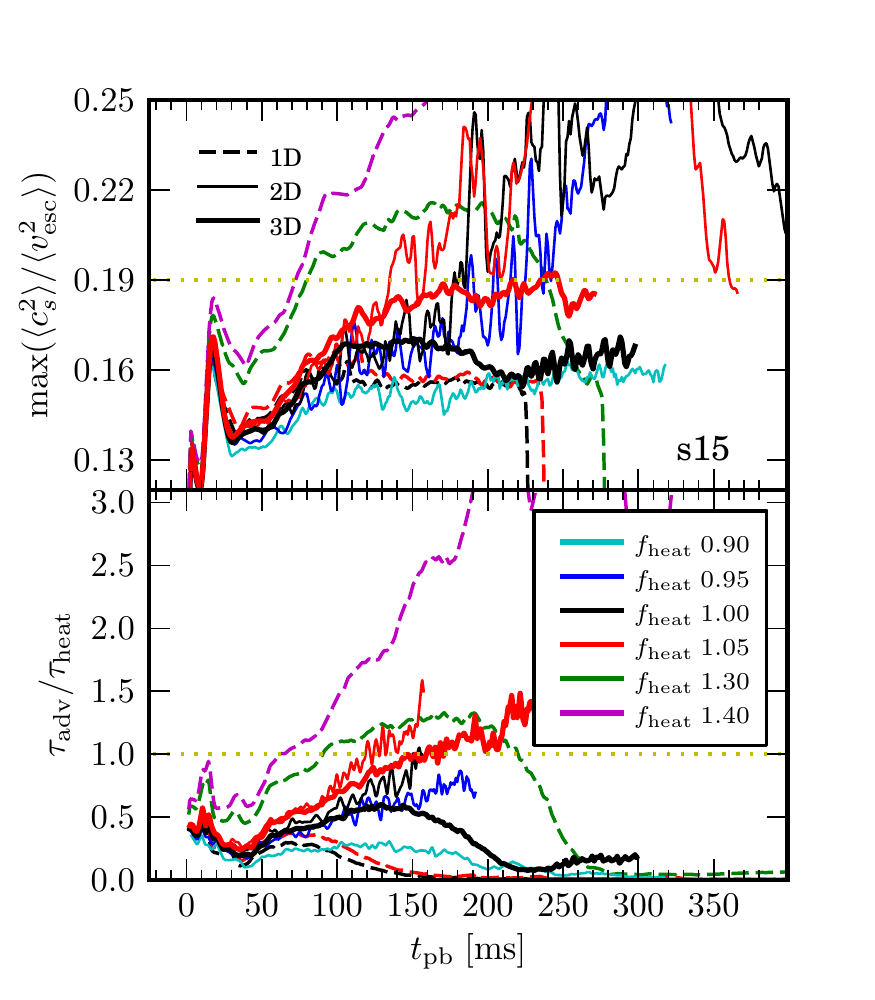} &
  \includegraphics[width=3.4in,trim= 0in 0.1in 0.3in 0.35in,clip]{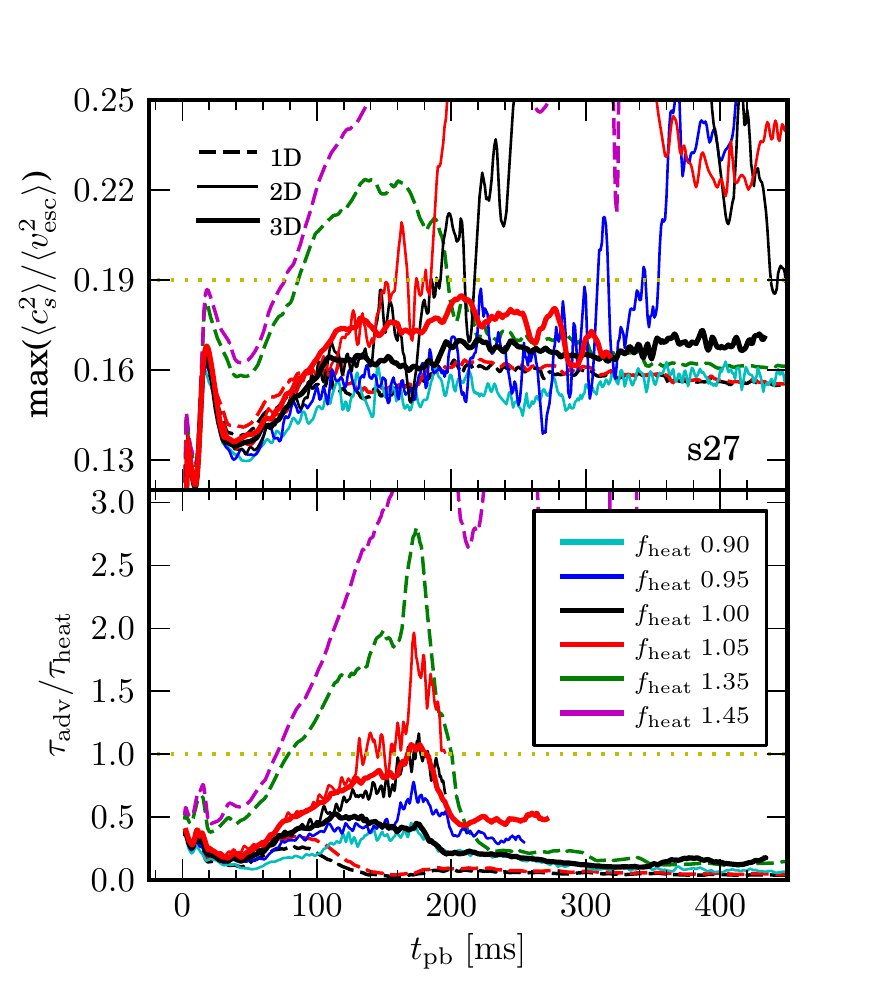}
\end{tabular}
\caption{
  Global explosion criteria for s15 (left) and s27 (right).  
  The top panel shows the antesonic ratio of \citet{Pejcha:2012cw}, and the bottom panel displays the ratio of advection time through the gain region to heating time \citep[e.g.,][]{{Thompson:2000gd}, {Janka:2001fp}, {Thompson:2005iw}}.
  Because the original derivation of the antesonic ratio only considered the post-shock accretion flow above the PNS, we only compute this criterion beyond a radius of 60 km.
  For multidimensional simulations, both criteria predict fairly well which models will explode.
  In particular, models that fail to explode remain well away from the critical thresholds, denoted by the yellow dotted lines in each plot.
}
\label{fig:expCrit}
\end{figure*}

We carry out a series of 1D, 2D, and 3D simulations in which the heat factor, $f_{\rm heat}$, and resolution were varied for both s15 and s27.  
Table \ref{table:results} summarizes the essential parameters and explosion metrics for our suite of simulations.
Hereafter, we refer to a given simulation with the convention [{\em progenitor}] \fheat [{\em value}] [1D/2D/3D], such that the 3D simulation in s15 with heat factor of 1.05 would be s15 \fheat 1.05 3D.
For the lower-resolution simulations we include the value of $\alpha$ so that this same simulation for the low-resolution case is referred to as s15 $\alpha$ 1.40 \fheat 1.05 3D.
We find that, for both high- and low-resolution cases in both s15 and s27, 2D explodes at smaller heat factors than 3D, and for the same heat factor 2D explodes more quickly and energetically.
For our fiducial high-resolution cases ($\alpha = 0.75$), 2D simulations explode for $f_{\rm heat} \ge 0.95$ while 3D simulations require $f_{\rm heat} \ge 1.05$ to achieve explosion.
In 1D, the minimum heat factors needed for explosion are of course much higher; 1.40 in s15 and 1.45 in s27.
Using the Newtonian variant of GR1D \citep{OConnor:2010bi} with the modifications to the leakage scheme discussed in \cite{Ott:2013gz}, we find a \fheat\ for s27 of 1.44 results in nearly {\it identical} shock evolution history as our 1D FLASH simulation with \fheat = 1.45.

We give the time-averaged values of the global heating efficiencies in Table \ref{table:results} (see Section \ref{sec:neutrinoEffects} for the definition of $\eta_\textrm{heat}$).
For exploding models, we only average up to the time at which the average shock radius exceeds 400 km.
As also reflected in the explosion times, the mean heating efficiences are higher in 2D than in 3D for the same \fheat.
The $\bar{\eta}_\textrm{heat}$ values for critical cases in both 2D and 3D are similar, whereas significantly larger mean heating efficiences are required for explosion in 1D.
Table \ref{table:results} also lists the maximum value of the diagnostic explosion energy, $E_{\rm diag} = E_{\rm int} + E_{\rm kin} + E_{\rm grav}$, where positive \citep[e.g.][]{{Buras:2006dl}, {Suwa:2010wp}, {Muller:2012gd}, {Bruenn:2013es}}.
The diagnostic energy is only a rough guide for the asymptotic explosion energies. 
In order to accurately assess the true explosion energies the simulations would need to be continued until much later postbounce times.
We see, however, that for the period of time we simulate, the diagnostic energy is generally greater in 2D explosions than in 3D for the same values of $f_{\rm heat}$.

In Figure \ref{fig:rshock} we show a number of shock diagnostics, along with the mass accretion rate histories, for both progenitors.
We evaluate the mass accretion rates at a constant radius of 500 km.
Along with the average shock radii, we show the maximum shock radii and the standard deviations of the shock radii, scaled by the average shock radii, $\tilde{\sigma}$.
Apparent is the more rapid ascension of the shock for explosions in 2D as compared with 3D, as well as the greater values of $\tilde{\sigma}$ for 2D, indicative of greater asymmetry.
In 3D failed explosions (models s15 \fheat 1.00 3D and s27 \fheat 1.00 3D) at late times, $\tilde{\sigma}$ increases and, for s15, shows obvious oscillatory behavior.
These oscillations are indicative of the strong SASI motions at these times.
We discuss in detail the development of the SASI in our simulations in Section \ref{sec:sasi}.

We also find clear evidence of strong neutrino-driven convection (see Section \ref{sec:convect}), particularly for exploding models.
In Section \ref{sec:convect}, we present quantitative evidence of convection, but convective motion is also obvious in the visualizations of our simulations.
In Figure \ref{fig:volRends} we show volume renderings of entropy for three of our simulations: s27 \fheat 1.00 3D, s27 \fheat 1.05 3D, and s27 \fheat 0.95 2D.
We show three different postbounce times for each simulation: 100 ms, 200 ms, and 300 ms.
Convective plumes are apparent, particularly in the movies of these visualizations\footnote{Available to view at \url{http://flash.uchicago.edu/~smc/movies}.}.
For the 2D simulation, we have wrapped the data around the $z$-axis, reflecting the axisymmetry imposed by 2D geometry.
In 2D, the convective plumes are in fact rings, in a ``3D'' sense.
Such rings do not appear in the 3D simulations because they would be manifestly unstable and would, therefore, break up into smaller structures, as is actually seen in our 3D models.
The forced axisymmetry of 2D geometry, and the unphysical prohibition of the break-up of these convective rings, has important implications for the behavior of convection in 2D as compared with 3D.
As discussed by \citet{Couch:2013fh}, these rings will tend to have larger ratios of buoyant-to-drag forces than their typically smaller 3D counterparts, and will, as a consequence, rise more quickly in the post-shock accretion flow \citep[see also][]{Dolence:2013iw}.
This results in a tendency for the average shock radius to expand more rapidly in 2D, all else being equal, at least for cases in which neutrino-driven convection is strong.
A more rapid expansion of the shock, even by a small amount, will bring more mass into the gain region and increase the efficiency at which neutrino radiation is absorbed by the gas.
This in turn increases the rate of shock expansion and, for the right conditions, a runaway of the shock can ensue (i.e., an explosion).
This tendency for larger-scale plumes in 2D is also related to the well-known ``inverse'' turbulent energy cascade in 2D \citep[][and Section \ref{sec:turb}]{Kraichnan:1967jk}.
In 2D, turbulent/convective motions begin at some driving scale and merge into larger structures, pumping kinetic energy from small scales to large.
For realistic 3D turbulence and convection, the opposite is seen.
Convective motion begins at the driving scale and cascades to small scales through the break up of these initially large structures, pumping kinetic energy from large to small scales.
We analyze the turbulent kinetic energy spectra of our simulations in Section \ref{sec:turb}.
A thoughtful and careful examination of turbulence in the CCSN context may be found in \citet{Murphy:2011ci}.

Also obvious from Figure \ref{fig:volRends} is the greater bipolar elongation of the shock structure in 2D, which has also been noted in other comparisons of 2D and 3D CCSNe \citep[e.g.,][]{{Nordhaus:2010ct}, {Hanke:2012dx}, {Burrows:2012gc}, {Takiwaki:2012ck}, {Dolence:2013iw}, Couch:2013fh}. 
This is undoubtedly an artifact of the symmetry axis, which unphysically amplifies the growth rate of important instabilities, both SASI and Rayleigh-Taylor/convection.  
This can bring extra mass into the gain region, enhancing the neutrino heating efficiency and aiding explosion (cf. Figure \ref{fig:heat}).  
We compare the details of SASI development between 2D and 3D simulations in Section \ref{sec:sasi}.

A number of critical conditions have been proposed as criteria for runaway CCSN shock expansion.
These criteria are usually derived by analyzing steady-state accretion flows to determine the limits at which the steady-state assumption breaks-down.
The balance of neutrino heating to advection time scales in the gain region has been suggested by a number of authors to be a critical quantity to the stability of the shock \citep[e.g.,][]{ {Janka:1998tx}, {Thompson:2000gd}, {Janka:2001fp}, {Thompson:2005iw}, {Fernandez:2012kg}}.
Steady-state solutions fail when the advection time, $\tau_{\rm adv}$, exceeds the heating time, $\tau_{\rm heat}$, allowing thermal energy to build up in the gain region eventually pushing the shock outward in search of a new equilibrium.
\citet{Fernandez:2012kg} presents a detailed study of the stability of CCSN shocks in simplified time-dependent 1D simulations that dissects the physics of this stability criterion.
The ratio of advection-to-heating times has been shown to be generally accurate in postdicting the likelihood of explosion in 2D simulations \citep{Buras:2006hl, {Marek:2009kc}, {Muller:2012gd}}.
\citet{Pejcha:2012cw} derived a new criterion for explosion, the ``antesonic'' condition, based on the global state of a neutrino-irradiated isothermal accretion flow.
This criterion can be viewed as a generalization of the Burrows-Goshy limit \citep{Burrows:1993ft} that states that for a given mass accretion rate there is a limiting neutrino luminosity for stable accretion.
The antesonic condition posits that there exists a critical {\em sound speed} that if achieved by the accretion flow no steady-state solution may be found, i.e., the accretion flow transitions to explosion.
The critical sound speed-squared is a fraction of the local escape speed-squared: $c_s^2 \approx 0.19 v_{\rm esc}^2$.
The critical value of 0.19 for the antesonic ratio is based on an isothermal EOS.
For EOSs more applicable to core-collapse supernovae, the critical sound speed is expected to be larger, perhaps $\sim 0.23$ (O. Pejcha 2013, private communication).

In Figure \ref{fig:expCrit}, we show both the antesonic ratio and the ratio of advection-to-heating time scales for our simulations.  
The respective critical values are denoted by the dotted yellow lines.
For the antesonic ratio, we first spherically-average our multidimensional results before computing the ratio.
We use the adiabatic sound speed, $\langle c_s^2 \rangle = \langle \gamma_c\rangle \langle P \rangle \langle \rho \rangle^{-1}$, and we calculate the escape speed from the 1D-averaged gravitational potential, $\langle v_{\rm esc}^2 \rangle = -2 \langle \Phi \rangle$.
For the heating time scale we divide the absolute value of the total specific energy (internal plus kinetic plus gravitational) in the gain layer, $\abs{E_{\rm gain}}$, by the net heating rate in the gain region, $Q_{\rm net}$, such that $\tau_{\rm heat} = \abs{E_{\rm gain}}/Q_{\rm net}$ \citep[][]{Marek:2009kc}.  
We estimate the advection time through the gain region in the same manner as, e.g., \citet{Muller:2012gd} and \citet{Ott:2013gz}, $\tau_{\rm adv} = M_{\rm gain} / \dot{M}$, where $M_{\rm gain}$ is the mass in the gain region and $\dot{M}$ is the mass accretion rate which, in our case, is evaluated at a fixed radius of 500 km.
Once the maximum shock radius has reached 500\,km, we stop the calculation as the mass accretion rate will be significantly altered by the shock dynamics.
However, in such a situation the explosions are all well underway. 

We see that the antesonic condition is a fairly good indicator of a model's likelihood for explosion.  
Our results suggest that the critical antesonic value in realistic 1D simulations is indeed higher, and our 1D results are consistent with a limit of 0.23, as suggested by Pejcha.
That our multidimensional simulations seem to correspond to the critical antesonic ratio value of 0.19 would then imply that the critical limit is {\em reduced} in multidimensional simulations.
It is well known that, for realistic time-dependent CCSN simulations, 2D models explode with much greater ease than 1D, suggesting that the criteria for explosion are less stringent in 2D \citep[e.g., ][]{{Murphy:2008ij}, {Marek:2009kc}, {Suwa:2010wp}, {Muller:2013kz}, {Bruenn:2013es}}.
Our analysis seem to indicate that the antesonic ratio captures this through a reduction in the critical limit.

Similarly, the ratio of advection-to-heating times is also generally reliable in identifying models that result in explosions.
The notable exception is the 2D models, in both progenitors, with \fheat = 0.95.  
At no point does the ratio of $\tau_{\rm adv} / \tau_{\rm heat}$ exceed unity yet this model explodes.
This model is marginal and our estimation of the advection time scale may be too crude for this case.
Specifically, measuring $\dot{M}$ at a fixed 500 km, rather than immediately exterior to the shock, may be skewing our estimate of the advection time through the gain region.
Using $\dot{M}$ itself is an approximation to the more correct integrated radial velocity through the gain region \citep[see, e.g.,][]{Buras:2006hl}.
All other explosions, in both 2D and 3D, at some point exceed the critical limit of one for this criterion, while failed explosions stay safely below it.

The advection-to-heating time criterion seems to break down for our 1D simulations.  
While the explosion cases handily exceed both critical limits, so do the critical non-exploding models (\fheat = 1.30 for s15 and \fheat = 1.35 for s27).
Since the criterion is based on essentially 1D dynamics, this is a curious result.
Exceeding this limit should mean that the pressure in the gain region continues to increase preventing shock contraction.
But this is not the case for our marginal 1D simulations in which this critical limit is met yet, subsequently, the shock does indeed recede.

\subsection{Neutrino Effects}
\label{sec:neutrinoEffects}

\begin{figure*}[htb]
\centering
\begin{tabular}{cc}
  \includegraphics[width=3.4in,trim= 0in 0.21in 0.3in 0.35in,clip]{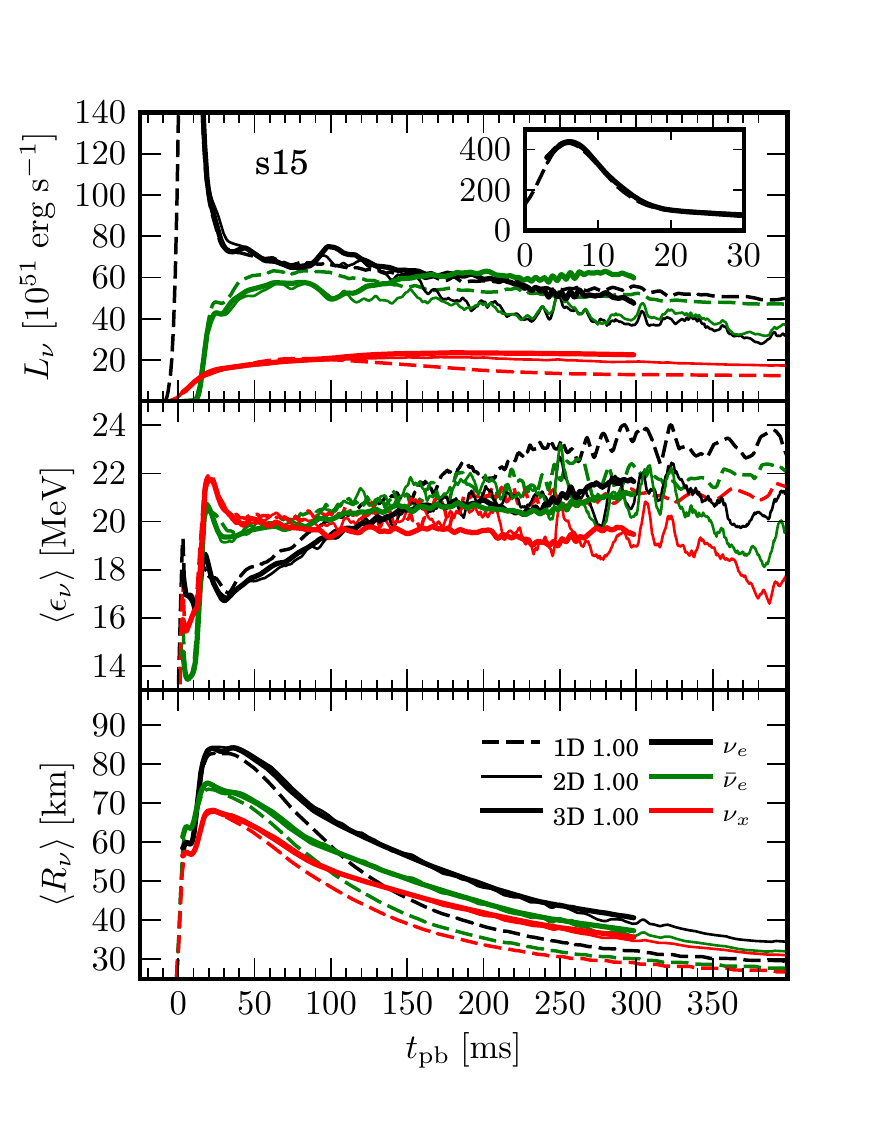} &
  \includegraphics[width=3.4in,trim= 0in 0.21in 0.3in 0.35in,clip]{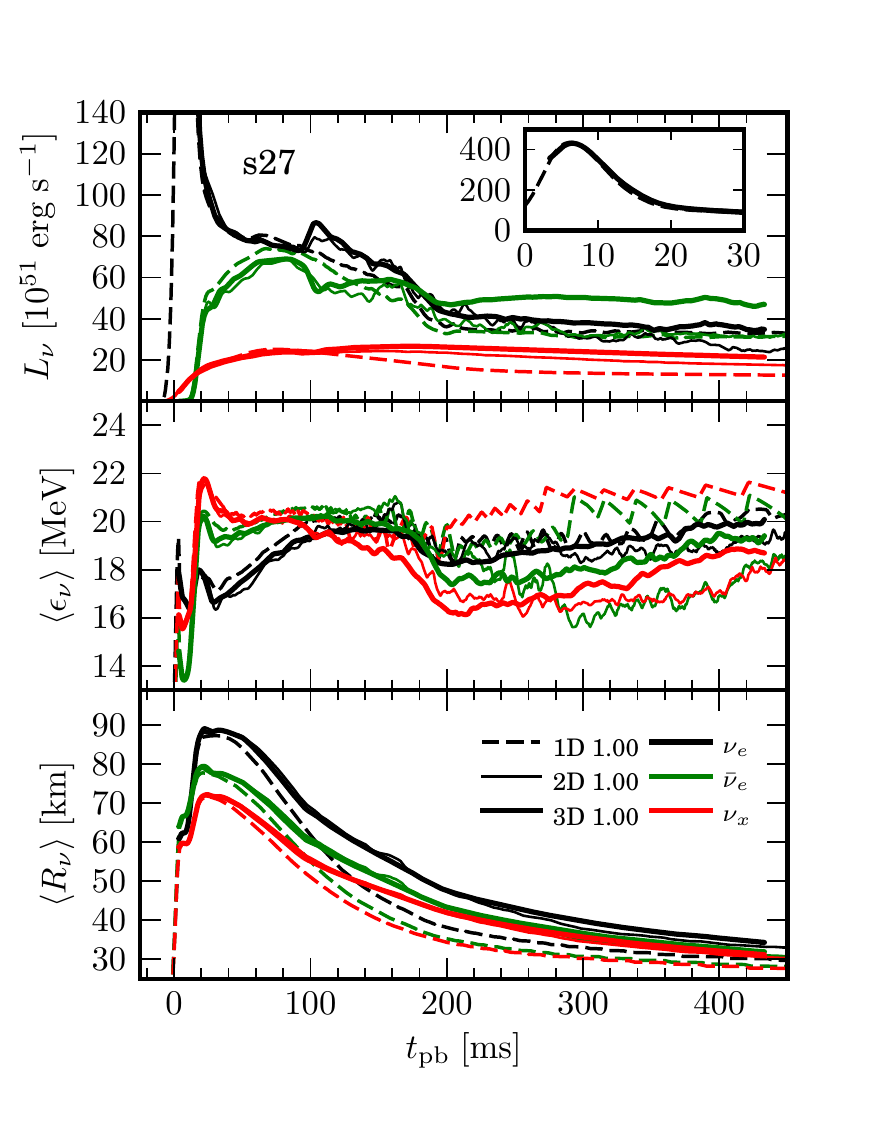}
\end{tabular}
\caption{
  Neutrino emission properties for s15 (left) and s27 (right). Shown are the luminosities (top), angle-averaged energies (middle), and neutrinospheric radii (bottom) for $\nu_e$, $\bar{\nu}_e$, and $\nu_x$ neutrino types. 
  Here we show only the \fheat = 1.00 models for 1D, 2D, and 3D.
  For clarity reasons, we omit the 1D mean energies from the plots.
  The neutrino emission properties between 2D and 3D are remarkably similar, up until about 100 ms.
  At this point, 2D transitions toward explosion while 3D fails and the shock begins to recede.
}
\label{fig:neut}
\end{figure*}

\begin{figure*}[htb]
\centering
\begin{tabular}{cc}
  \includegraphics[width=3.4in,trim= 0in 0.21in 0.3in 0.35in,clip]{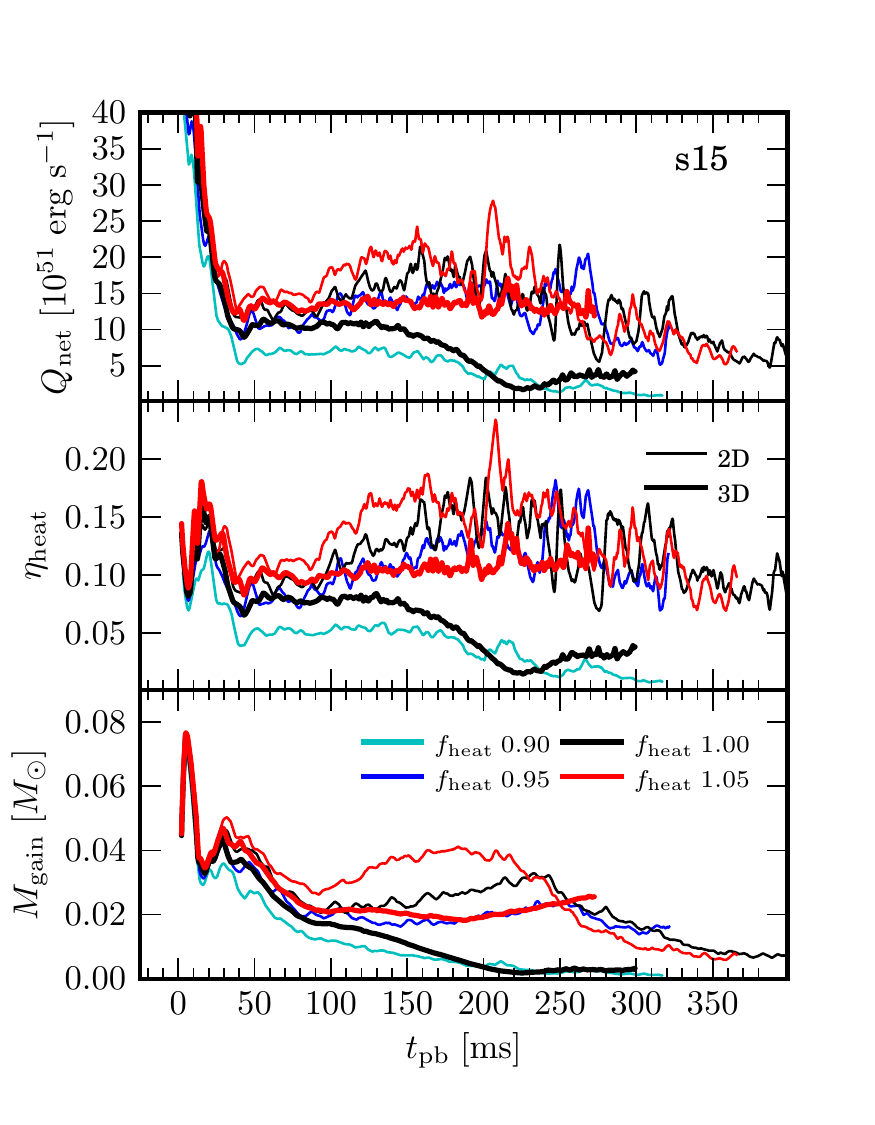} &
  \includegraphics[width=3.4in,trim= 0in 0.21in 0.3in 0.35in,clip]{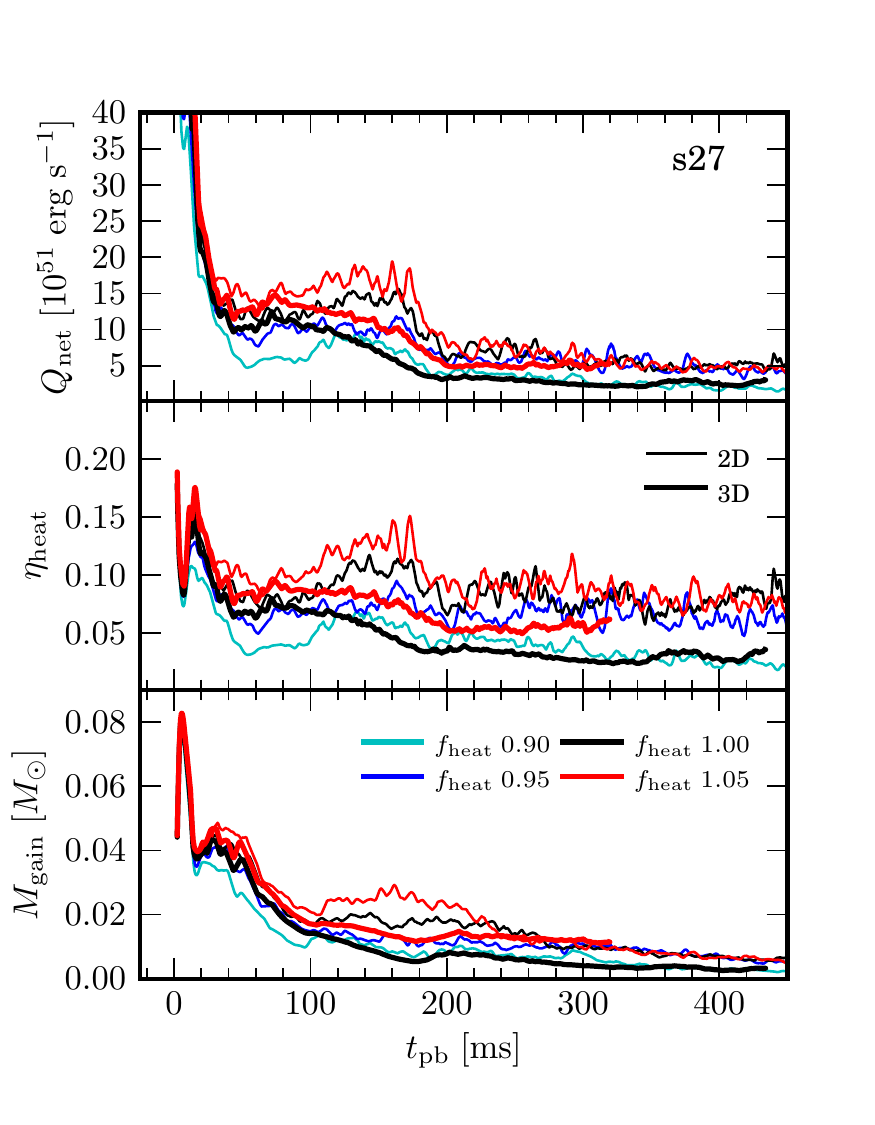}
\end{tabular}
\caption{
  Net neutrino heating in the gain region (top), heating efficiency, $\eta_{\rm heat} = Q_{\rm net} (L_{\nu_e,\mathrm{gain}}+L_{\bar{\nu}_e,\mathrm{gain}})^{-1}$, (middle), and mass in the gain region (bottom) for s15 (left) and s27(right).
  At early times ($t_{\rm pb} < 50$ ms), before significant non-spherical motion as developed in 2D and 3D, these metrics are very similar for all models with the same heat factor.
}
\label{fig:heat}
\end{figure*}

Neutrino heating and cooling is fundamental to the CCSN mechanism.
Our leakage scheme has significant advantages over the Murphy-Burrows ``lightbulb'' scheme \citep[][]{Murphy:2008ij} employed by a number of other 3D CCSN simulations \citep[e.g.,][]{{Nordhaus:2010ct}, {Hanke:2012dx}, {Burrows:2012gc}, {Murphy:2013eg}, {Dolence:2013iw}, {Couch:2013fh}}.
The multispecies leakage approach captures the essential features of PNS cooling and contraction, postbounce deleptonization, self-consistently computed neutrino luminosities that vary in space and time, as well as variable mean neutrino energies.
As detailed in \citet{Ott:2013gz}, who employ the same neutrino leakage scheme as we do, this approach reproduces the global features of realistic spectral neutrino transport very well, while requiring a mere fraction of the computational cost.

We present the basic features of the neutrino emission from our simulations in Figure \ref{fig:neut}, which shows the spherically averaged emergent neutrino luminosities, angle-averaged mean energies, and neutrinospheric radii as functions of time.
For each dimensionality, we show only the \fheat = 1.00 cases.
The salient effects of the leakage scheme are as follows.  
The rapid electron capture on free protons liberated during the dissociation of heavy nuclei produces the characteristic $\nu_e$ neutronization burst soon after bounce. 
The shape and magnitude of this breakout $\nu_e$ burst in s15 and s27 models are very similar, and are essentially identical between 1D, 2D, and 3D as expected from the universal and highly-spherical nature of core collapse.
After the initial burst but before the explosion, the electron-type neutrino luminosities are high.
This is due to the large accretion rate of fresh shock-heated material.
These ``accretion'' luminosities essentially mirror the mass accretion rate (see Figure \ref{fig:rshock}).
The s27 electron-type neutrino luminosities are initially higher than the s15, but drop below when a sharp drop in the accretion rate occurs at $\sim 170\,$ms postbounce (due to the advection of the silicon-oxygen interface through the shock).
For the s15 model, the late time ($t_\mathrm{postbounce} \gtrsim 0.2\,$s) neutrino emission of the 2D simulation is significantly lower than the 1D and 3D counterparts.
This is because the 2D simulation is well into the explosion phase and mass accretion onto the protoneutron star has ceased.  
This difference between the exploding/non-exploding neutrino luminosities is not as large in the s27 set of simulations since the late-time mass accretion rate is much lower than in the s15 model. 

In all models, the neutrinosphere radii, defined as the radius where the optical depth is 2/3, peak between 30 and 50\,ms after bounce.
The heavy-lepton neutrinos, which lack charged-current interactions have the lowest optical depth and therefore the deepest neutrinosphere (peaking at $\sim$70\,km).  
The main contribution to the $\bar{\nu}_e$ opacity is charged-current absorption on free protons.
Free protons are less abundant in the neutron-rich postshock environment leading to a lower $\bar{\nu}_e$ neutrinosphere which peaks at ($\sim$75\,km).
$\nu_e$ have the largest opacity and therefore decouple from the matter at the largest radii ($\sim 85\,$km at 50\,ms).  
As the protoneutron star cools and contracts, the neutrinosphere radii recede.
The lepton-gradient-driven protoneutron star convection present in 2D and 3D maintains the neutrinosphere radii at larger values than seen in 1D, in agreement with more sophisticated multi-dimensional core-collapse supernova simulations (e.g., \citealt{Muller:2012gd})

As also pointed out by \citet{Ott:2012ib} and \citet{Ott:2013gz}, the leakage scheme generally overestimates the mean neutrino energies, and the energy hierarchy becomes inverted after around 100 ms postbounce. 
This has a feedback on the neutrino heating term (see Equation (\ref{eq:heating_rate})). 
However, as we discuss below, the more important quantities when discussing the neutrino mechanism are the net heating rates and heating efficiencies.  
These global neutrino heating metrics depend not solely on the microscopic choice for the heating rate, but on the highly non-linear coupling between the microscopic heating rate and the core-collapse supernova hydrodynamics.

The SASI can leave an imprint on the temporal history of the neutrino emission from CCSNe.  
As mentioned above and described in detail in the following subsection, model s15 \fheat 1.00 3D develops a strong $\ell = 1$ SASI.
Figure \ref{fig:neut} shows that at late times ($t_{\rm pb} > 200$ ms) for this model the neutrino luminosities of all three types are oscillating sympathetic with the SASI activity (see Section \ref{sec:sasi}).
\citet{Tamborra:2013uf} discuss in detail similar modulation of the neutrino signal by the SASI in the context of the 3D full-transport CCSN simulation of \citet{Hanke:2013kf}.
We note that since our neutrino luminosities are spherically averaged, the overall magnitude of the SASI induced variation in the neutrino luminosities is not as large as it could be in any given direction.
Given the approximate nature of our neutrino treatment, we refrain from a more detailed analysis of the neutrino signal associated with the SASI activity.

We show global heating metrics in Figure \ref{fig:heat} for both the s15 simulations (left panels) and the s27 simulations (right panels).  
In Figure \ref{fig:heat} we plot the net neutrino heating in the gain region, $Q_{\rm net}$ (top panels), the heating efficiency\footnote{Our heating efficiences are systematically higher than those of \citet{Ott:2013gz} because in that work the heavy-lepton neutrino luminosity was mistakenly included in the denominator for calculating the heating efficiency (C.D. Ott 2013, private communication).} $\eta_{\rm heat} = Q_{\rm net} (L_{\nu_e,\mathrm{gain}} + L_{\bar{\nu}_e,\mathrm{gain}})^{-1}$ (middle panels), and the mass in the gain layer, $M_{\rm gain}$ (bottom panels).
For clarity, we omit the 1D cases from Figure \ref{fig:heat}.
In addition, we include the average heating efficiency (averaged between bounce and when the average shock passes 400\,km for the last time) in Table \ref{table:results}. 
 As early as 30 ms postbounce, the amount of matter in the gain region is greater in 2D than in 3D.
This sets off a self-sustaining chain of events: with greater $M_{\rm gain}$, more neutrino energy is absorbed, i.e., $Q_{\rm net}$ is larger in 2D.
With a greater neutrino heating rate, the shock is pushed to larger extents in 2D, thus bringing into the gain region even more mass, and so on.
\citet{{Dolence:2013iw}} also find greater mass in the gain region, and concomitant larger net heating rates, in 2D as compared with their 3D simulations using the lightbulb approach of \citet{Murphy:2008ij}.

Finally, we draw the readers attention to an artificial effect in our current implementation of neutrino leakage.
Soon after bounce, there is significant heating (as seen in Figure \ref{fig:heat}) that is not present in more detailed neutrino treatments (eg. \citealt{Marek:2009kc}). 
This excess heating excites small, mildly convectively-stable regions behind the shock that advect down through the cooling region and eventually merge with the protoneutron star convection region causing it to expand abruptly. 
This is most clearly seen in Figures \ref{fig:s15vaniso} and \ref{fig:s27vaniso} and will be discussed in \S \ref{sec:convect}.
The abrupt expansion of the protoneutron star convectively unstable region at occurs at $\sim$100\,ms in both the s15 and s27 2D and 3D simulations. 
Concurrently, there is a rapid change in the character of the neutrino luminosities and neutrinosphere radii compared to 1D.
The extended convection zone levels out the density and temperature gradients, causing material at higher mass coordinates (and lower optical depths) to be both more electron degenerate, reducing the $\bar{\nu}_e$ luminosity, and hotter, increasing the $\nu_e$ luminosity.
We note similar behaviour was seen in \cite{Ott:2013gz} (their Figure 9), but the influence on the neutrino luminosities is not as apparent in that work as it is here.

\subsection{SASI in 2D and 3D}
\label{sec:sasi}

\begin{figure*}[htb]
\centering
\begin{tabular}{cc}
  \includegraphics[width=3.4in,trim= 0in 0.0in 0.3in 0.3in,clip]{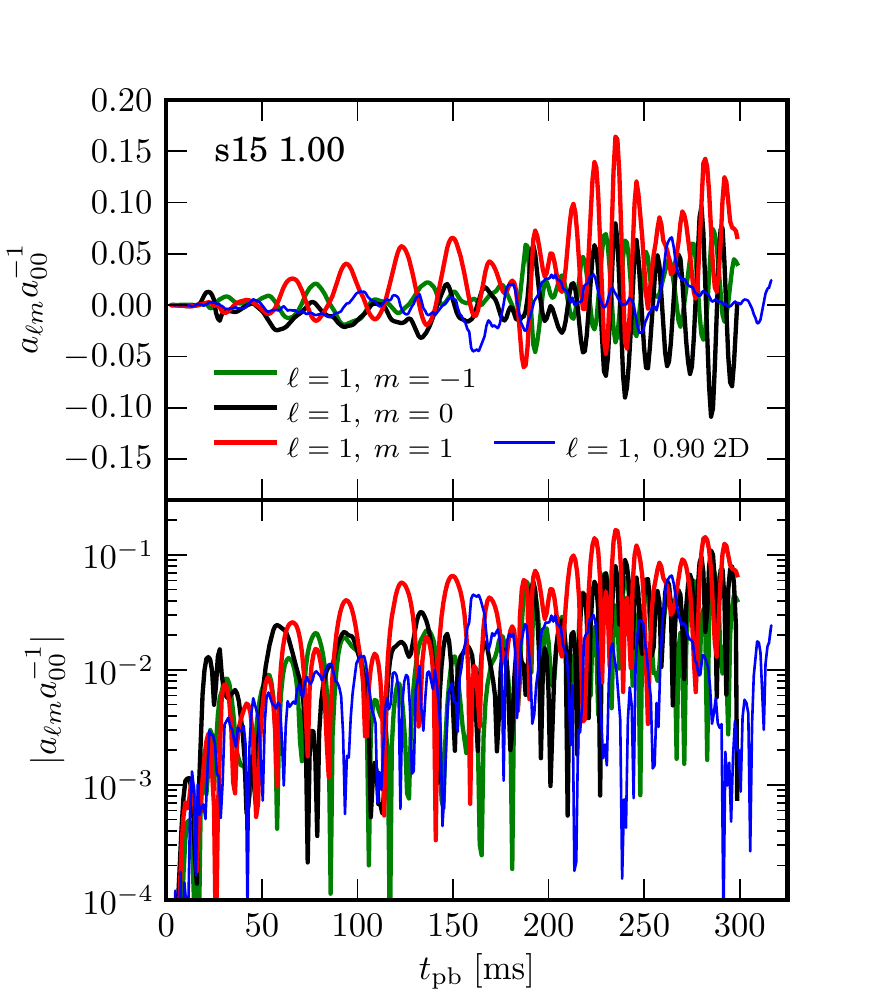} &
  \includegraphics[width=3.4in,trim= 0in 0.0in 0.3in 0.3in,clip]{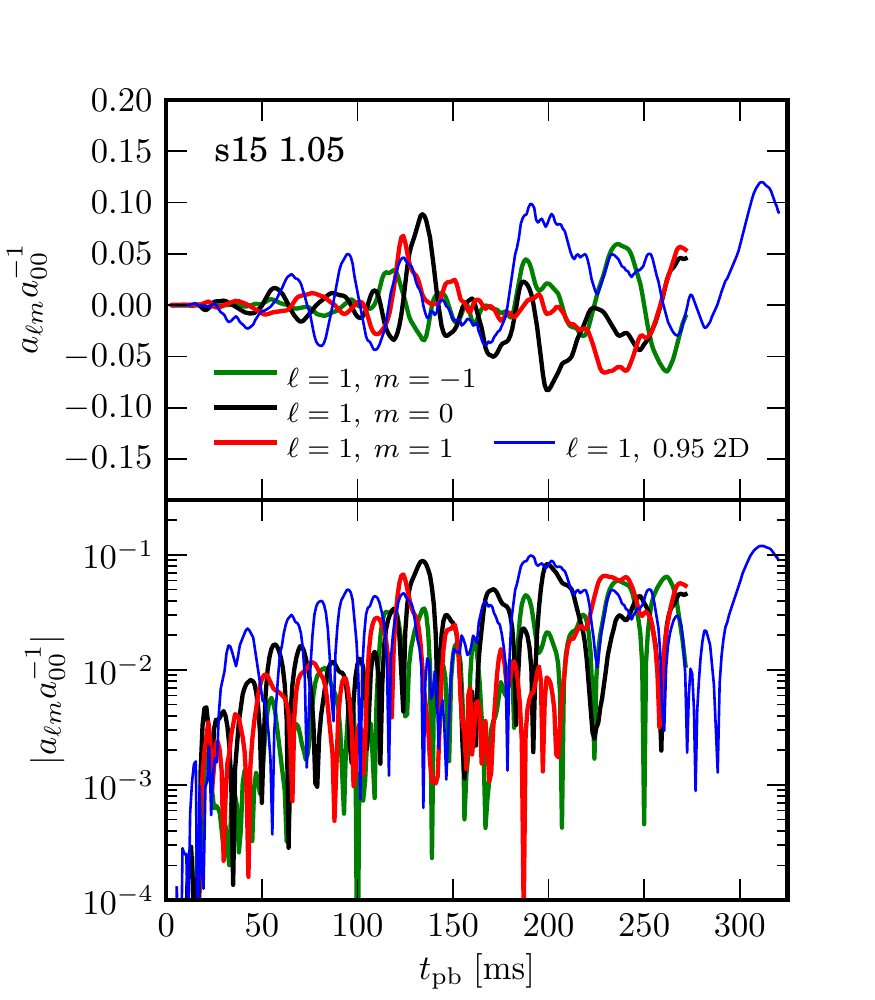} \\
  \includegraphics[width=3.4in,trim= 0in 0.0in 0.3in 0.3in,clip]{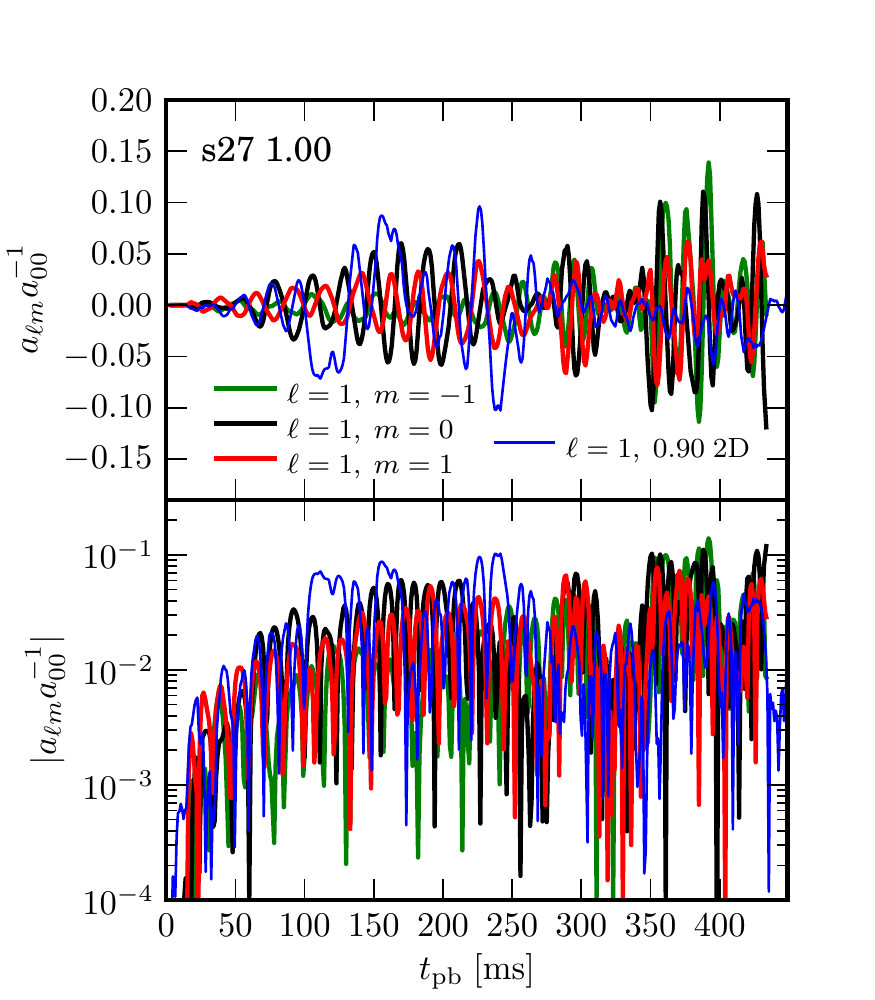} &
  \includegraphics[width=3.4in,trim= 0in 0.0in 0.3in 0.3in,clip]{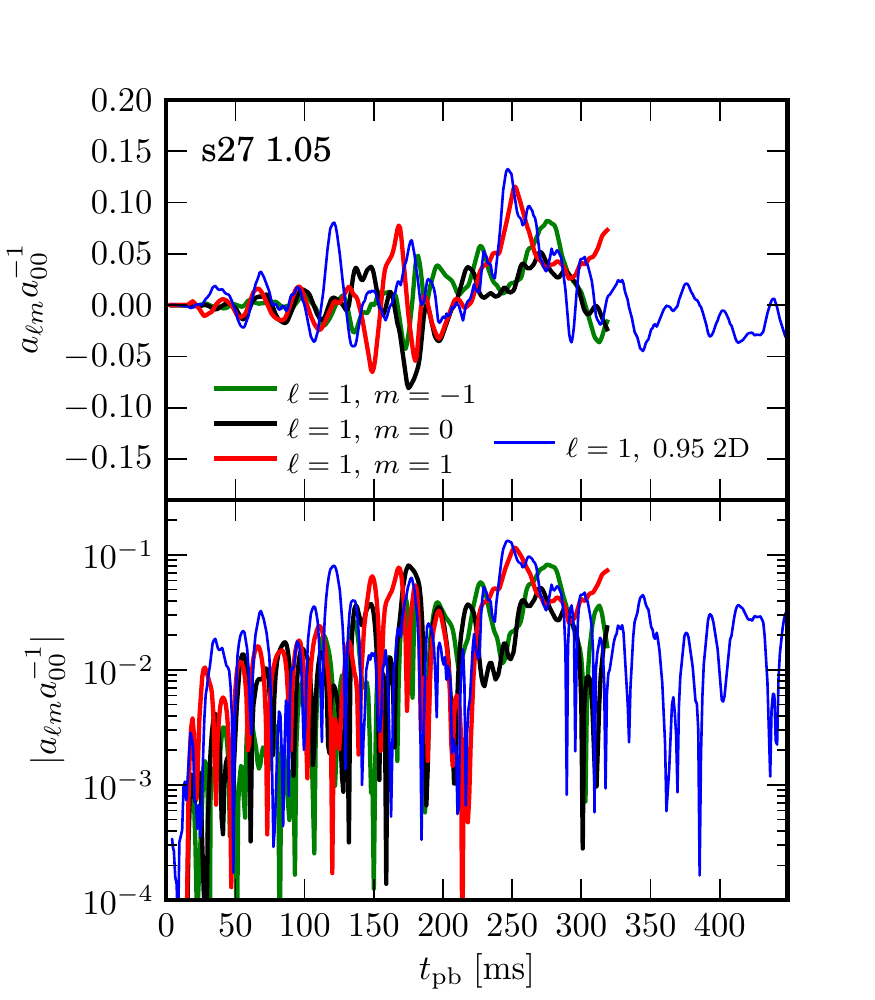}
\end{tabular}
\caption{
  Spherical harmonic coefficients, $a_{\ell m}$, for the first Legendre order, $\ell = 1$, in the 2D and 3D non-exploding simulations ($f_{\rm heat} = 1.00$, left) and critical-case explosion simulations ($f_{\rm heat} = 1.05$, right) for s15 (top) and s27 (bottom).
  For the 2D cases (thin blue lines) we show the data for $f_{\rm heat} = 0.90$ (left) and $f_{\rm heat} = 0.95$ (right).
  Strong periodicity with growing amplitude is seen at certain times in all of these simulations, strongly indicative of SASI activity.  
  For s15 in the non-exploding cases (top, left) the SASI activity is clearly stronger in 3D than in 2D.
  In the exploding case for s15 (top, right), the SASI is dramatically diminished in 3D yet {\em strengthened} in 2D.
  s27 in the non-exploding cases (bottom, left) is very favorable to the SASI during two distinct epochs in 3D: early-on while the mass accretion rate is still quite high and later ($t_{\rm pb} > 300$ ms) at which point the shock has receded to small radii.
  For s27 \fheat 0.90 2D, only the early epoch shows clear SASI activity, while the later, post-shock-recession SASI is extremely muted relative to 3D.
  In the exploding cases for s27, both 2D and 3D show low-frequency SASI activity up until $\sim 200$ ms, after which the models transition to explosion and the SASI is no longer in evidence.
}
\label{fig:lm15}
\end{figure*}

\begin{figure*}[htb]
\centering
\begin{tabular}{cccc}
  \includegraphics[width=1.75in,trim= 5.in 3.5in 4.2in 2.7in, clip]{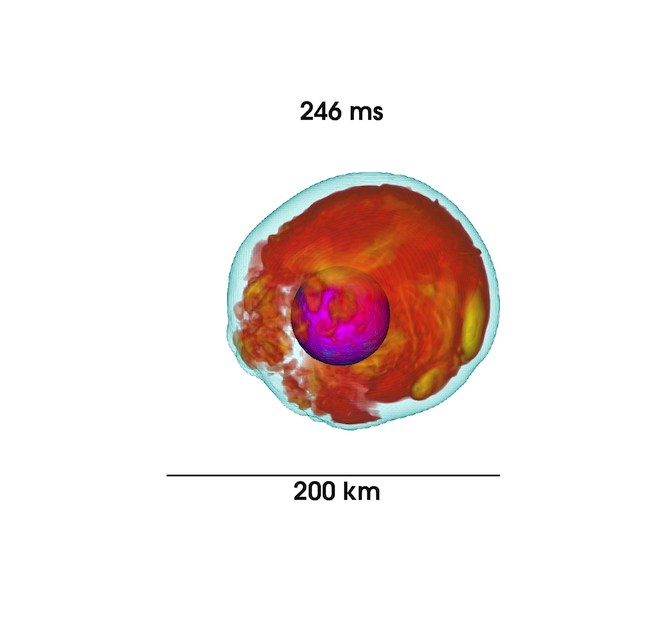}  
  \includegraphics[width=1.75in,trim= 5.in 3.5in 4.2in 2.7in, clip]{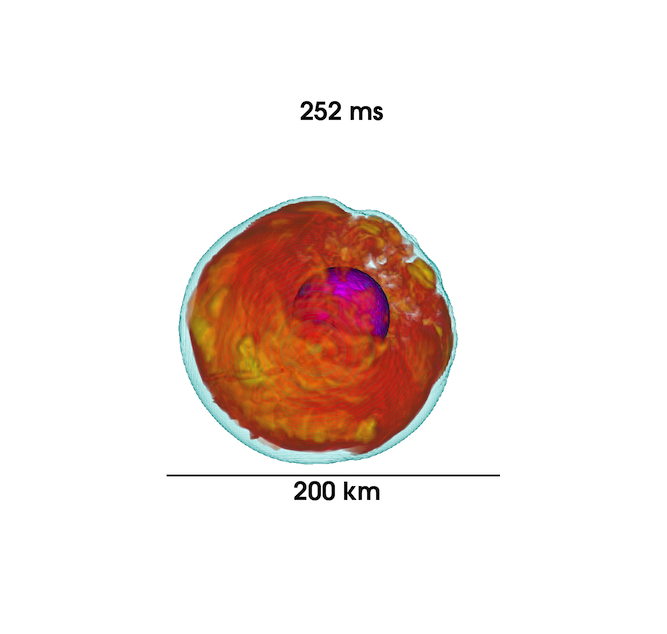}  
  \includegraphics[width=1.75in,trim= 5.in 3.5in 4.2in 2.7in, clip]{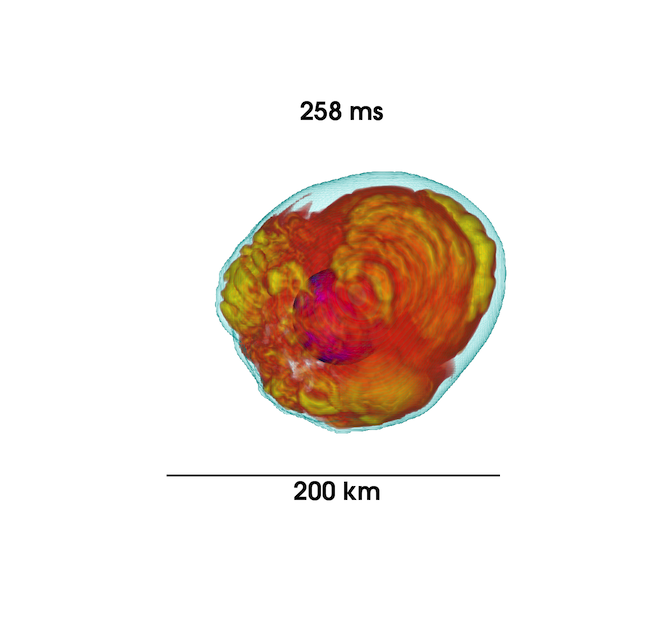}  
  \includegraphics[width=1.75in,trim= 5.in 3.4in 4.2in 2.8in, clip]{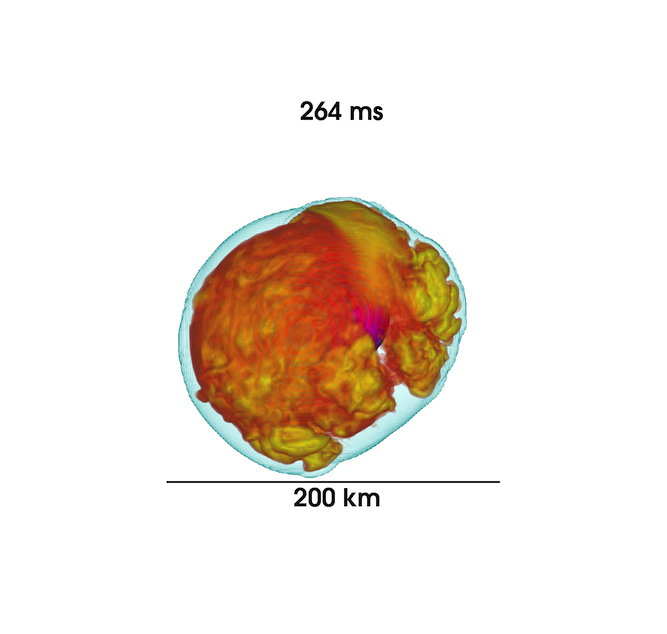}  
\end{tabular}
\caption{
  Entropy volume renderings for s15 \fheat 1.00 3D at late times showing the development of the SASI.
  Four different postbounce times are shown, spaced 6 ms apart, from left to right.
  The shock itself is visible as the pale blue surfaces, as in Figure \ref{fig:volRends}, and constant-density contours with a value of $10^{12}$ g cm$^{-3}$ (magenta) mark the edge of the PNS.
  At this stage the shock shows clear spiral motion, as indicated by the spherical harmonic components shown in Figure \ref{fig:lm15}.
}
\label{fig:sasi}
\end{figure*}

\begin{figure}
  \centering
  \includegraphics[width=3.4in,trim= 0in 0.0in 0.3in 0.3in,clip]{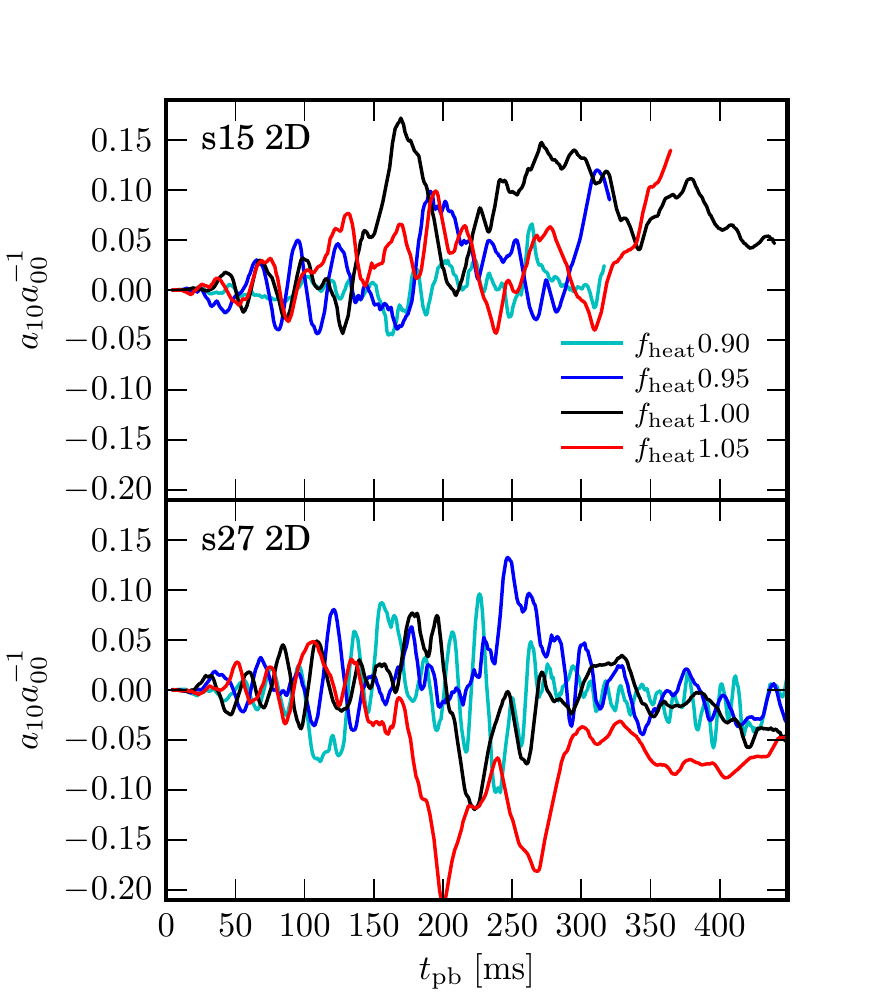}
  \caption{
    Spherical harmonic coefficient $a_{10}$ of the shock surface for 2D simulations at different \fheat.
    The top panel shows results for s15 and the bottom for s27.
  }
  \label{fig:lm2d}
\end{figure}

\begin{figure*}[htb]
\centering
\begin{tabular}{cc}
  \includegraphics[width=3.in,trim= 0.in 0.15in 0.1in 0.2in,clip]{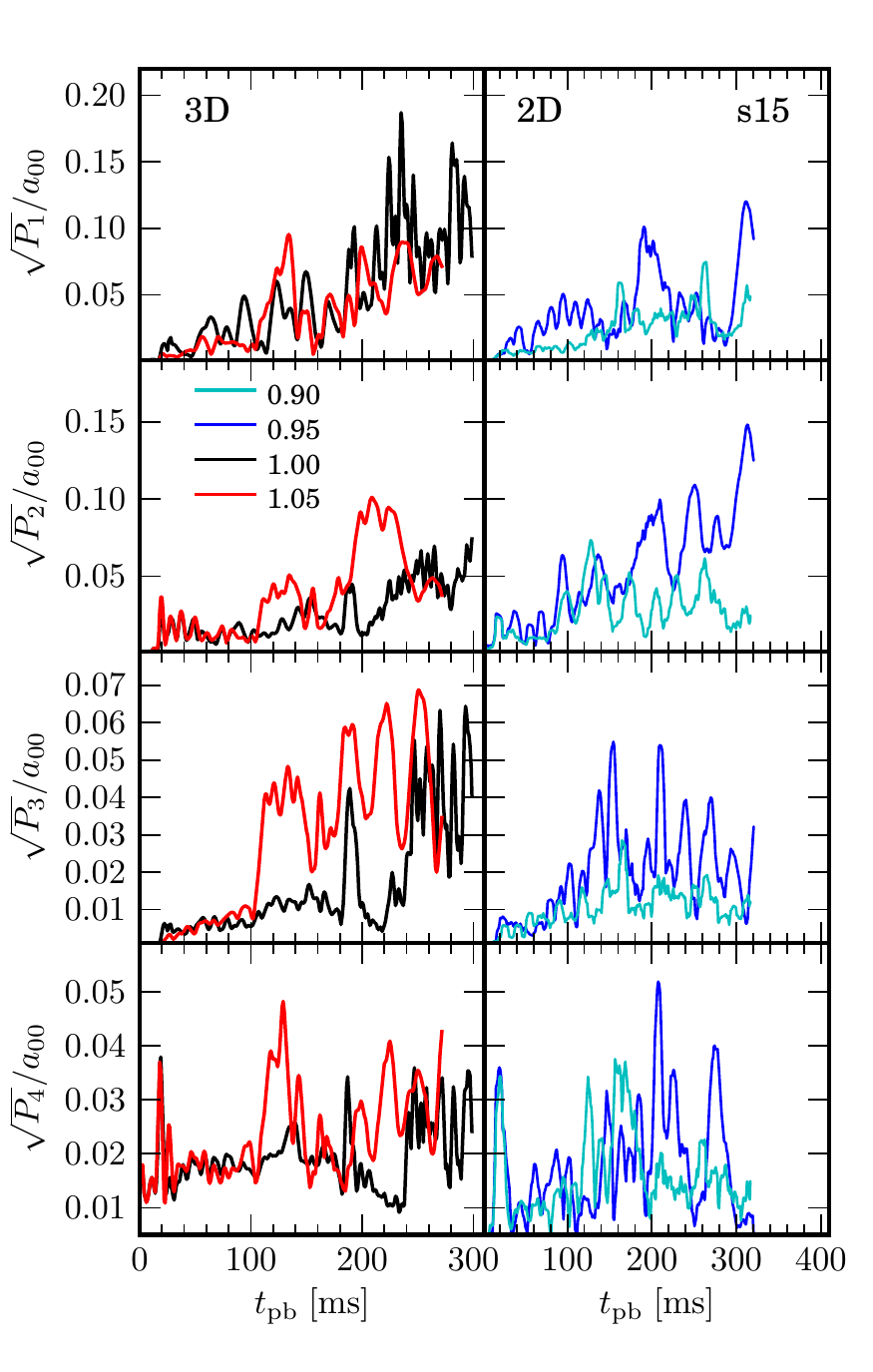} &
  \includegraphics[width=3.in,trim= 0.in 0.15in 0.1in 0.2in,clip]{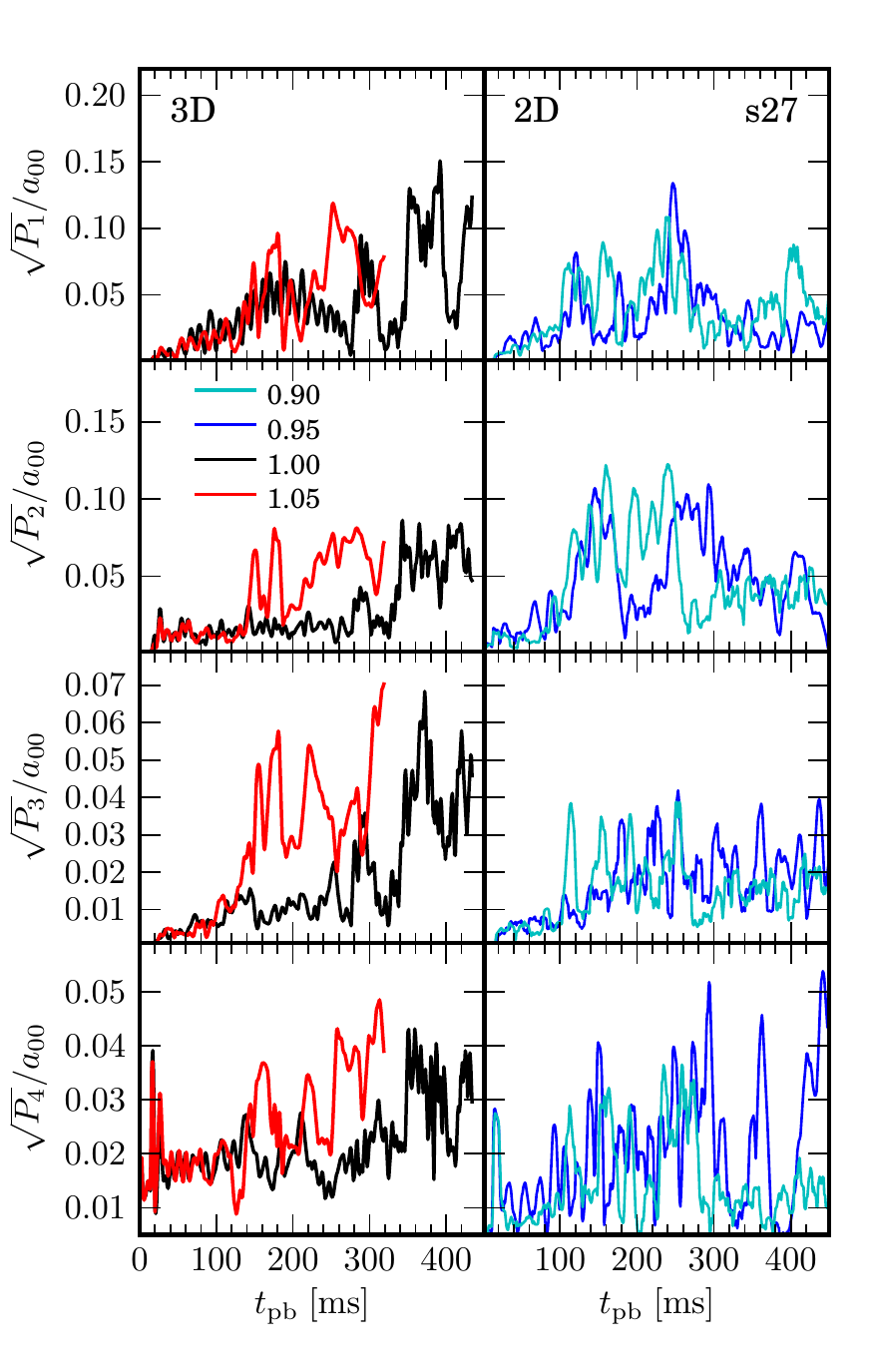}
\end{tabular}
\caption{
  Square-root of the powers, $P(\ell) = \sum_{m=-\ell}^{\ell} a_{\ell m}^2$, in the first four spherical harmonics of the shock surface, weighted by the mean shock radius, as functions of time, for s15 (left) and s27 (right).
  Late time SASI activity in s15 \fheat 1.00 3D is apparent, as is the early ($t_{\rm pb} < 300$ ms) SASI activity in both s27 \fheat 1.00 3D and s27 \fheat 1.05 3D.
  Notably, there is not so a great a difference in the $P_\ell$'s between 2D and 3D as reported in some previous 2D vs. 3D studies \citep{{Burrows:2012gc}, Couch:2013fh}.
}
\label{fig:spPow}
\end{figure*}

The role of the SASI in the CCSN mechanism is a topic of much current debate.
On the one hand, its presence and growth has been unmistakably identified in many multidimensional CCSN simulations \citep{{Scheck:2008ja}, {Marek:2009kc}, {Ott:2008gn}, {Suwa:2010wp}, {Muller:2012gd}, {Muller:2012kq}, {Hanke:2013kf}}, and both theory and experiment have shown that the SASI is a fundamental hydrodynamic instability of accretion shocks that is not unique to the core-collapse context \citep[see, e.g.,][]{Foglizzo:2012kl}.
On the other, its importance in the CCSN context has been questioned on the basis of some 3D simulations \citep{{Nordhaus:2010ct}, {Burrows:2012gc}, {Murphy:2013eg}}.
On the basis of 3D neutrino-lightbulb simulations, these authors argue that neutrino-driven convection must dominate the SASI in neutrino-driven explosions.

We examine the growth of the SASI in our simulations using the common approach of decomposing the shock surface into spherical harmonics \citep{{Blondin:2006dv}, {Marek:2009kc}, {Burrows:2012gc}, {Muller:2012kq}, {Ott:2013gz}, {Couch:2013fh}}.
The coefficients of this decomposition are
\beq 
a_{\ell m} = W(\ell,m) \oint r_{\rm shock}(\theta,\phi)
Y_\ell^m(\theta,\phi) d\Omega,
\label{eq:alm}
\eeq
where $r_{\rm shock}(\theta,\phi)$ is the shock radius as a function of angular coordinates and the weights are $W(\ell,m) = (-1)^{|m|}/\sqrt{4\pi(2\ell+1)}$.  
The weights are chosen so that $a_{00} = \langle r_{\rm shock} \rangle$, $a_{11} = \langle x_{\rm shock} \rangle$, $a_{1-1} = \langle y_{\rm shock} \rangle$, and $a_{10} = \langle z_{\rm shock} \rangle $. 
The spherical harmonics are
\beq
Y_\ell^m = 
\begin{cases} 
  \sqrt{2} N_\ell^m P_\ell^m(\cos\theta) \cos m\phi &		m>0,\\
  N_\ell^0 P_\ell^0(\cos \theta) &				m=0,\\
  \sqrt{2} N_\ell^{|m|} P_\ell^{|m|}(\cos\theta) \sin |m|\phi &	m<0,
\end{cases}
\label{eq:ylm}
\eeq 
with 
\beq N_\ell^m = \sqrt{\frac{2\ell + 1}{4\pi}
  \frac{(\ell-m)!}{(\ell+m)!}}.  
\eeq 
In Equations (\ref{eq:ylm}), $P_\ell^m$ are the usual associated Legendre polynomials.
From the coefficients of Equation (\ref{eq:alm}) we compute the ``power'' in the $\ell^{\rm th}$ Legendre order:
\beq 
P(\ell) = \sum_{m=-\ell}^{\ell} a_{\ell m}^2.
\eeq 
The only non-canceling coefficients in 2D are $a_{\ell 0}$'s. 

The linear theory of the SASI shows that the dipole $\ell = 1$ is the fastest-growing mode of the instability \citep{{Blondin:2006dv}, Foglizzo:2007cq}.
In 2D, the dipole mode has only one degree of freedom: up and down the symmetry axis (i.e., $m=0$).
In 3D, however, the additional degree of freedom allows the development of $m \ne 0$ modes \citep{{Blondin:2007fk}, {Blondin:2007bf}, {Iwakami:2008dw}} that can be represented as the superposition of multiple aphasic $\ell = 1,\ m=0$ modes \citep{Fernandez:2010ko}.
These additional modes sap energy from the singular $\ell=1,\ m=0$ mode found in 2D, resulting in a reduction in the radial shock extensions along the axis, even in cases for which the total power in $\ell = 1$ modes is similar between 2D and 3D.  
Given this, we expect evidence of strong SASI in our simulations to be manifest in $\ell=1$-dominant oscillations whose amplitudes in the linear regime grow exponentially in time.  

We find unmistakable evidence of the presence and growth of the SASI in our 2D and 3D simulations for both s15 and s27 exploding and non-exploding models.
In Figure \ref{fig:lm15} we show the spherical harmonic coefficients for the first Legendre order, $a_{1m}$, for both s15 and s27. 
In the left (right) panels we show non-exploding (exploding) models.
For s15 \fheat 1.00 3D, there is two clear periods of exponential amplitude growth of $a_{1m}$, from $\sim$50-150 ms and then from $\sim$175 - 225 ms.
The first growth period is dominated by $a_{11}$, with the other coefficients growing but to smaller amplitude and slightly out-of-phase with $a_{11}$.
This is indicative of a weak sprial SASI mode.
The second period of linear growth occurs after the shock has receded substantially and shows growth in all three $a_{1m}$ modes.
During this late period of growth in s15 \fheat 1.00 3D, the $a_{10}$ and $a_{11}$ modes are in phase, resulting in a distinctive ``sloshing'' motion that is the hallmark of the SASI in 2D.
The $a_{1-1}$ mode is out-of-phase with the others, imparting an additional spiral motion to the shock.
This behavior is especially apparent in the movies of this simulation.\footnote{\url{http://flash.uchicago.edu/~smc/movies}}
In Figure \ref{fig:sasi} we show four volume renderings of s15 \fheat 1.00 3D at late times demonstrating the clear, strong SASI development.

Comparing these two periods of exponential SASI growth in s15 \fheat 1.00 3D, we note that the later epoch oscillations are of higher frequency and reach larger amplitude.
The SASI growth time is shorter due to the small amount of mass in the gain region at this epoch, which makes the advective-acoustic cycle frequency larger.
The advective-acoustic time will be proportional to the advection time through the gain region, $\tau_{\rm SASI} \propto \tau_{\rm adv} \sim M_\textrm{gain} / \dot{M}$ \citep{Foglizzo:2007cq}.
Considering s27 \fheat 1.00 3D (bottom-left panel of Figure \ref{fig:lm15}), we also see two distinct periods of exponential SASI growth, analogous to s15 \fheat 1.00 3D.
The frequency of the SASI oscillations during the first period of growth in s27, 50 - 225 ms, is greater than for the similar epoch in s15.
This is thanks to the larger mass accretion rate in s27 at these times.
As in s15, following the period of shock recession in s27 \fheat 1.00 3D the SASI again grows exponentially with a short growth time.
As seen in Figure \ref{fig:lm15}, the $\ell=1$ coefficients are also out-of-phase, giving similar spiral shock motion as is observed in s15 \fheat 1.00 3D.
Thus, both 3D failed-explosion simulations show robust development of the SASI, including the spiral mode.
The growth is faster in s27, implying this progenitor exhibits conditions more favorable to the SASI \citep[i.e.,][]{Muller:2012kq, Hanke:2013kf}, though the peak amplitudes of the $a_{\ell m}$ are similar between the two models.

The blue lines in Figure \ref{fig:lm15} represent critical non-exploding (left panels) and exploding (right panels) 2D simulations.
For s27, the $\ell = 1$ spherical harmonic coefficient in 2D reaches larger amplitudes than in 3D while showing a similar growth time scale.
This is true for both the critical exploding and non-exploding cases in s27.
The situation is different for s15, where the non-exploding 2D simulation, s15 \fheat 0.90 2D, shows small peak amplitudes of $a_{10}$ and lacks clear periodicity, evidencing a very weak SASI.
The 3D non-exploding simulation shows much stronger development of the SASI than its 2D counterpart.
The critical exploding 2D case in s15 (top-left panel of Figure \ref{fig:lm15}), however, shows clear growth of the SASI and larger peak amplitude of the $\ell =1$ coefficient than for s15 \fheat 1.05 3D.
This highlights our finding that the growth of the SASI is dependent on the neutrino heating factor, \fheat.

Figure \ref{fig:lm2d} shows $a_{10}$ for our 2D simulations at multiple heat factors.
Focusing on the period prior to runaway shock expansion, $\lesssim$150 ms, we see that every model, with perhaps the exception of s15 \fheat 0.90 2D, shows exponential growth of the SASI.
We also find a clear dependence of the period of SASI oscillations on \fheat: larger heat factors result in longer SASI periods.
This is because larger heat factors yield more rapid shock expansion and growth of the mass in the gain region (see Figure \ref{fig:heat}), increasing $\tau_{\rm adv}$.
After the explosion begins, each 2D model shows a larger spike in $a_{10}$, indicative of a strongly asymmetric explosion.
For 2D exploding models, it is difficult to separate SASI motions from convective motions.
We find neutrino-driven convection in all of our simulations, particularly for exploding models (see Section \ref{sec:convect}).
In 2D, buoyant plumes merge into large plumes preferentially on the symmetry axis. 
These large plumes are able to rise quickly \citep[e.g.,][]{Dolence:2013iw, Couch:2013fh}, contributing to the $\ell=1$ deformation of the shock.
This might be interpreted as SASI growth, but the time scale of the convection, on which a plume will rise and fall, is different from that of the advective-acoustic cycle that characterizes the growth of the SASI.
Thus, these large axial plumes that occur only in 2D can break the coherence of the advective-acoustic cycle, hampering the development of the SASI, while still causing large, quasi-periodic variation in $a_{10}$.

We also find a heat factor dependence in 3D.
For the s15 \fheat 1.05 3D model the growth of the $a_{\ell m}$’s is substantially slowed and the peak amplitudes reduced relative to the failed explosion.
For the s27 \fheat 1.05 3D case, the SASI amplitudes are similar to the non-exploding case while the frequencies are slightly lower.
This behavior is present until around 100 ms post-bounce.
At this point, neutrino-driven convection becomes dominant (see Fig. \ref{fig:chi} and discussion in Section \ref{sec:convect}).
Our s27 \fheat 1.00 3D and s27 \fheat 1.05 3D results are in remarkable agreement with the 3D GR simulations of \citet{Ott:2013gz} (compare their Figure 12 with our Figure \ref{fig:lm15} and their Figure 11 with our Figure \ref{fig:spPow}).
Up to the time they simulate ($\sim 160$ ms), we observe about the same number of oscillatory cycles (seven to their ten), and similar peak amplitudes, though ours may be a bit larger. 
\citeauthor{Ott:2013gz} find that increasing \fheat\ beyond 1.05 severely curtails the growth of the SASI and reduces the oscillatory frequency in a manner consistent with what we observe for s15 \fheat 1.05 3D.
Thus, we reach the conclusion that for exploding 3D simulations the early time SASI motions are weakened as the neutrino heating increases.
As the s15 \fheat 1.05 3D case has markedly higher neutrino heating (c.f. Table \ref{table:results} and Figure \ref{fig:heat}), this model has a stronger SASI suppression than s27 \fheat 1.05 3D.
The late time SASI motions in exploding models are non-existent.
For explosions, strong neutrino-driven convection is evident (see Section \ref{sec:convect}).
We, therefore, find two different instability regimes, one in which SASI is dominant and one in which convection is dominant.
This delineation, on the basis of progenitor structure, was first suggested by \citet{Muller:2012kq} and recently studied in detail by \citet{Fernandez:2013um}.
We see that the different regimes can occur in the same progenitor model, depending on the heat factor.
Critically, we find that, in 3D especially, the SASI is only dominant to convection in non-exploding models.
The secular outward motion of the shock, and strong convection, that occurs in exploding models appears to be unfavorable for a dominant SASI.
We find this is particularly true in 3D.

In Figure \ref{fig:spPow} we show the square root of the spherical harmonic power for the first four $\ell$'s, weighted by the mean shock radius, $a_{00}$, for both 2D and 3D simulations.
The amplitudes of $P_1$ in s15 \fheat 0.95 2D are comparable with those of s15 \fheat 1.05 3D, while the $P_1$ amplitudes in the non-exploding s15 \fheat 0.90 2D are smaller than the comparable 3D simulation, s15 \fheat 1.00 3D.
In the period leading up to explosion (0-150 ms), s15 \fheat 0.95 2D shows clear growing, oscillatory $P_1$.
The amplitudes of these oscillations are, however, relatively meager.
For the non-exploding 2D model, s15 \fheat 0.90, the amplitudes of $P_1$ and $P_2$ are similar, and more obvious oscillation is seen for $P_2$.
We also note that in both 2D and 3D, we see a large spike early-on ($t_{\rm pb} \sim 20$ ms) in $P_4$ reflecting the asymmetry induced by the Cartesian grid.
A comparable spike in $P_4$ was also seen in the 3D GR simulations of \citet{Ott:2013gz}.

Previous investigations of the differences between 2D and 3D SASI development \citep[e.g.,][]{{Burrows:2012gc}, {Couch:2013fh}} have found reduced power in the low-order spherical harmonics in 3D as compared with 2D.
For simulations that demonstrate SASI development, we find that there is not so much difference between the amplitudes of the power in a given Legendre order, $\ell$, between 2D and 3D.
This could be due, in part, to the progenitor-dependence of the SASI \citep{Muller:2012kq}, but also due to the less-realistic lightbulb treatment used in other 2D vs. 3D studies.  
The lack of PNS cooling and contraction for lightbulb schemes could negatively impact the strength of the SASI and the spatially-constant neutrino luminosity assumed in such schemes may be over-driving convection.

\subsection{Neutrino-Driven Convection}
\label{sec:convect}

\begin{figure*}[htb]
\centering
\begin{tabular}{cc}
  \includegraphics[width=3.4in,trim= 0.6in 1.0in 1.in 0.4in,clip]{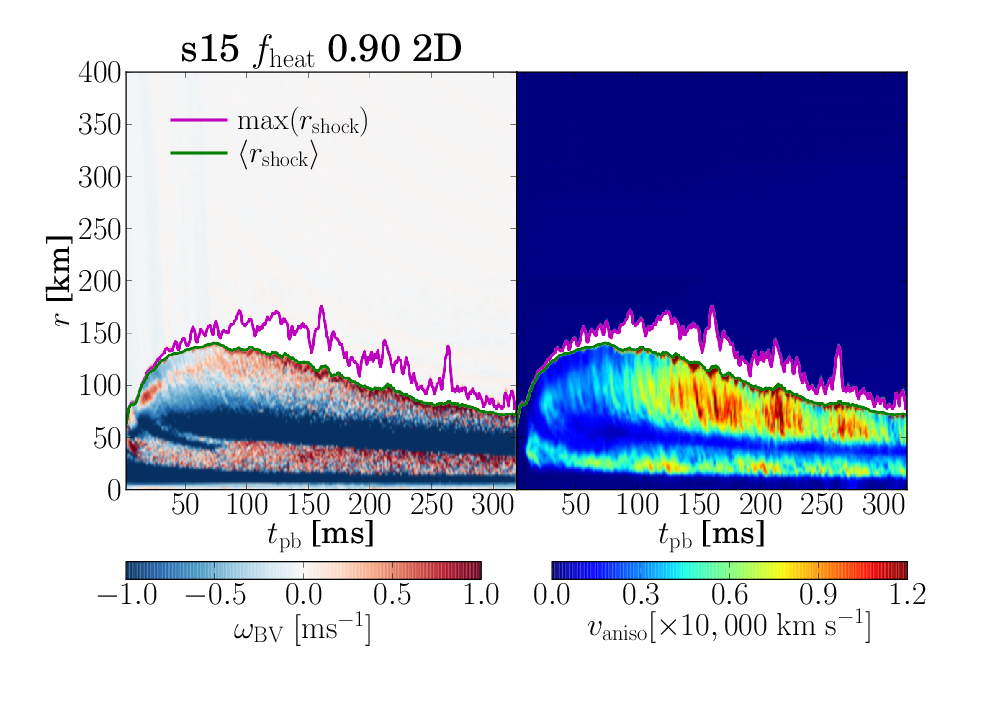} &
  \includegraphics[width=3.4in,trim= 0.6in 1.0in 1.in 0.4in,clip]{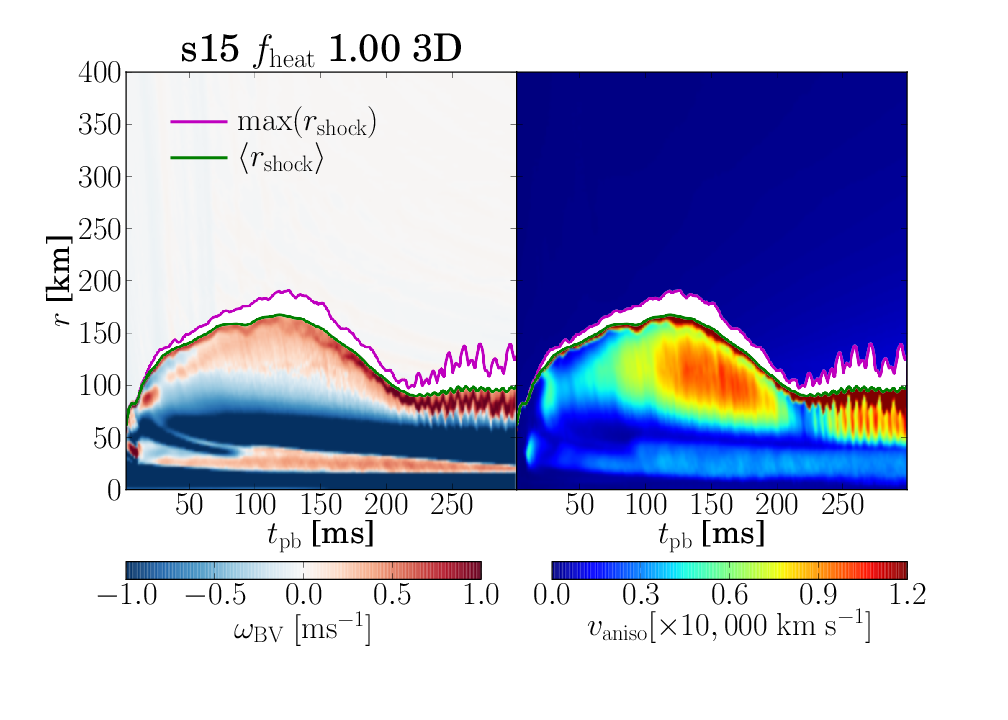} \\
  \includegraphics[width=3.4in,trim= 0.6in 1.0in 1.in 0.4in,clip]{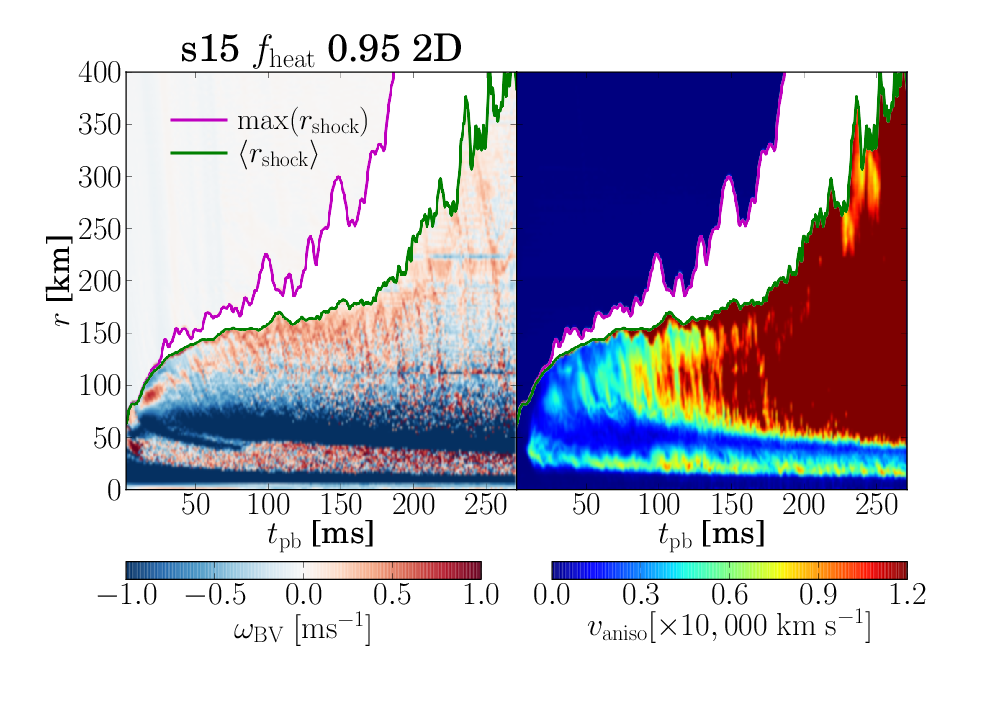} &
  \includegraphics[width=3.4in,trim= 0.6in 1.0in 1.in 0.4in,clip]{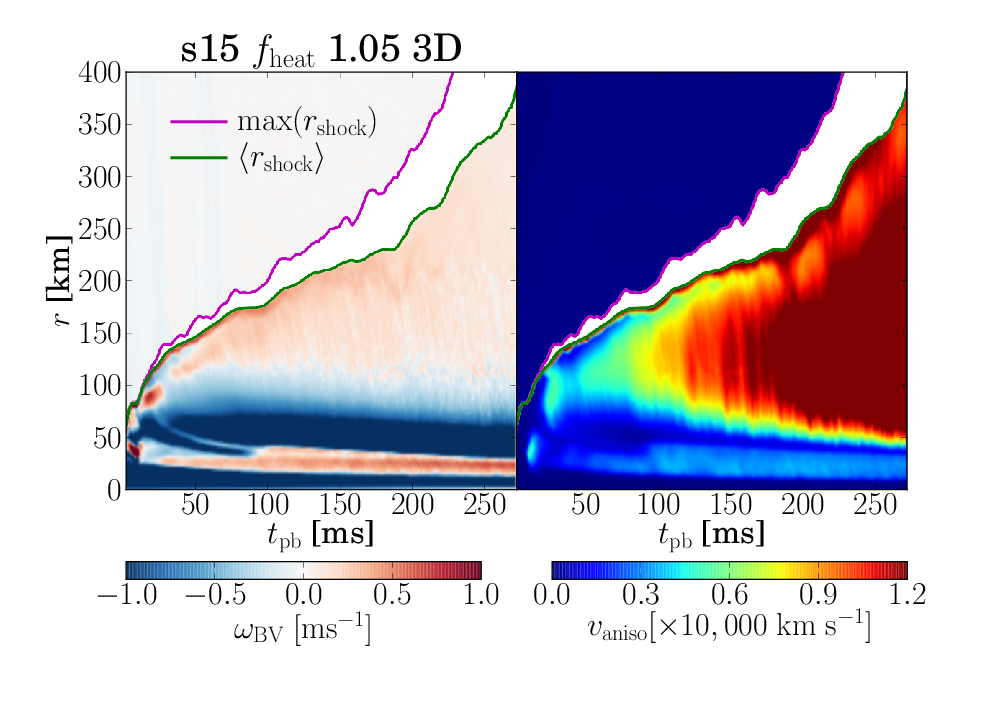} 
\end{tabular}
\caption{ 
  Pseudo-color plots of the spherically-averaged Brunt-V\"ais\"al\"a frequencies, $\omega_{\rm BV}$ (left half of each panel) and spherically-averaged anisotropic velocities, $v_{\rm aniso}$ (right half of each panel) as functions of time for s15 in both 2D and 3D.
  The top two panels display the critical non-exploding models, s15 \fheat 0.90 2D and s15 \fheat 1.00 3D, and the bottom panels show the critical exploding models, s15 \fheat 0.95 2D and s15 \fheat 1.05 3D.
  Over-plotted are the average (green lines) and maximum (magenta lines) shock radii.
  In between the maximum and average shock radii we zero-out both $\omega_{\rm BV}$ and $v_{\rm aniso}$ since their calculation is unreliable in this region.
}
\label{fig:s15vaniso}
\end{figure*}

\begin{figure*}[htb]
\centering
\begin{tabular}{cc}
  \includegraphics[width=3.4in,trim= 0.6in 1.0in 1.in 0.4in,clip]{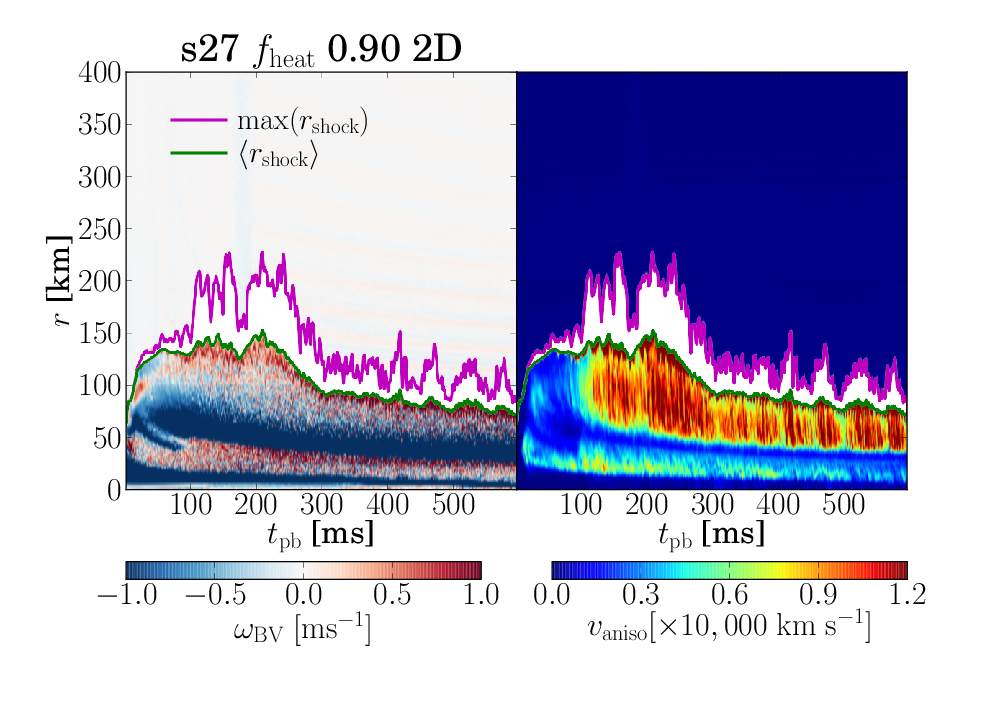} &
  \includegraphics[width=3.4in,trim= 0.6in 1.0in 1.in 0.4in,clip]{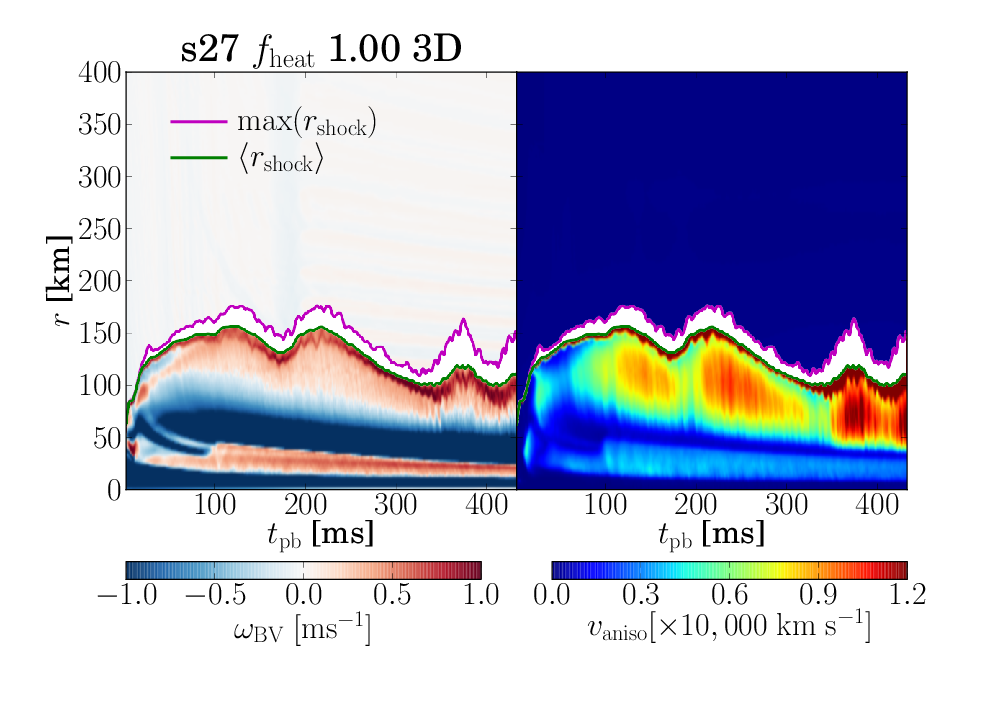} \\
  \includegraphics[width=3.4in,trim= 0.6in 1.0in 1.in 0.4in,clip]{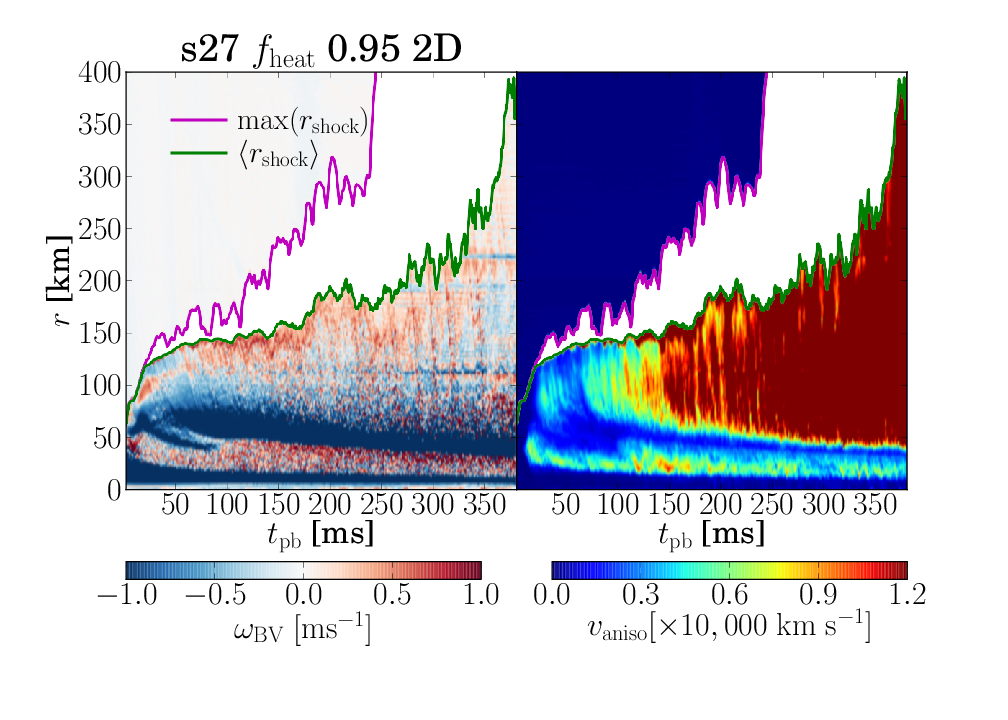} &
  \includegraphics[width=3.4in,trim= 0.6in 1.0in 1.in 0.4in,clip]{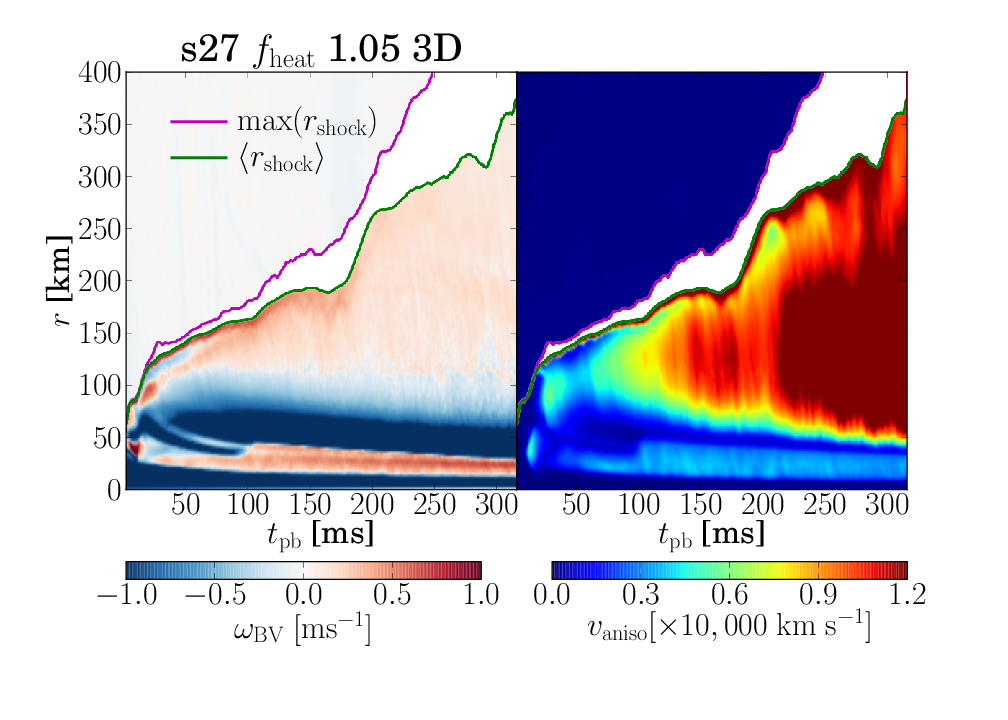} 
\end{tabular}
\caption{Same as Figure \ref{fig:s15vaniso} but for s27.}
\label{fig:s27vaniso}
\end{figure*}

\begin{figure}[htb]
\centering
\includegraphics[width=3.4in,trim= 0in 0in 0in 0.2in,clip]{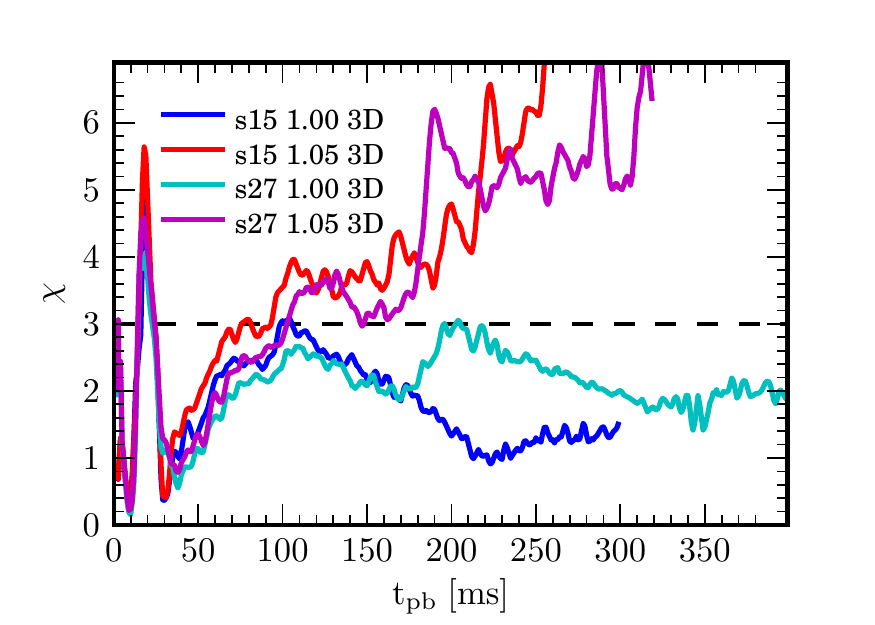}
\caption{
  Convective $\chi$ parameter of \citet{Foglizzo:2006js} for our 3D simulations in both progenitors.
  Marked as the dashed black line is $\chi=3$ which \citeauthor{Foglizzo:2006js} suggest is the critical threshold above which conditions are favorable for convection.
  Successful explosions result in $\chi>3$ after $\sim 100$ ms, while the failed explosions show smaller values of $\chi$ that remain below $\sim 3$.
}
\label{fig:chi}
\end{figure}

\begin{figure*}[htb]
  \centering
  \begin{tabular}{cccc}
    \includegraphics[width=1.92in,trim= 1in 0.75in 0.6in 0.5in, clip]{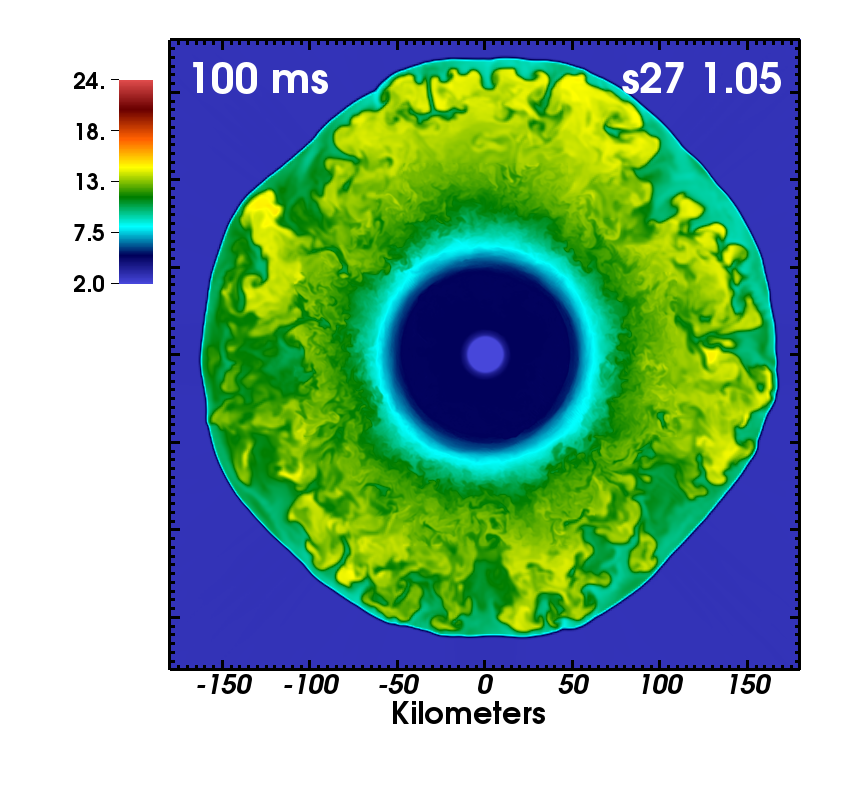} & \hspace{-0.18in}
    \includegraphics[width=1.7in,trim= 2.22in 0.75in 0.6in 0.5in, clip]{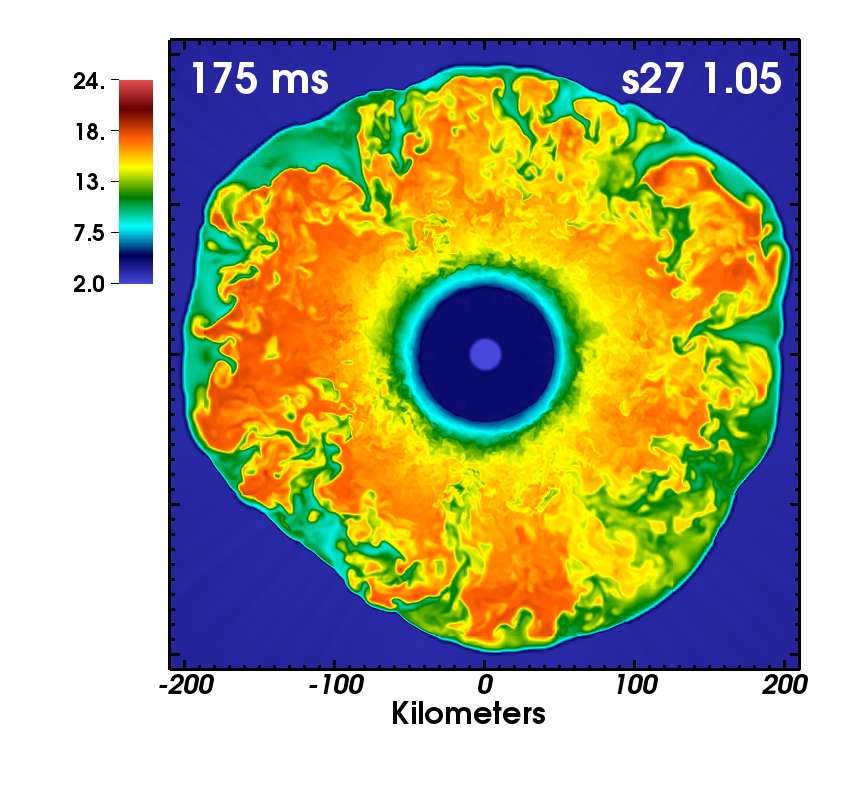} & \hspace{-0.18in}
    \includegraphics[width=1.7in,trim= 2.2in 0.75in 0.6in 0.5in, clip]{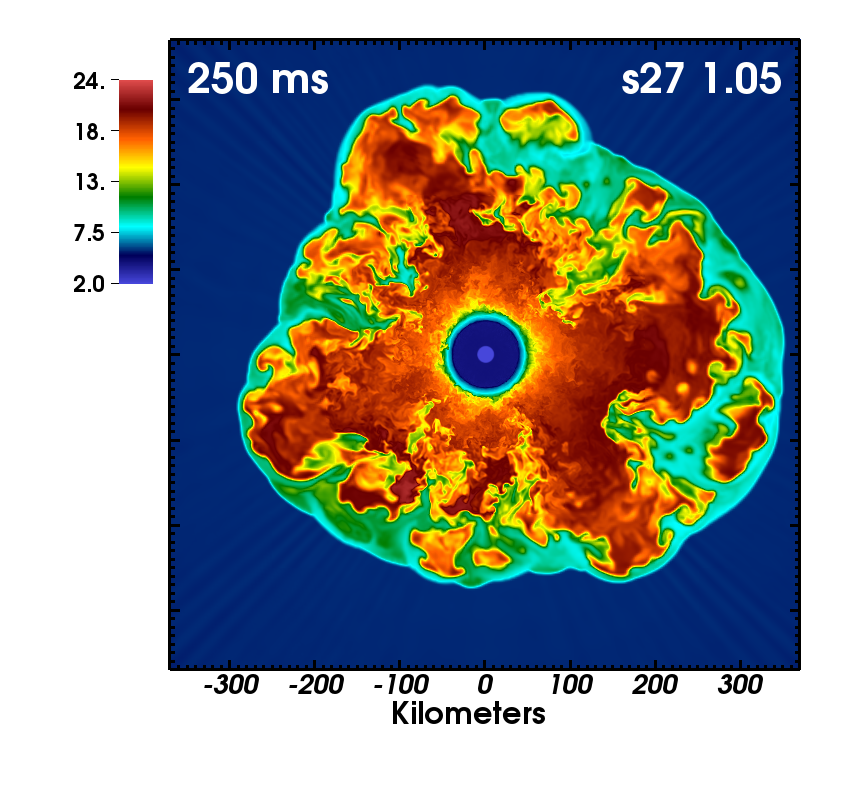} & \hspace{-0.18in}
    \includegraphics[width=1.7in,trim= 2.2in 0.75in 0.6in 0.5in, clip]{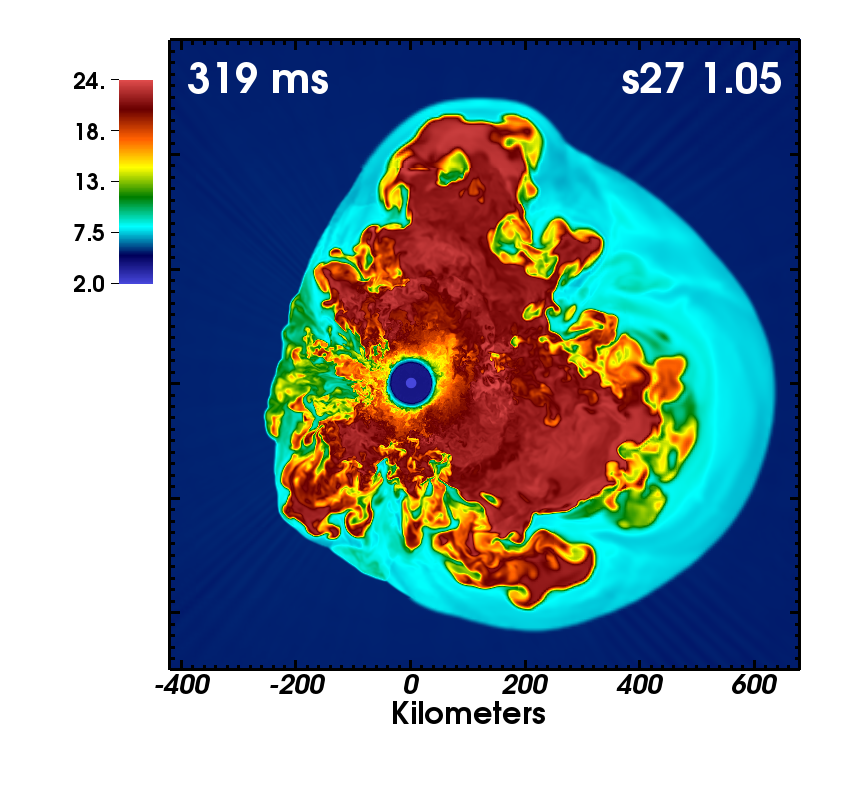} 
  \end{tabular}
  \caption{
    Pseudo-color slices of entropy at four postbounce times for s27 \fheat\ 1.05 3D. 
    The colormap and limits are indicated on the left and kept fixed for each time.
    Convection is already strong by 100 ms, as is indicated in Figures \ref{fig:s27vaniso} \& \ref{fig:chi}.
    As explosion sets in (right two panels), the convection becomes volume-filling and large, high-entropy bubbles emerge that push the shock outward.
    The explosion begins in an asymmetrical fashion (right-most panel).
    The development of convection in our simulations is very similar to that of \citet{Ott:2013gz}.
  }
  \label{fig:slices}
\end{figure*}

In Section \ref{sec:sasi}, we demonstrated the clear presence of the SASI in our simulations, in both progenitors, as well as in 2D {\em and} 3D exploding and non-exploding cases.
We argued that the strength of the SASI was substantially diminished in 3D explosions and suggested that this is due to the dominance of neutrino-driven convection, quenching the growth of the SASI.
In this subsection we seek to justify this assertion by quantifying the neutrino-driven convection in our simulations.  

In the linear regime, the growth rate of convection is given by the Brunt-V\"ais\"al\"a (BV) frequency,
\beq
\omega_{\rm BV} = {\rm sgn}(C_L) \sqrt{\left \lvert \frac{C_L}{\rho} \frac{d\Phi}{dr} \right \rvert },
\label{eq:bv}
\eeq
where $d\Phi / dr$ is the local gradient of the gravitational potential, i.e., the gravitational acceleration.  The Ledoux criterion is,
\begin{align}
  &C_L = \nonumber \\ 
  &- \left ( \frac{\partial \rho}{\partial P} \right )_{s,Y_e} 
  \left [ \left ( \frac{\partial P}{\partial s} \right )_{\rho,Y_e} 
    \left ( \frac{ds}{dr} \right ) + 
    \left ( \frac{\partial P}{\partial Y_e} \right )_{\rho,s}
    \left ( \frac{dY_e}{dr} \right ) \right ],
\label{eq:cl}
\end{align}
where we have substituted the electron fraction, $Y_e$ for the usual lepton fraction, $Y_l$.
We do this because the neutrino fractions are not readily available within our leakage scheme, but this approximation will only result in quantitative differences in the PNS, where neutrinos are either partially or fully trapped.
The thermodynamic derivatives in Equation (\ref{eq:cl}) are computed by finite differences based on data from the EOS.
We compute the BV frequency by first spherically averaging the 2D/3D data onto 1D radial grids, and then use these profiles to compute the Ledoux criterion via Equation (\ref{eq:cl}) and $\omega_{\rm BV}$ via Equation (\ref{eq:bv}).  
For $\omega_{\rm BV} < 0$ the flow is stable to convection and for $\omega_{\rm BV} > 0$ we expect convection to develop on roughly a time scale of $\tau_{\rm conv} \sim \omega_{\rm BV}^{-1}$.
In the CCSN context, positive BV frequency is an insufficient criterion for the development of convection.  
As pointed out by \citet{Foglizzo:2006js}, small perturbations that seed convection will be advected out of the gain region due to the background accretion flow.
If the advection time is sufficiently smaller than the convective growth time the development of convection will be curtailed.
As a measure of active convection, we define the anisotropic velocity, $v_{\rm aniso}$ \citep{{Takiwaki:2012ck}, {Ott:2013gz}}:
\beq
v_{\rm aniso} = \sqrt{ \frac{\langle \rho [(v_r - \langle v_r \rangle_{4\pi} )^2 
    + v_\theta^2 + v_\phi^2] \rangle_{4\pi}}{\langle \rho \rangle_{4\pi}} },
\eeq
where all the $\langle ... \rangle_{4\pi}$ are spherical averages.  
This measure quantifies the velocity scale of the motion not associated with the background radial velocity profile.

In Figures \ref{fig:s15vaniso} and \ref{fig:s27vaniso} we show pseudo-color plots of $\omega_{\rm BV}$ and $v_{\rm vaniso}$ for s15 and s27, respectively.
Plotted also are the maximum and average shock radii. 
Comparable plots can be found in \citet{Ott:2013gz}.
The top rows of both Figures show the critical non-exploding cases in both 2D and 3D.
The bottom rows show the critical exploding cases for the same.
For s15 \fheat 0.90 2D, the region of positive $\omega_{\rm BV}$ is especially small, reflective of the meager neutrino heating in this model.
For s15 \fheat 1.00 3D, more vigorous anisotropic motion is seen during the epoch of shock ascent, indicative of post-shock convection.
As also mentioned in Section \ref{sec:sasi}, following the recession of the shock in 3D the SASI develops and grows to significant amplitude while in s15 \fheat 0.90 2D there is very little evidence of such clear SASI behavior.
Also worth noting is the larger positive values of $\omega_{\rm BV}$ just behind the shock during the period of strong SASI motion for s15 \fheat 1.00 3D. 
During this time ($t_{\rm pb} > 200$ ms) there is very little neutrino heating (see Figure \ref{fig:heat}). 
The large values of the BV frequency at these times are clearly driven by the entropy gradients established by the SASI.

The exploding models in Figure \ref{fig:s15vaniso}, s15 \fheat 0.95 2D and s15 \fheat 1.05 3D, are more similar in terms of $\omega_{\rm BV}$ and $v_{\rm vaniso}$.  
Apparent, especially in 2D, is the stronger anisotropic motion during the stalled-shock phase, $t_{\rm pb} < 100$ ms.
This is caused by stronger neutrino-driven convection in the exploding models.

The situation is somewhat similar for s27, as shown in Figure \ref{fig:s27vaniso}.
A notable difference from s15 is for the 2D cases.
We see larger typical values of $v_{\rm vaniso}$ for the s27 \fheat 0.90 2D from 100-150 ms as compared with the exploding case, s27 \fheat 0.95 2D.
Combined with the strength of SASI-like motions for this progenitor (Section \ref{sec:sasi}), this suggests that the explosions in s27 may be aided by the SASI.
This is in qualitative agreement with \citet{Muller:2012kq} and \citet{Fernandez:2013um}, with the quantitative difference that we find smaller amplitude SASI motion.

For our 3D simulations, successful explosions seem to be attended by vigorous neutrino-driven convection.
To further diagnose the strength of convection in our simulations, we consider the integrated convective growth paramater \citep{Foglizzo:2006js},
\begin{equation}
\chi = \int_{r_\textrm{gain}}^{r_{\rm shock}} \frac{\omega_{\rm BV}}{\abs{v_r}} dr.
\end{equation}
\citet{Foglizzo:2006js} argue that for $\chi < 3$, perturbations that seed convection would be advected out of the gain region before convective motion could become well-established.
The 2D simulations of \citet{Buras:2006hl} provide experimental support for this critical value of $\chi$, although these authors also show that for larger perturbations entering the gain region, the $\chi$ necessary for strong convection is reduced.
In Figure \ref{fig:chi}, we show the time evolution of the $\chi$ parameter for our 3D simulations in both s15 and s27.  
All models show an early, $< 30$ ms, spike in $\chi$ that is associated with the overly-strong prompt convection that our leakage scheme produces, as mentioned in Section \ref{sec:neutrinoEffects}.
Precisely the same behavior is found in \citet{Ott:2013gz}.
This strong prompt convection is also visible in Figures \ref{fig:s15vaniso} and \ref{fig:s27vaniso}.
These figures show brief epochs of very large $\omega_{\rm BV}$ in the gain region on the same time scales as the spikes in $\chi$ shown in Figure \ref{fig:chi}.
This early over-driving of convection quickly subsides and Figures \ref{fig:s15vaniso} and \ref{fig:s27vaniso} also show that this convective region is advected away from the gain region and eventually joins the region of PNS convection, around 100 ms in all models.
Figure \ref{fig:chi} shows that the $\chi$ parameter is greater for the exploding 3D models, and in both exploding and non-exploding cases s15 has larger typical values of $\chi$ than s27.
This is the case in the non-exploding models until the shock in s15 has receded to small radii while the shock in s27 \fheat 1.00 3D remains at larger radius due to the infall of the Si/O interface, around 200 ms.
For both non-exploding 3D models, $\chi$ does not exceed the critical value of 3, except at the time of maximum shock extension, and then only barely.
It is important to note that the critical $\chi$ value of 3 is based on a quasi-steady-state, stalled-shock scenario.
In our simulations, this is a poor approximation of the conditions during the phase of shock expansion, $t_{\rm pb} < 100$ ms.
During this epoch, we see clear evidence for convective motion in Figures \ref{fig:s15vaniso} and \ref{fig:s27vaniso}, and particularly in the visualizations of these simulations, yet $\chi < 3$, in agreement with \citet{Ott:2013gz}.
This highlights the complication of ambiguous definition for $\chi$ in multidimensional simulations.
Here, we base our calculation of $\chi$ on angle averages, but this can skew the results in a manner that makes interpretation of $\chi$ troublesome.
The difficulty of defining $\chi$ is also discussed by \citet{Fernandez:2013um}, who find very different values for $\chi$ depending on the method of calculation.

In Figure \ref{fig:slices}, we demonstrate the development of convection in s27 \fheat 1.05 3D with entropy pseudo-color plots.
We show four meridional slices at different postbounce times, beginning at 100 ms and ending with the final state of the simulation at 319 ms.
We see 3D convection that is strikingly similar to \citet{Ott:2013gz}.
The high-entropy buoyant plumes tend to be much smaller than their 2D counterparts and are shredded by parasitic instabilities as they rise through the post-shock accretion flow (see Section \ref{sec:turb}).
Large buoyant plumes are able to survive long enough to rise all the way to the shock, stochastically pushing the shock outward, aiding explosion \citep[see also,][]{{Dolence:2013iw}, {Couch:2013fh}, Ott:2013gz}.
As explosion sets in (right two panels of Figure \ref{fig:slices}), larger buoyant plumes appear and remain coherent.
The right-most panel of Figure \ref{fig:slices} shows that the explosion occurs asymmetrically, in an initially $\ell =1$ fashion, similar to the results of \citet{Dolence:2013iw}.

\subsection{Turbulence}
\label{sec:turb}

\begin{figure}[tb]
  \includegraphics[width=3.4in,trim= 0in 0in 0in 0.2in,clip]{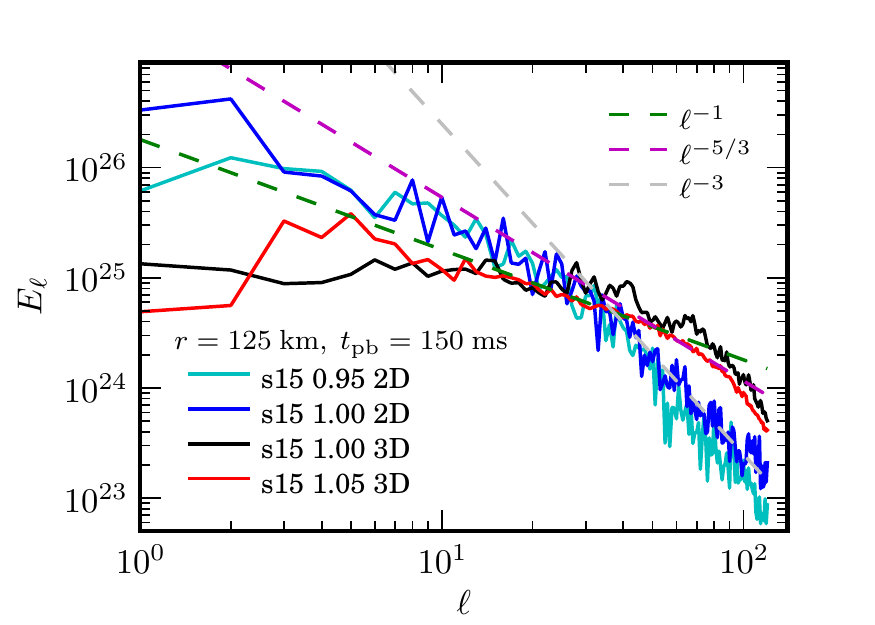} \\
  \includegraphics[width=3.4in,trim= 0in 0in 0in 0.2in,clip]{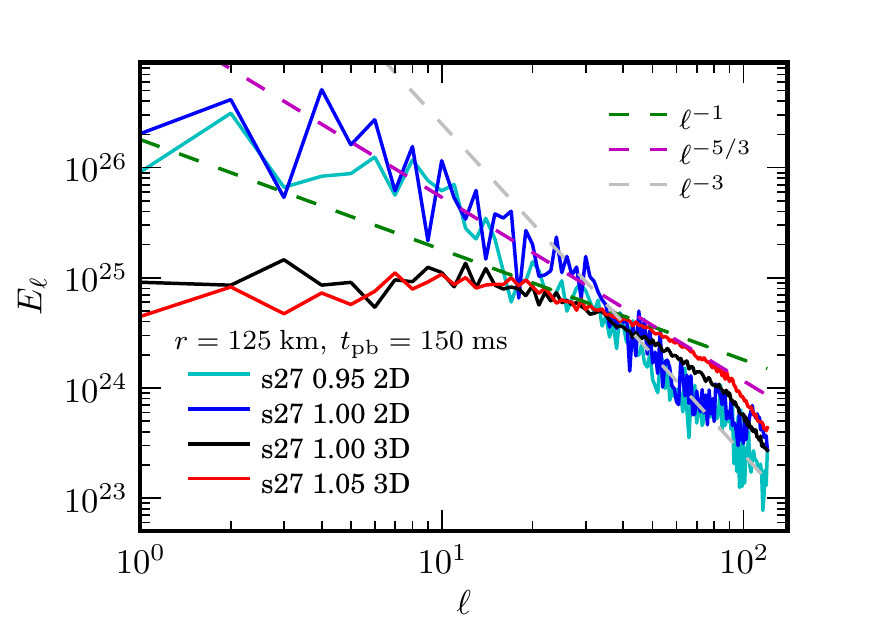}
  \caption{Turbulent kinetic energy spectra, as measured by the non-radial component of the velocity.
    The top panel shows 2D and 3D spectra for s15 and the bottom panel displays the same for s27.
    The $E_\ell$ are averaged over a 10 km-wide shell, centered on a radius of 125 km, and over 10 ms, centered at 150 ms postbounce.
    In all cases, 2D simulations result in much greater kinetic energy density on large scales than 3D. 
    Kinetic energy on large scales has been suggested to be conducive to explosion \citep{Hanke:2012dx}.
  }
  \label{fig:enerSpec}
\end{figure}

Turbulence in the CCSN context can be driven by convection \citep{Murphy:2013eg}, the SASI \citep{Endeve:2012ht}, or by the magnetorotational instability \citep{{Obergaulinger:2009fv}, {Sawai:2013er}}.
Turbulent stresses can aid shock expansion in multidimensional simulations of CCSNe \citep{Murphy:2013eg}.
The presence of strong turbulent motions behind the forward shock during the explosion phase may even effect collective neutrino flavor oscillations \citep{Lund:2013dv}.
Based on the global CCSN turbulence model developed by \citet{Murphy:2011ci}, \citet{Murphy:2013eg} argue that the turbulence in neutrino-powered CCSNe explosions is primarily the result of neutrino-driven convection. 
Here, rather than focus on the primary driver of turbulence in our simulations, we address the differences in the development of turbulence between 2D and 3D.

Following a number of previous studies, we examine turbulent motion by decomposing the non-radial component of the kinetic energy density in terms of spherical harmonics \citep[e.g.,][]{{Hanke:2012dx}, {Dolence:2013iw}, {Couch:2013fh}, {Fernandez:2013um}}.
We define coefficients,
\begin{equation}
  \epsilon_{\ell m} = \oint \sqrt{\rho (\theta, \phi)} v_t (\theta, \phi) Y_\ell^m (\theta, \phi) d\Omega,
\end{equation}
where the transverse velocity magnitude is $v_t = [v_\theta^2 + v_\phi^2]^{1/2}$.
The non-radial kinetic energy density as a function of $\ell$ is then
\begin{equation}
  E_\ell = \sum_{m=-\ell}^\ell \epsilon_{\ell m}^2\ [{\rm erg\ cm^{-3}}].
\end{equation}
In Figure \ref{fig:enerSpec}, we show the $E_\ell$ spectra for s15 (top) and s27 (bottom) in both 2D and 3D.
The spectra are computed in a 10 km-wide spherical shell centered on a radius of 125 km and at a postbounce time of 150 ms.  
This time and radius were chosen to coincide with the initial development of strong non-radial motion yet prior to onset of significant shock expansion or contraction (see Figs. \ref{fig:s15vaniso} \& \ref{fig:s27vaniso}).
Immediately apparent is that 2D simulations have {\it much} greater turbulent kinetic energy on large scales (small $\ell$) than 3D.
This is the case even when comparing the 2D $f_\textrm{heat} = 0.95$ cases with the 3D $f_\textrm{heat} = 1.05$ cases.
Similar behavior is found in other comparisons of turbulence in 2D and 3D \citep{Hanke:2012dx, Dolence:2013iw, Couch:2013fh}.
These studies also found that non-radial kinetic energy on large scales correlated with vigor of explosion.
\citet{Hanke:2012dx} even suggest that non-radial kinetic energy on large scales, by significantly increasing matter dwell times in the gain region, could be key to the success of the neutrino mechanism. 
Our results also support this conclusion; the closer a model is to explosion, the larger the turbulent kinetic energy on large scales.

It is well-known that turbulence in 2D exhibits very different behavior than in 3D.
The most significant difference, particularly for the present discussion, is the so-called ``inverse energy cascade'' in 2D.
According to Kolmogorov's theory of turbulence, turbulent energy is injected on large scales and subsequently is transfered via the turbulent cascade to small scales \citep{Kolmogorov:1941vt}.
In 2D, turbulent energy is still injected at the large, {\it driving} scale, but from there cascades to {\rm large} scales instead.
Enstrophy, the integrated squared-vorticity, experiences the forward cascade to small scales in 2D \citep{Kraichnan:1967jk}.
Independent of dimension, for fully-developed, isotropic, incompressible turbulence, the energy and enstrophy cascades have canonical power-law scalings of -5/3 and -3, respectively.
These three qualifications are not always satisfied in the CCSN context, but it is still interesting to examine the scaling behavior of the turbulent energies in our simulations.

Figure \ref{fig:enerSpec} displays three power-law scalings, -1, -5/3, and -3, along with the turbulent kinetic energy spectra.
At small scales in 2D, we see the $\ell^{-3}$ behavior expected from turbulence theory.
Going to larger scales in 2D, the spectra tend to follow a scaling law somewhere between -1 and -5/3, also roughly in agreement with expectations.
The 3D spectra, as anticipated, show very different scaling than for 2D, though the agreement with turbulence theory is less clear.
Over a broad range in $\ell$ ($\sim 10 - 40$), the 3D spectra are roughly consistent with a power-law slope of -1, as was also found in similar analysis by \citet{Dolence:2013iw}.
The spectra become steeper at larger $\ell$, following the -5/3 power-law up to $\ell \sim 80$.
It should be stressed, that while this broken power-law behavior of the 3D spectra is suggestive, there are not clear {\it breaks} in the spectra.
Instead the smooth steepening is more indicative of {\it exponential} scaling with $\ell$ than of a simple power-law.
We are also unaware of theoretical support for a turbulent cascade that scales as $\ell^{-1}$.
It is worth noting that our 2D and 3D turbulent energy spectra are qualitatively very similar to those of \citet{Dolence:2013iw}.

The transition of the 2D spectra from a -5/3 power-law to a -3 power-law occurs around $\ell \sim 40$, which also corresponds to the scale at which the 3D spectra take up the -5/3 behavior.
Based on this, we might expect $\ell \sim 40$ to be the scale at which the turbulence is being driven, but this $\ell$ corresponds to rather small linear scales ($dx^i \sim 2\pi r (\ell+1)^{-1} \sim 19$ km).
\citet{Hanke:2012dx} argue on the basis of similar kinetic energy spectra that the turbulent driving scale in their simulations is $\ell \sim 10$, which would correspond to half the width of the gain region.
If the turbulence is the result of neutrino-driven convection in the gain region, this is the scale at which we should expect the driving to occur.
The breakdown of the agreement of our simulations with turbulence theory may be the result of the aforementioned differences between the turbulence in CCSNe and Kolmogorov-type turbulence.
The turbulent speeds in CCSN can reach appreciable fractions of the sound speed, breaking the assumption of incompressibility.
The background radial velocity field is also problematic for the isotropy of the turbulence.
And, importantly, the turbulence we analyze here is not truly full-developed, quasi-steady-state turbulence.

\begin{figure}[tb]
  \includegraphics[width=3.4in,trim= 0in 0in 0in 0.2in,clip]{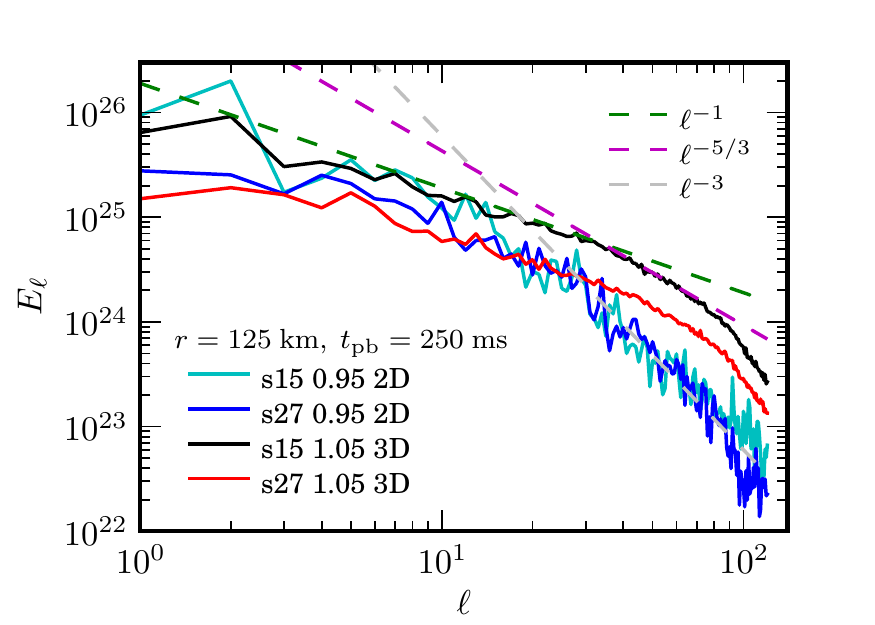}
  \caption{
    Turbulent kinetic energy spectra as in Figure \ref{fig:enerSpec}, but at $t_{\rm pb} = 250$ ms.
    Shown are 2D and 3D results for both s15 and s27.
    At this time, large-scale radial outflow has set in as the models transition to explosion.
    Notably, the 3D cases show relatively more energy at large scales than for the earlier time shown in Figure \ref{fig:enerSpec}.
    The overall normalization of the spectra is lower than for the earlier epoch, reflecting that a large amount of the earlier, non-radial kinetic energy has gone into powering radial outflow.
  }
  \label{fig:enerSpec2}
\end{figure}

To demonstrate this last point, we consider the turbulent energy spectra at a later time in our simulations.
Figure \ref{fig:enerSpec2} shows the turbulent kinetic energy density spectra at 250 ms postbounce for four exploding simulations: s15 \fheat 0.95 2D, s27 \fheat 0.95 2D, s15 \fheat 1.05 3D, and s27 \fheat 1.05 3D.
The 2D spectra are fairly similar to their earlier counterparts shown in Figure \ref{fig:enerSpec}, but the 3D spectra have clearly evolved.
Specifically, much greater turbulent energy is found on larger scales in 3D at this later time.
The power-law scalings discussed above still roughly hold, and the 3D spectra seem to following the $\ell^{-1}$ scaling down to much smaller $\ell$.
The enormous difference in the energy density on large scales between 2D and 3D is now also absent, which is unsurprising since all four simulations are in the midst of large scale radial outflow, i.e., explosion (see Figures \ref{fig:s15vaniso}, \ref{fig:s27vaniso}, \& \ref{fig:slices}).

The salient point of our turbulence analysis is that, prior to the initiation of explosion, the 2D inverse energy cascade artificially pumps turbulent energy to large scales where it aids in shock expansion. 
In simple, practical terms, convective plumes start out at the driving scale in 2D and, in subsequent evolution, merge together to form larger, even more buoyant plumes, thus increasing the turbulent kinetic energies on large scales.
Due to the constraint of the symmetry axis, these large plumes tend to develop along this axis and can mimic the $\ell = 1$ motion of the SASI.
This helps 2D simulations with lower heat factors to achieve explosion by enhancing the maximal extensions of the shock.
In 3D, the opposite occurs.
Convective plumes that reach sufficient size to be buoyant (roughly the driving scale of the turbulence), rise and are torn apart into smaller plumes, cascading turbulent energy to ever smaller scales.

\subsection{Resolution Dependence}

\begin{figure*}[tb]
  \centering
  \begin{tabular}{cc}
    \includegraphics[width=3.4in,trim= 0in 0.15in 0.3in 0.45in,clip]{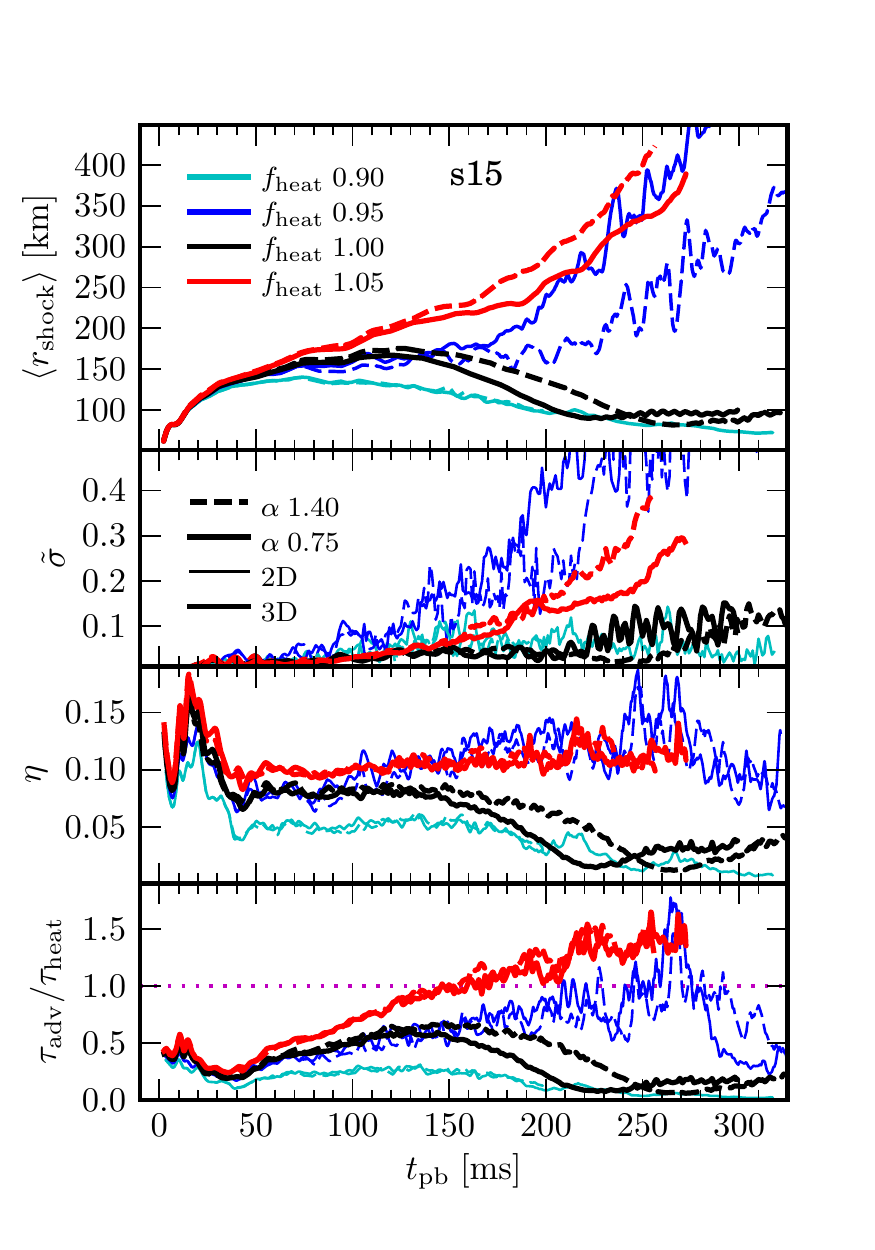} &
    \includegraphics[width=3.4in,trim= 0in 0.15in 0.3in 0.45in,clip]{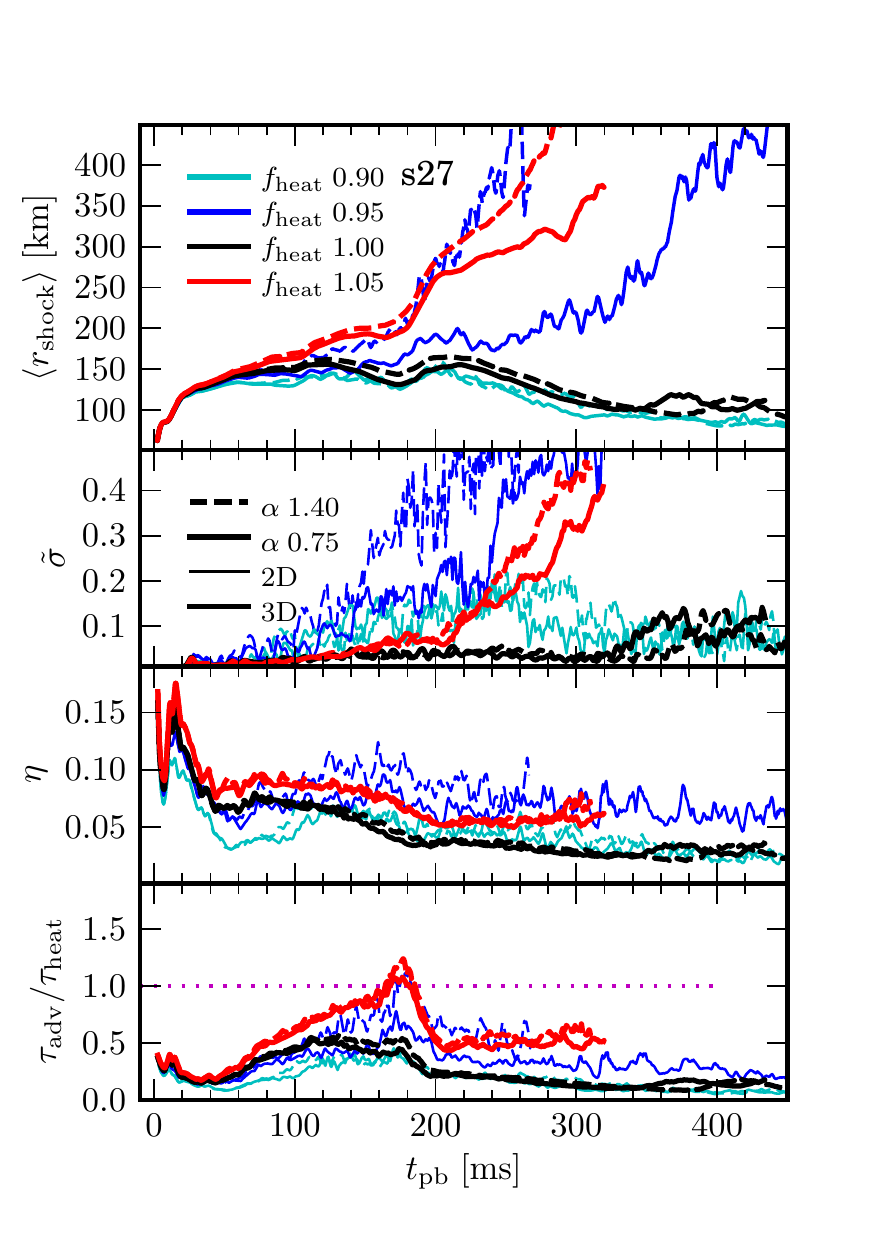} 
  \end{tabular}
  \caption{
    Various global metrics for 2D and 3D simulations at different resolutions for s15 (left) and s27 (right).
    The top panels are the average shock radius, the second panels are the normalized standard deviation of the shock radius, the third panels is the net heating efficiency, and the bottom panels are the ratio of advection-to-heating time scales.
    Shown only are the critical exploding and non-exploding cases.
    There is very little significant difference in the global metrics between the 3D simulations at different resolutions.
    The differences are larger for 2D, although as shown in Table \ref{table:results}, these differences are reduced for larger \fheat.
  }
  \label{fig:resolution}
\end{figure*}

In order to assess the sensitivity of our results to resolution, we have carried out a number of 2D and 3D simulations at reduced resolution for multiple heat factors.
The results of this resolution study are summarized in Table \ref{table:results}.
For the low-resolution simulations, we increase the value of the effective angular resolution parameter, $\alpha$, from our fiducial value of 0.75 to 1.40.
This gives a typical `angular' grid spacing of $dx^i / r \sim 0.81^\circ$.
The minimum grid spacing covering the PNS is kept the same at 0.49 km, and the first decrement in refinement level occurs at a radius of 60 km, instead of 100 km.
This gives a reasonably large change in resolution in the gain region and beyond of nearly a factor of two as compared with our fiducial resolution.

Table \ref{table:results} lists the postbounce times at which the shock radius exceeds 400 km for each of the exploding simulations, $t_{400}$.
In 3D, we find that for both progenitors decreasing the resolution results in faster shock expansion, i.e. decreased $t_{400}$.
The change in $t_{400}$ for the 3D simulations is, however, rather modest.
For s15 decreasing the resolution gives a change in $t_{400}$ of $-23$ ms, and for s27 the change is $-51$ ms.
The 2D simulations are more sensitive to resolution, and the sign of the change in $t_{400}$ is different between the two progenitors.
For s27 in 2D, decreasing the resolution yields smaller $t_{400}$, as in 3D.
The opposite occurs for s15 where decreased resolution delays explosion, increasing $t_{400}$.
The magnitude of the differences, as given in Table \ref{table:results}, decreases with larger \fheat, but the sign of the change is consistent as \fheat\ changes.
For all but the critical \fheat\ values, the differences in $t_{400}$ for 2D simulations are also modest, typically $\lesssim 25$ ms.
Importantly, for the heat factors we simulate, decreasing the resolution significantly does not result in a change in the critical \fheat\ for explosion, nor does it significantly change the heating efficiency in either 2D or 3D.

In Figure \ref{fig:resolution} we show some global metrics of our simulations for different resolutions.
For both 2D and 3D, we display only the critical exploding and non-exploding simulations for both progenitors.
As can be seen, the differences in these metrics between simulations of different resolution are modest, particularly in 3D.
For 2D, greater average shock radius is accompanied by larger values of the standard deviation of the shock surface, indicating greater shock elongation along the symmetry axis.
In summary, while there is some level of dependence of our results on resolution, more so in 2D than in 3D, changing the resolution by nearly a factor of two does not appear to significantly alter the stability of a given model, nor are the main conclusions of this work changed.

\section{Conclusions}
\label{sec:conclusion}

We have carried out a series of 1D, 2D, and 3D CCSN simulations in two different progenitors using the FLASH simulation framework.
Our simulations employ very-high resolution (effective angular resolution of 0.43$^\circ$) and an efficient multispecies neutrino leakage scheme that captures the salient aspects of more detailed neutrino transport simulations.
We find that, for all cases, 2D simulations explode more easily and more vigorously than comparable 3D simulations, confirming the results of \citet{Couch:2013fh}.
In addition, we find that without additional heating, beyond that which our neutrino leakage scheme would nominally predict, our 3D simulations fail to explode.
The 3D neutrino transport simulations of \citet{Hanke:2013kf} lend support to this conclusion; all else being equal in their simulations they find that s27 does not explode in 3D yet does explode in 2D.
The possibility that 3D explosions were harder to obtain than 2D was raised by \citet{Hanke:2012dx}, but the strong resolution dependence of their results precluded any definite conclusions.
The comparative ease of achieving 2D explosions as compared with 3D has also been reported recently by \citet{Takiwaki:2013ui}.

Our work goes beyond previous detailed comparisons of 2D and 3D CCSN simulations \citep{{Nordhaus:2010ct}, Hanke:2012dx, {Dolence:2013iw}, {Burrows:2012gc}, {Murphy:2013eg}, Couch:2013fh} which employed a parameterized neutrino heating and cooling scheme based on that of \citet{Murphy:2008ij}, approximate or no postbounce evolution of the electron fraction, monopole gravity, and a single progenitor, s15s7b2 of \citet{Woosley:1995jn}.\footnote{\citet{Hanke:2012dx} also consider a 11.2-\Msun~progenitor.}
In the present study, we have greatly extended the basic approach of \citet{Couch:2013fh} with the inclusion of multispecies neutrino leakage, a new multipole gravity approach \citep{Couch:2013ws}, the use of two modern progenitors, and higher resolution.
The neutrino leakage scheme, which has also been used in the 3D simulations of \citet{{Ott:2012ib}, {Ott:2013gz}}, self-consistently calculates the cooling and lepton number evolution due to electron (anti-) neutrinos and heavy-lepton neutrinos, as well as the heating due to electron-type neutrinos and antineutrinos. 
This scheme reproduces a number of the key global results of realistic neutrino transport calculations, though there are still important differences in the detailed results from the leakage scheme and from transport (see Section \ref{sec:neutrinoEffects}).

We have examined a number of crucial differences that arise between 2D and 3D simulations.
In 2D, the presence of the symmetry axis artificially exacerbates the growth of both convection and the SASI.
This results in an enhancement of the shock extension along the axis, bringing additional mass into the gain layer.
Due to the strong non-linearities of the problem, this sets off a chain reaction wherein the additional mass in the gain region results in a greater net neutrino heating, which in turn pushes the shock outward, bringing even more mass into the gain region (see Figure \ref{fig:heat}).

In Sections \ref{sec:sasi} - \ref{sec:turb}, we measure the differences in the development of the SASI, convection, and turbulence between 2D and 3D.
When considering 2D and 3D models near the critical instability threshold, we find that the peak amplitudes of the SASI motion, as measured by the spherical harmonic coefficients $a_{\ell m}$, tend to be larger in 2D than in 3D.
The total {\it power} in the $\ell = 1$ mode of the shock is similar between these 2D and 3D models, but in 3D this power is shared amongst more degrees of freedom than in 2D, where only the $m=0$ mode is available.
This results in greater periodic extension of the shock along the symmetry axis.
Some previous studies of the SASI in 2D and 3D \citep{Burrows:2012gc, Couch:2013fh} had found that the power in $\ell = 1$ modes was reduced in 3D as compared with 2D, which is somewhat different than we find here.
This could be due to the use of the neutrino ``lightbulb'' in these previous studies, or to peculiarities of the s15s7b2 progenitor model they used, since there is progenitor dependence to the SASI \citep{{Muller:2012kq}, {Fernandez:2013um}}.
We unmistakably identify the SASI in both s15 and s27.
\citet{Hanke:2013kf} also show clear evidence of the SASI in their 3D full-transport simulations.
The SASI grows especially strong at late times in failed explosions, after the shock has receded to small radii.
As we increase the heat factor to obtain explosions in 3D, we find that the strength of the SASI is either stagnated at low values for the s27 models, or significantly reduced in the s15 models. (see Figure \ref{fig:lm15}).
Similar behavior was also found for s27 in the 3D general-relativistic simulations of \citet{Ott:2013gz}.
As in that study, we find that convection is strong for 3D explosions (see Section \ref{sec:convect} and Figure \ref{fig:chi}).
Caution must be taken, however, when interpreting the implications of this result for the significance of the SASI in the explosion mechanism; since in order to obtain explosions in 3D we must artificially enhance the neutrino heating, convection could be over-driven in a manner that incorrectly suppresses the SASI and that does not occur in Nature.

We have analyzed the behavior of turbulence in our simulations in Section \ref{sec:turb}, where we present turbulent kinetic energy spectra of our results.
There is clear evidence of the 2D ``inverse'' turbulent energy cascade pumping energy to large scales in our simulations (see Figure \ref{fig:enerSpec}), as has been observed in other works \citep{Hanke:2012dx, Dolence:2013iw, Couch:2013fh}.
Such behavior is not observed for 3D simulations prior to explosion where turbulent energy instead cascades to small scales. 
This inverse cascade is a reflection of the tendency for turbulent/convective eddies to merge into larger eddies in 2D.
These larger eddies, which preferentially form along the symmetry axes, are then excessively buoyant and rise more quickly than smaller eddies \citep[e.g.,][]{{Dolence:2013iw}, Couch:2013fh}.
The collective effect of this is to contribute to a faster shock expansion in 2D than in 3D where, rather than merge, turbulent eddies are shredded apart into smaller eddies.
The inverse cascade, being in some sense a reflection of the forced symmetry in 2D, is artificial.
We have demonstrated that this artificial behavior of 2D turbulence can have a critical impact on the qualitative results of CCSN simulations.
Additionally, the differences in the development of the SASI, convection, and turbulence between 2D and 3D could have enormous impacts on predicted neutrino \citep[e.g.,][]{Tamborra:2013uf, Lund:2013dv} and gravitational wave \citep[e.g.,][]{Muller:2012bz} signals.

In summary, we have illustrated significant differences between 2D and 3D simulations, particularly with respect to the development of critical instabilities and turbulence.  
These differences result in artificially-favorable conditions for explosion in 2D.
\citet{Dolence:2013iw} find quite a different result.
While they do not find a difference between the critical luminosities between 2D and 3D, as was reported by \citet{Nordhaus:2010ct}, they find that 3D explosions explode more rapidly than comparable 2D simulations.
We find quite the opposite, as did \citet{Couch:2013fh} who used a much more similar numerical approach to \citeauthor{Dolence:2013iw} than we do here.
The reason for the persistent disagreement is unclear.
Other 2D/3D comparisons are, however, seemingly in agreement with our result that 2D explodes more easily than 3D \citep{Hanke:2012dx, Hanke:2013kf, {Takiwaki:2013ui}}.

The difficulty of explosion in 3D has troubling implications for the resilience of the handful of successful 2D simulations now in the literature \citep[e.g.,][]{{Marek:2009kc}, {Suwa:2010wp}, {Muller:2012gd}, {Bruenn:2013es}}.
Given that most of these 2D explosions are already somewhat marginal and under-energetic, comparable simulations in 3D may fail to explode entirely.
The first 3D full neutrino transport simulation accomplished \citep{Hanke:2013kf} seems to imply just this:  the comparable 2D simulation explodes while the 3D simulation fails.
This is not the end of the story, however, as the enormous expense of this simulation required the use of rather coarse resolution, $\Delta \theta \sim 2^\circ$, much larger than even the low-resolution cases we considered.
Although, in 3D we consistently find that coarser resolution favors explosion.
The poor prospects of robust neutrino-driven explosions in 3D may be signaling that our simulations {\it are missing key physics}, or are otherwise not capturing the included physics with sufficient accuracy.
After all, massive stars explode robustly in Nature all the time.
More realistic, multi-dimensional progenitor structures is an enticing prospect for important missing details that could have a positive impact on the potential for neutrino-driven explosions in 3D.
As demonstrated recently by \citet{Couch:2013tn}, pre-collapse asphericity in the progenitor, that is a natural outcome of realistic convective burning, results in more favorable conditions for explosion in 3D.
The impact of magnetic fields and rotation in 3D CCSN simulations also merits further exploration.
Irrespective of the favorability for explosion, the preponderance of current evidence indicates that studying CCSN in 3D is a {\it necessity}.

\acknowledgements 
We thank Ernazar Abdikamalov, Adam Burrows, Rodrigo Fernandez, Thomas Janka, Eric Lentz, Christian Ott, Chris Thompson, Todd Thompson, and Craig Wheeler for many helpful conversations.
We are also grateful for the support and hospitality of the Institute for Nuclear Theory at the University of Washington, where parts of this work began.
SMC is supported by NASA through Hubble Fellowship grant No. 51286.01 awarded by the Space Telescope Science Institute, which is operated by the Association of Universities for Research in Astronomy, Inc., for NASA, under contract NAS 5-26555.  
The software used in this work was in part developed by the DOE NNSA-ASC OASCR Flash Center at the University of Chicago.  
This research used computational resources at ALCF at ANL, which is supported by the Office of Science of the US Department of Energy under Contract No. DE-AC02-06CH11357.  

\bibliography{extraRefs,papersDB}

\begin{thebibliography}{81}
\expandafter\ifx\csname natexlab\endcsname\relax\def\natexlab#1{#1}\fi

\bibitem[{Baade \& Zwicky(1934)}]{Baade:1934ed}
Baade, W., \& Zwicky, F. 1934, in Proceedings of the National Academy of
  Sciences of the United States of America, 254--259

\bibitem[{Bethe(1990)}]{Bethe:1990he}
Bethe, H.~A. 1990, Reviews of Modern Physics, 62, 801

\bibitem[{Bethe \& Wilson(1985)}]{Bethe:1985da}
Bethe, H.~A., \& Wilson, J.~R. 1985, \apj, 295, 14

\bibitem[{Blondin \& Mezzacappa(2006)}]{Blondin:2006dv}
Blondin, J.~M., \& Mezzacappa, A. 2006, \apj, 642, 401

\bibitem[{Blondin \& Mezzacappa(2007)}]{Blondin:2007fk}
------. 2007, Nature, 445, 58

\bibitem[{Blondin \& Shaw(2007)}]{Blondin:2007bf}
Blondin, J.~M., \& Shaw, S. 2007, \apj, 656, 366

\bibitem[{Bruenn {et~al.}(2013)Bruenn, Mezzacappa, Hix, Lentz, Bronson~Messer,
  Lingerfelt, Blondin, Endeve, Marronetti, \& Yakunin}]{Bruenn:2013es}
Bruenn, S.~W. {et~al.} 2013, \apj, 767, L6

\bibitem[{Buras {et~al.}(2006{\natexlab{a}})Buras, Janka, Rampp, \&
  Kifonidis}]{Buras:2006hl}
Buras, R., Janka, H.-T., Rampp, M., \& Kifonidis, K. 2006{\natexlab{a}}, \aap,
  457, 281

\bibitem[{Buras {et~al.}(2006{\natexlab{b}})Buras, Rampp, Janka, \&
  Kifonidis}]{Buras:2006dl}
Buras, R., Rampp, M., Janka, H.-T., \& Kifonidis, K. 2006{\natexlab{b}}, \aap,
  447, 1049

\bibitem[{Burrows(2013)}]{Burrows:2013hp}
Burrows, A. 2013, Reviews of Modern Physics, 85, 245

\bibitem[{Burrows {et~al.}(2012)Burrows, Dolence, \& Murphy}]{Burrows:2012gc}
Burrows, A., Dolence, J.~C., \& Murphy, J.~W. 2012, \apj, 759, 5

\bibitem[{Burrows \& Goshy(1993)}]{Burrows:1993ft}
Burrows, A., \& Goshy, J. 1993, \apj, 416, L75

\bibitem[{Burrows \& Lattimer(1986)}]{Burrows:1986kx}
Burrows, A., \& Lattimer, J.~M. 1986, \apj, 307, 178

\bibitem[{Colella \& Woodward(1984)}]{Colella:1984cg}
Colella, P., \& Woodward, P.~R. 1984, JCoPh, 54, 174

\bibitem[{Colgate \& White(1966)}]{Colgate:1966cl}
Colgate, S.~A., \& White, R.~H. 1966, \apj, 143, 626

\bibitem[{Couch(2013{\natexlab{a}})}]{Couch:2013fh}
Couch, S.~M. 2013{\natexlab{a}}, \apj, 775, 35

\bibitem[{Couch(2013{\natexlab{b}})}]{Couch:2013df}
------. 2013{\natexlab{b}}, \apj, 765, 29

\bibitem[{Couch {et~al.}(2013)Couch, Graziani, \& Flocke}]{Couch:2013ws}
Couch, S.~M., Graziani, C., \& Flocke, N. 2013, \apj, 778, 181

\bibitem[{Couch \& Ott(2013)}]{Couch:2013tn}
Couch, S.~M., \& Ott, C.~D. 2013, \apjl, 778, L7

\bibitem[{Dolence {et~al.}(2013)Dolence, Burrows, Murphy, \&
  Nordhaus}]{Dolence:2013iw}
Dolence, J.~C., Burrows, A., Murphy, J.~W., \& Nordhaus, J. 2013, \apj, 765,
  110

\bibitem[{Dubey {et~al.}(2009)Dubey, Antypas, Ganapathy, Reid, Riley, Sheeler,
  Siegel, \& Weide}]{Dubey:2009wz}
Dubey, A., Antypas, K., Ganapathy, M.~K., Reid, L.~B., Riley, K., Sheeler, D.,
  Siegel, A., \& Weide, K. 2009, Parallel Computing, 35, 512

\bibitem[{Endeve {et~al.}(2012)Endeve, Cardall, Budiardja, Beck, Bejnood,
  Toedte, Mezzacappa, \& Blondin}]{Endeve:2012ht}
Endeve, E., Cardall, C.~Y., Budiardja, R.~D., Beck, S.~W., Bejnood, A., Toedte,
  R.~J., Mezzacappa, A., \& Blondin, J.~M. 2012, \apj, 751, 26

\bibitem[{Fern{\'a}ndez(2010)}]{Fernandez:2010ko}
Fern{\'a}ndez, R. 2010, \apj, 725, 1563

\bibitem[{Fern{\'a}ndez(2012)}]{Fernandez:2012kg}
------. 2012, \apj, 749, 142

\bibitem[{Fern{\'a}ndez {et~al.}(2013)Fern{\'a}ndez, Mueller, Foglizzo, \&
  Janka}]{Fernandez:2013um}
Fern{\'a}ndez, R., Mueller, B., Foglizzo, T., \& Janka, H.-T. 2013, submitted
  to MNRAS, arXiv:1310.0469

\bibitem[{Foglizzo {et~al.}(2007)Foglizzo, Galletti, Scheck, \&
  Janka}]{Foglizzo:2007cq}
Foglizzo, T., Galletti, P., Scheck, L., \& Janka, H.-T. 2007, \apj, 654, 1006

\bibitem[{Foglizzo {et~al.}(2012)Foglizzo, Masset, Guilet, \&
  Durand}]{Foglizzo:2012kl}
Foglizzo, T., Masset, F., Guilet, J., \& Durand, G. 2012, \prl, 108, 51103

\bibitem[{Foglizzo {et~al.}(2006)Foglizzo, Scheck, \& Janka}]{Foglizzo:2006js}
Foglizzo, T., Scheck, L., \& Janka, H.-T. 2006, \apj, 652, 1436

\bibitem[{Fryxell {et~al.}(2000)Fryxell, Olson, Ricker, Timmes, Zingale, Lamb,
  MacNeice, Rosner, Truran, \& Tufo}]{Fryxell:2000em}
Fryxell, B. {et~al.} 2000, \apjs, 131, 273

\bibitem[{Hanke {et~al.}(2012)Hanke, Marek, M{\"u}ller, \&
  Janka}]{Hanke:2012dx}
Hanke, F., Marek, A., M{\"u}ller, B., \& Janka, H.-T. 2012, \apj, 755, 138

\bibitem[{Hanke {et~al.}(2013)Hanke, M{\"u}ller, Wongwathanarat, Marek, \&
  Janka}]{Hanke:2013kf}
Hanke, F., M{\"u}ller, B., Wongwathanarat, A., Marek, A., \& Janka, H.-T. 2013,
  \apj, 770, 66

\bibitem[{Hempel {et~al.}(2012)Hempel, Fischer, Schaffner-Bielich, \&
  Liebend{\"o}rfer}]{Hempel:2012bh}
Hempel, M., Fischer, T., Schaffner-Bielich, J., \& Liebend{\"o}rfer, M. 2012,
  \apj, 748, 70

\bibitem[{Iwakami {et~al.}(2008)Iwakami, Kotake, Ohnishi, Yamada, \&
  Sawada}]{Iwakami:2008dw}
Iwakami, W., Kotake, K., Ohnishi, N., Yamada, S., \& Sawada, K. 2008, \apj,
  678, 1207

\bibitem[{Janka(2001)}]{Janka:2001fp}
Janka, H.-T. 2001, \aap, 368, 527

\bibitem[{Janka {et~al.}(2012)Janka, Hanke, H{\"u}depohl, Marek, M{\"u}ller, \&
  Obergaulinger}]{Janka:2012cb}
Janka, H.-T., Hanke, F., H{\"u}depohl, L., Marek, A., M{\"u}ller, B., \&
  Obergaulinger, M. 2012, Progress of Theoretical and Experimental Physics,
  2012

\bibitem[{Janka \& Keil(1998)}]{Janka:1998tx}
Janka, H.-T., \& Keil, W. 1998, Supernovae and cosmology, -1, 7

\bibitem[{Janka {et~al.}(2007)Janka, Langanke, Marek, Mart{\'\i}nez-Pinedo, \&
  M{\"u}ller}]{Janka:2007cz}
Janka, H.-T., Langanke, K., Marek, A., Mart{\'\i}nez-Pinedo, G., \& M{\"u}ller,
  B. 2007, \physrep, 442, 38

\bibitem[{Kitaura {et~al.}(2006)Kitaura, Janka, \&
  Hillebrandt}]{Kitaura:2006gm}
Kitaura, F.~S., Janka, H.-T., \& Hillebrandt, W. 2006, \aap, 450, 345

\bibitem[{Kolmogorov(1941)}]{Kolmogorov:1941vt}
Kolmogorov, A. 1941, Doklady Akademiia Nauk SSSR, 30, 301

\bibitem[{Kraichnan(1967)}]{Kraichnan:1967jk}
Kraichnan, R.~H. 1967, Physics of Fluids, 10, 1417

\bibitem[{Lattimer \& Swesty(1991)}]{Lattimer:1991fz}
Lattimer, J.~M., \& Swesty, D.~F. 1991, \nphysa, 535, 331

\bibitem[{Lee(2013)}]{Lee:2013cd}
Lee, D. 2013, JCoPh

\bibitem[{Lee \& Deane(2009)}]{Lee:2009kq}
Lee, D., \& Deane, A.~E. 2009, JCoPh, 228, 952

\bibitem[{Lee {et~al.}(2013)Lee, Tzeferacos, Couch, \& et~al.}]{Lee:2013flash}
Lee, D., Tzeferacos, P., Couch, S.~M., \& et~al. 2013, in preparation

\bibitem[{Liebend{\"o}rfer(2005)}]{Liebendorfer:2005ft}
Liebend{\"o}rfer, M. 2005, \apj, 633, 1042

\bibitem[{Liebend{\"o}rfer {et~al.}(2001)Liebend{\"o}rfer, Mezzacappa,
  Thielemann, Messer, Hix, \& Bruenn}]{Liebendorfer:2001fl}
Liebend{\"o}rfer, M., Mezzacappa, A., Thielemann, F.-K., Messer, O., Hix, W.,
  \& Bruenn, S. 2001, \prd, 63, 103004

\bibitem[{{Lund} \& {Kneller}(2013)}]{Lund:2013dv}
{Lund}, T., \& {Kneller}, J.~P. 2013, \prd, 88, 023008, 1304.6372

\bibitem[{MacNeice {et~al.}(2000)MacNeice, Olson, Mobarry, de~Fainchtein, \&
  Packer}]{MacNeice:2000fc}
MacNeice, P., Olson, K.~M., Mobarry, C., de~Fainchtein, R., \& Packer, C. 2000,
  Computer Physics Communications, 126, 330

\bibitem[{Marek \& Janka(2009)}]{Marek:2009kc}
Marek, A., \& Janka, H.-T. 2009, \apj, 694, 664

\bibitem[{M{\"u}ller {et~al.}(2012{\natexlab{a}})M{\"u}ller, Janka, \&
  Heger}]{Muller:2012kq}
M{\"u}ller, B., Janka, H.-T., \& Heger, A. 2012{\natexlab{a}}, \apj, 761, 72

\bibitem[{M{\"u}ller {et~al.}(2012{\natexlab{b}})M{\"u}ller, Janka, \&
  Marek}]{Muller:2012gd}
M{\"u}ller, B., Janka, H.-T., \& Marek, A. 2012{\natexlab{b}}, \apj, 756, 84

\bibitem[{M{\"u}ller {et~al.}(2013)M{\"u}ller, Janka, \& Marek}]{Muller:2013kz}
------. 2013, \apj, 766, 43

\bibitem[{M{\"u}ller {et~al.}(2012{\natexlab{c}})M{\"u}ller, Janka, \&
  Wongwathanarat}]{Muller:2012bz}
M{\"u}ller, E., Janka, H.-T., \& Wongwathanarat, A. 2012{\natexlab{c}}, \aap,
  537, 63

\bibitem[{Murphy \& Burrows(2008)}]{Murphy:2008ij}
Murphy, J.~W., \& Burrows, A. 2008, \apj, 688, 1159

\bibitem[{Murphy {et~al.}(2013)Murphy, Dolence, \& Burrows}]{Murphy:2013eg}
Murphy, J.~W., Dolence, J.~C., \& Burrows, A. 2013, \apj, 771, 52

\bibitem[{Murphy \& Meakin(2011)}]{Murphy:2011ci}
Murphy, J.~W., \& Meakin, C. 2011, \apj, 742, 74

\bibitem[{Nordhaus {et~al.}(2010)Nordhaus, Burrows, Almgren, \&
  Bell}]{Nordhaus:2010ct}
Nordhaus, J., Burrows, A., Almgren, A., \& Bell, J. 2010, \apj, 720, 694

\bibitem[{Obergaulinger {et~al.}(2009)Obergaulinger, Cerd{\'a}-Dur{\'a}n,
  M{\"u}ller, \& Aloy}]{Obergaulinger:2009fv}
Obergaulinger, M., Cerd{\'a}-Dur{\'a}n, P., M{\"u}ller, E., \& Aloy, M.~A.
  2009, \aap, 498, 241

\bibitem[{O'Connor \& Ott(2010)}]{OConnor:2010bi}
O'Connor, E., \& Ott, C.~D. 2010, CQGra, 27, 114103

\bibitem[{O'Connor \& Ott(2011)}]{OConnor:2011hk}
------. 2011, \apj, 730, 70

\bibitem[{Ott {et~al.}(2013)Ott, Abdikamalov, M{\"o}sta, Haas, Drasco,
  O'Connor, Reisswig, Meakin, \& Schnetter}]{Ott:2013gz}
Ott, C.~D. {et~al.} 2013, \apj, 768, 115

\bibitem[{Ott {et~al.}(2012)Ott, Abdikamalov, O'Connor, Reisswig, Haas, Kalmus,
  Drasco, Burrows, \& Schnetter}]{Ott:2012ib}
------. 2012, \prd, 86, 24026

\bibitem[{Ott {et~al.}(2008)Ott, Burrows, Dessart, \& Livne}]{Ott:2008gn}
Ott, C.~D., Burrows, A., Dessart, L., \& Livne, E. 2008, \apj, 685, 1069

\bibitem[{Pejcha \& Thompson(2012)}]{Pejcha:2012cw}
Pejcha, O., \& Thompson, T.~A. 2012, \apj, 746, 106

\bibitem[{Rosswog \& Liebend{\"o}rfer(2003)}]{Rosswog:2003fu}
Rosswog, S., \& Liebend{\"o}rfer, M. 2003, Monthly Notice of the Royal
  Astronomical Society, 342, 673

\bibitem[{Ruffert {et~al.}(1996)Ruffert, Janka, \& Schaefer}]{Ruffert:1996te}
Ruffert, M., Janka, H.-T., \& Schaefer, G. 1996, \aap, 311, 532

\bibitem[{Sawai {et~al.}(2013)Sawai, Yamada, \& Suzuki}]{Sawai:2013er}
Sawai, H., Yamada, S., \& Suzuki, H. 2013, \apj, 770, L19

\bibitem[{Scheck {et~al.}(2008)Scheck, Janka, Foglizzo, \&
  Kifonidis}]{Scheck:2008ja}
Scheck, L., Janka, H.-T., Foglizzo, T., \& Kifonidis, K. 2008, \aap, 477, 931

\bibitem[{Shen {et~al.}(1998)Shen, Toki, Oyamatsu, \& Sumiyoshi}]{Shen:1998kx}
Shen, H., Toki, H., Oyamatsu, K., \& Sumiyoshi, K. 1998, \nphysa, 637, 435

\bibitem[{Steiner {et~al.}(2013)Steiner, Hempel, \& Fischer}]{Steiner:2013hi}
Steiner, A.~W., Hempel, M., \& Fischer, T. 2013, \apj, 774, 17

\bibitem[{Suwa {et~al.}(2010)Suwa, Kotake, Takiwaki, Whitehouse,
  Liebend{\"o}rfer, \& Sato}]{Suwa:2010wp}
Suwa, Y., Kotake, K., Takiwaki, T., Whitehouse, S.~C., Liebend{\"o}rfer, M., \&
  Sato, K. 2010, \pasj, 62, L49

\bibitem[{{Suwa} {et~al.}(2013){Suwa}, {Takiwaki}, {Kotake}, {Fischer},
  {Liebend{\"o}rfer}, \& {Sato}}]{Suwa:2012ug}
{Suwa}, Y., {Takiwaki}, T., {Kotake}, K., {Fischer}, T., {Liebend{\"o}rfer},
  M., \& {Sato}, K. 2013, \apj, 764, 99, 1206.6101

\bibitem[{Takiwaki {et~al.}(2012)Takiwaki, Kotake, \& Suwa}]{Takiwaki:2012ck}
Takiwaki, T., Kotake, K., \& Suwa, Y. 2012, \apj, 749, 98

\bibitem[{Takiwaki {et~al.}(2013)Takiwaki, Kotake, \& Suwa}]{Takiwaki:2013ui}
------. 2013, arXiv:1308.5755

\bibitem[{Tamborra {et~al.}(2013)Tamborra, Hanke, Mueller, Janka, \&
  Raffelt}]{Tamborra:2013uf}
Tamborra, I., Hanke, F., Mueller, B., Janka, H.-T., \& Raffelt, G. 2013,
  arXiv:1307.7936

\bibitem[{Thompson(2000)}]{Thompson:2000gd}
Thompson, C. 2000, \apj, 534, 915

\bibitem[{Thompson {et~al.}(2005)Thompson, Quataert, \&
  Burrows}]{Thompson:2005iw}
Thompson, T.~A., Quataert, E., \& Burrows, A. 2005, \apj, 620, 861

\bibitem[{Ugliano {et~al.}(2012)Ugliano, Janka, Marek, \&
  Arcones}]{Ugliano:2012cr}
Ugliano, M., Janka, H.-T., Marek, A., \& Arcones, A. 2012, \apj, 757, 69

\bibitem[{Woosley \& Heger(2007)}]{Woosley:2007bd}
Woosley, S., \& Heger, A. 2007, \physrep, 442, 269

\bibitem[{Woosley {et~al.}(2002)Woosley, Heger, \& Weaver}]{Woosley:2002ck}
Woosley, S.~E., Heger, A., \& Weaver, T.~A. 2002, Reviews of Modern Physics,
  74, 1015

\bibitem[{Woosley \& Weaver(1995)}]{Woosley:1995jn}
Woosley, S.~E., \& Weaver, T.~A. 1995, \apjs, 101, 181

\end{thebibliography}

\end{document}